\documentclass[a4paper,12pt]{article}
\pdfoutput=1

\usepackage{amsmath,amssymb,amsfonts}
\usepackage[dvips]{graphicx}
\usepackage{epsfig}

\usepackage{color}
\usepackage{amsmath}
\usepackage{amsfonts}
\usepackage{verbatim}
\usepackage{amssymb}
\usepackage{graphicx}
\usepackage{amsmath,amssymb,calc}
\usepackage{bbm}
\usepackage{setspace}
\usepackage{color}
\usepackage{amsmath}
\usepackage{amsfonts}
\usepackage{verbatim}
\usepackage{amssymb}
\usepackage{graphicx}
\usepackage{array}
\usepackage{epstopdf}
\usepackage{amsmath,amssymb,amsfonts,graphicx}
\usepackage{epsfig}
\usepackage[left=2.4cm,top=3.4cm,right=2.4cm,bottom=3.4cm,bindingoffset=0cm]{geometry}

\newcommand{\beq}{\begin{eqnarray}}
\newcommand{\eeq}{\end{eqnarray}}
\newcommand{\bea}{\begin{eqnarray}}
\newcommand{\eea}{\end{eqnarray}}
\newcommand{\be}{\begin{equation}}
\newcommand{\ee}{\end{equation}}

\def\1{\mathbbm{1}}

\numberwithin{equation}{section}

\begin{document}

\title{
\begin{flushright}\ \vskip -1.5cm {\small {IFUP-TH-2016}}\end{flushright}
\vskip 10pt
\bf{ \Large  String pair production in non homogeneous backgrounds }
\vskip 5pt}
\author{\large S. Bolognesi${}^{a}$, E. Rabinovici${}^b$ and G. Tallarita${}^{c} $\\[20pt]
{\em \normalsize
${}^{a}$Department of Physics � E. Fermi� University of Pisa
and INFN Sezione di Pisa}\\[0pt]
{\em \normalsize
Largo Pontecorvo, 3, Ed. C, 56127 Pisa, Italy}\\[3pt]
{\em \normalsize 
${}^{b}$Racah Institute of Physics, The Hebrew University of Jerusalem, 91904, Israel}\\[3pt]
{\em \normalsize ${}^{c}$ Departamento de Ciencias, Facultad de Artes Liberales, }\\[0pt]
{\em \normalsize
Universidad Adolfo Ib\'a\~nez, Santiago 7941169, Chile.}\\[0pt]
}
\vskip 0pt
\date{March   2016}
\maketitle
\vskip 0pt

\begin{abstract}

We consider string pair production in non homogeneous electric backgrounds. 
We study several particular configurations which can be addressed with the Euclidean world-sheet instanton technique, the analogue of the world-line instanton for particles.
In the first case  the string is suspended between two D-branes in flat space-time, in the second case the string lives in AdS and terminates on one D-brane (this realizes the holographic Schwinger effect). 
In some regions of parameter space the result is well approximated by the known analytical formulas, either the particle pair production in non-homogeneous background or the string pair production in homogeneous background. In other cases we see effects which are intrinsically stringy and related to the non-homogeneity of the background. The pair production is enhanced already for particles in time dependent electric field backgrounds. The string nature enhances this  even further. 
For spacial varying electrical background fields the string pair production is less suppressed than the  rate of  particle pair production.
We discuss in some detail  how the critical field is affected by the non-homogeneity, for both time and space dependent electric field backgrouds. We also comment on what could be an interesting new prediction for the small field limit.
The third case we consider is pair production in holographic confining backgrounds with homogeneous and non-homogeneous fields.

\end{abstract}
\newpage

\section{Introduction}

In QED a constant electric field can trigger non-perturbative pair production of electron-positron pairs out of the vacuum.   This effect becomes significant when the electric field is of order $E \simeq m^2/e$. 
In the hope of observing this effect directly, significant theoretical and experimental effort has been made recently in considering the pair production in non-homogeneous backgrounds, such as the ones produced by strong laser pulses. 
This can alter the pair production probability and in some cases enhance it. This may bring nearer the date of the direct observation of the Schwinger effect which has long eluded us (see for example \cite{Brezin:1970xf,Schutzhold:2008pz,Bulanov:2010ei}).

The Schwinger effect has also been considered in the context of string theory. For small fields the behavior is very similar to what happens in a quantum field theory.  For high fields one enters a stringy regime. In particular, a critical field, $ E_{cr} = T/e $ where $T$ is the string tension, is present  when the barrier for the pair production vanishes  and the vacuum becomes unstable \cite{Fradkin:1985qd,Burgess:1986dw,Bachas:1992bh}. Only few works have addressed the problem of string pair production  in non-homogeneous backgrounds \cite{Durin:2003gj} and many aspects of this phenomenon are yet to be uncovered.

The Schwinger effect has been considered in the context of holography \cite{Semenoff:2011ng}. A quantum field theory at strong coupling can, in some cases, be described by a weakly coupled string theory. 
In the presence of particles there is no known critical field while  in the presence of strings such a field emerges. It was thus  interesting to find out which of these behaviors would be chosen when the system has a holographic description.
In the simplest prototype example of $N=4$ supersymmetric Yang-Mills theory in the Higgs phase, it has been found that charged W bosons are  pair produced by a background electric field.  When the  electric field is large enough the phenomenon  exhibits  typical string features, including the existence of a critical electric field when the barrier for pair production drops to zero. This effect has been subsequently studied in various cases such as a general electro-magnetic field, the presence of a finite temperature, and confining bulk backgrounds \cite{Bolognesi:2012gr,Sato:2013hyw,Kawai:2013xya,Kawai:2015mha,Hashimoto:2014yya,Hashimoto:2014yza,Dietrich:2014ala}

In this paper we  consider the pair production in string theory for non-homogeneous backgrounds. 
Various methods have been used to address this problem in field theory. The case of alternating fields was discussed already in reference \cite{Brezin:1970xf}, they noted both the perturbative channel that opened up and some alterations in the non perturbative
channel. Using some simplifying assumptions they obtained formulas which interpolated between the low field perturbative regime and the high field regime which is dominated by the non perturbative effects.
In this work we consider mainly the high field regime case and we make brief comments on the perturbative regime.  
One method  in particular for dealing with the non perturbative effect in field theory is the world-line instanton technique.
It is very suitable for the generalization to string theory studies. 
This is the method used in this paper. The world-line instanton method, introduced in \cite{Affleck:1981bma} for the case of pair particle creation in a constant background electric field, has been used in \cite{Dunne:2005sx} for non homogeneous backgrounds where it was found to give the same results found by other methods, such as WKB.  We shall see that a  ``world-line instanton'' technique method can be used in two specific string situations. One is that of a string suspended between two D-branes and the other is for the holographic Schwinger effect. 
Unfortunately we do not have a non-constant electric field background  for which the world-sheet instanton can be analytically solved. We restrict our discussion here to one-dimensional, temporal or spatial, electric backgrounds and  provide numerical solutions in some accessible regions of the parameter space.

There are two overall features involved in this phenomenon. One is the rate of pair production and the other is the value of the critical electric field. One could expect to find that the first is altered by the inhomogeneity while the second could well remain unchanged. The Born-Infeld action is set up ab-initio for non-homogeneous (although slowly varying) electro-magnetic fields.

The paper is organized as follows. In Section \ref{due} we give a brief review of the world-line instanton technique for particle pair production. In Section \ref{tre} we discuss pair production for strings suspended between D-branes in flat space-time. In Section \ref{quattro} we discuss the holographic Schwinger effect for non-homogeneous electric field  and  also the generalization to a confining background.  We conclude in Section \ref{conclusion}.

\section{Particle pair production}
\label{due}

We now review the main results of the world-line instanton technique applied to the study of the particle pair production \cite{Affleck:1981bma,Dunne:2005sx}. The particle production probability in this method is  obtained from a semi-classical expansion method. At leading order the probability of pair production is given by
\beq
P \propto \exp{\left( -S_E \right)}
\eeq
where $S_E$ the Euclidean action evaluated on its stationary solution.

One searches for a stationary solution of the Euclidean action. For a particle the action is the world-line invariant length plus the Euclidean electromagnetic coupling
\beq
S_E = m \int d\tau \sqrt{\dot{x}^{\mu}\dot{x}^{\mu}}  + i q \int d\tau  \dot{x}^{\mu} A^{\mu} \ .
\eeq
One parameterizes the world-line loop by $\tau$ in the range $(0,1)$. The values of the fields at $\tau=0,1$ coincide. 
The  equation of motion is
\beq
\frac{m\ddot{x}^{\mu} }{2 \sqrt{\dot{x}^{\mu}\dot{x}^{\mu}} } = i q F_{\mu\nu}\dot{x}^{\nu}\ ,
\eeq
and the velocity modulus is conserved
\beq
\label{constderqua}
\dot{x}^{\mu}\dot{x}^{\mu} = {\rm const} = L^2 \ ,
\eeq
where $L$ is the total length of the loop in the physical space. 
This conservation is valid for any background field. 
The equation is then reduced to 
\beq
\frac{m\ddot{x}^{\mu} }{ L } = i q F_{\mu\nu}\dot{x}^{\nu} \ .
\eeq
This is the same equation as that of a particle  moving in a constant Euclidean magnetic field $F_{\mu\nu}$.

For a constant field the solution is given by a circular trajectory
\beq
x_3(\tau) =\frac{m}{ q E} \cos{(2\pi\tau)}  \qquad x_4(\tau) = \frac{m}{ q E} \sin{(2\pi \tau)} \ ,
\eeq
for which  the action is
\beq
S_E = \frac{\pi m^2}{q E} \ .
\eeq
This is indeed the local maximum of the action evaluated on generic circle of radius $R$:
\beq
S_E = m 2 \pi R - q E \pi R^2  \ .
\eeq

The non-homogeneous background we consider first is that of a  single pulse which depends only on time
\beq
\label{realpulse1}
E_3(t) = \frac{E}{\cosh^2{(\omega t)}} \ .
\eeq
In the Euclidean formulation this corresponds to a magnetic field in the $34$ plane
\beq
\label{pulse1}
F_{34} = \frac{-i E}{\cos^2{(\omega x_4)}} \ .
\eeq
Note that while the electric field in Minkowski space (\ref{realpulse1}) has its absolute value maximum at $t=0$, its Euclidean counterpart has actually a minimum at $x_4 =0$.  
The absolute value is larger for $x_4 \neq 0$. 
An instanton with non zero size thus experiences an electric field which is on the average bigger then the one at $x_4 = 0$.
One can  thus expect to have an enhancement in the pair production rate relative to the case when the field is constant.

For the non-homogeneous field of the pulse type (\ref{pulse1}) the equations of motion are
\beq 
\label{eqpulse}
\frac{m\ddot{x_3} }{ L } = -\frac{ q E }{ \cos^2{(\omega x_4)}}\dot{x_4}  \nonumber \\ 
\frac{m\ddot{x_4} }{ L } = \frac{ q E }{ \cos^2{(\omega x_4)}}  \dot{x_3} \ .
\eeq
We  discuss first some qualitative aspects of the solution.
Note that the Euclidean field (\ref{pulse1}) diverges at a finite value of the Euclidean time
\beq
\label{div}
x_4 = \pm \frac{\pi}{2 \omega} \ .
\eeq 
So any particle trajectory approaching this Euclidean time experiences a very large field $F_{34}$ and thus it will have a high radius of curvature. 
The value (\ref{div})  depends only on $\omega$ and does  not depend on $E$. 
Decreasing the field $E$ causes the particle trajectory to become flat, but at the same time infinitely curved at (\ref{div}). 
The result is that in the $E \to 0$ limit the trajectory has to become very thin in $x_3$ with the maximum thickness in $x_4$ given by (\ref{div}).

The explicit analytic solution of equations (\ref{eqpulse}) was found in \cite{Dunne:2005sx} and is 
\bea
\label{looppulse}
x_3(\tau) &=& \frac{1}{\omega} \frac{1}{ \sqrt{1+ \gamma^2}}\,  {\rm arcsinh}{\left( \gamma \cos{(2\pi\tau)} \right) } \nonumber \\
 x_4(\tau) &=& \frac{1}{\omega}  \arcsin{ \left(\frac{\gamma}{ \sqrt{1+ \gamma^2}} \sin{(2\pi \tau)}\right)} \ ,
\eea
where 
\beq
\gamma = \frac{m \omega}{q E} \ .
\eeq
The corresponding action is 
\beq
S_E =\frac{2 \pi m^2}{ q E \left(1 + \sqrt{1+\gamma^2} \right)}  
\eeq
The solutions have different characteristics for  small and large values of $\gamma$
\beq
S_E \simeq  \frac{\pi m^2}{q E}  \qquad {\rm for} \qquad \gamma \ll 1 \ , \nonumber \\
S_E \simeq \frac{2 m \pi}{ \omega} \qquad {\rm for} \qquad \gamma \gg 1 \ . 
\eeq
In the small $ \gamma $ regime the instanton is approximatively a circle and its size is very small compared to the scale $1/\omega$ at which the electric field is changing.  There is no significant variation in $F_{34}$ and over this small size the solution is  well approximated by a circular trajectory obtained with a locally constant approximation. 
In the  large $\gamma$ regime the instanton is very thin in $x_3$ and of maximal thickness in $x_4$ as predicted by the qualitative analysis.

\begin{figure}[h!t]
\centerline{
\begin{tabular}{ccc}
\epsfxsize=3.6cm \epsfbox{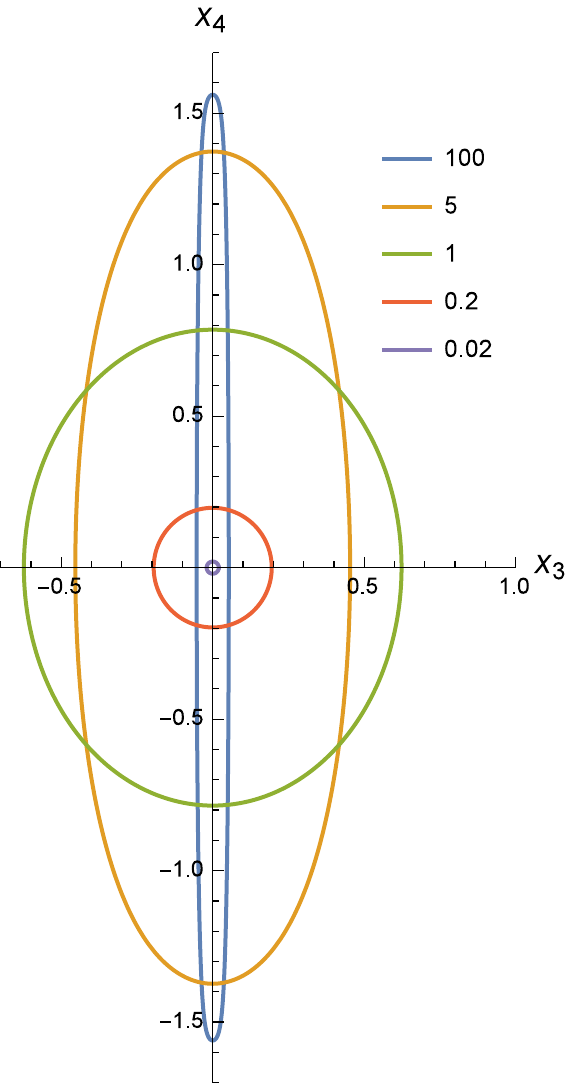}& &
\epsfxsize=8cm \epsfbox{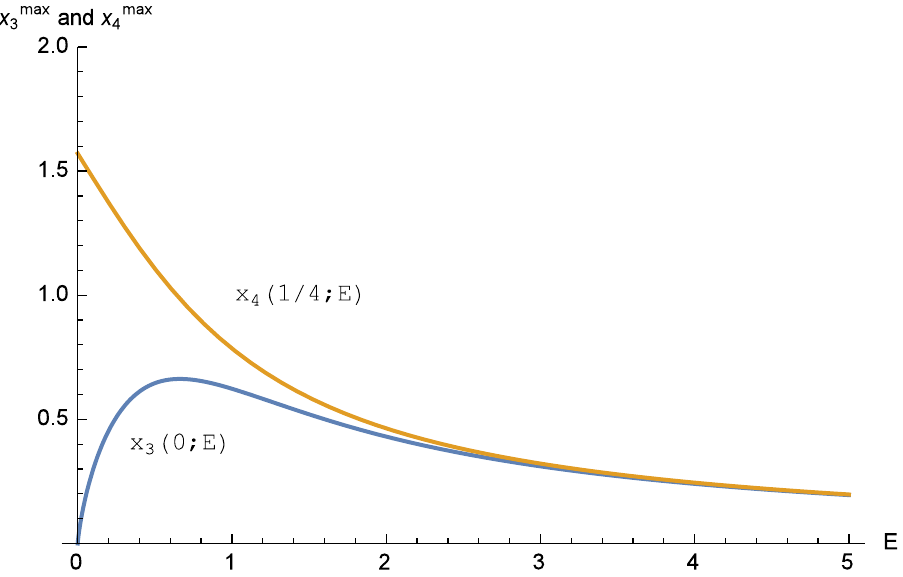}
\end{tabular}}
\caption{{\footnotesize Some particle trajectories (left panel ) and the $x_3$ and $x_4$ maximal values of the trajectories (right panel) for the pulse temporal background  for different values of $E$  are plotted. On the left the values of $E$ are listed in the legend.  We use the normalization $m=1$, $q=1$ and $\omega=1$.}}
\label{plotmax1}
\end{figure}

Some trajectories are plotted for different values of $E$ in Figure \ref{plotmax1} (left panel). Keeping $q$, $\omega$ and $m$ all constant,  one gets a sense of the change of shape and the transition which happens at $E \simeq m \omega /q $. 
We also plot in Figure \ref{plotmax1} (right panel) the maximal values of the trajectory in $x_3$ and $x_4$, namely $x_3^{\rm max} = x_3(0)$ and $x_4^{\rm max} = x_4(1/4)$. We can identify the turning point in the general solution at the place where $x_3^{\rm max}$ reaches its maximum.
Note that in the small $E$ regime the area enclosed in the loop, which  can roughly be estimated as  $A \propto x_3^{\rm max} x_4^{\rm max}$,  is increasing with $E$ while it is  instead  decreasing in the regime   of large $E$.  The area is vanishing in both limits $ E \to \infty$ and $E \to 0$.

\begin{figure}[h!t]
\centerline{
\begin{tabular}{ccc}
\epsfxsize=3.2cm \epsfbox{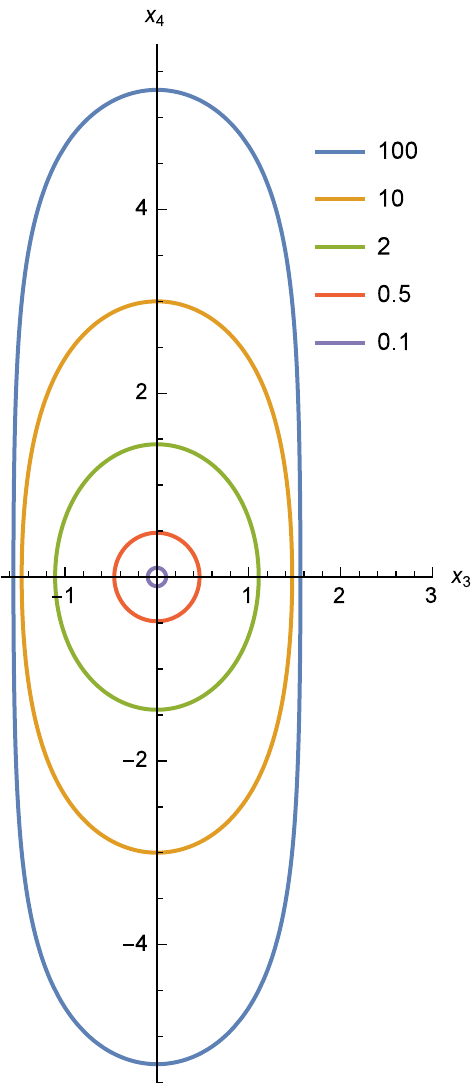} & &
\epsfxsize=8cm \epsfbox{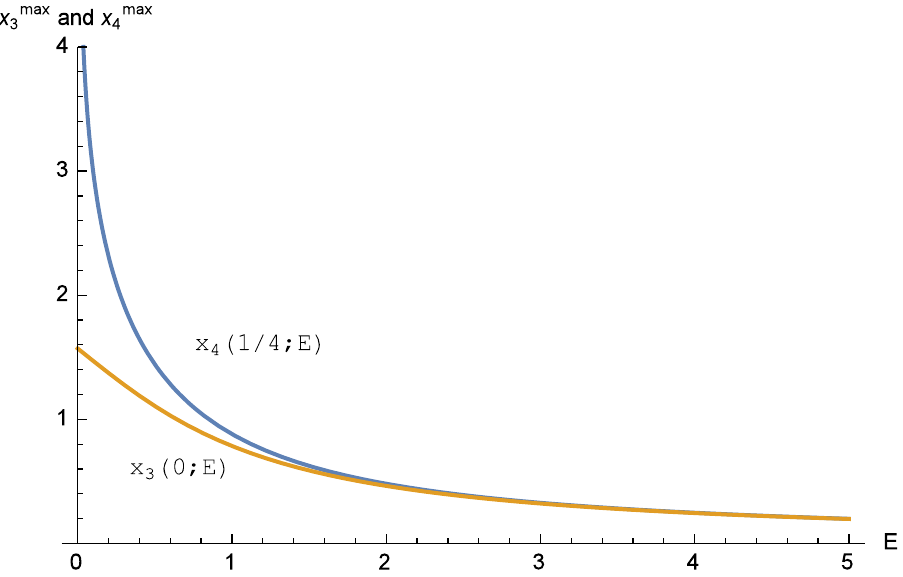}
\end{tabular}}
\caption{{\footnotesize Same as Figure \ref{plotmax1} for the  oscillatory temporal background.}}
\label{plotmax2}
\end{figure}
Another background which admits an analytic solution is the oscillating field with constant frequency
\beq
\label{realpulse}
E_3(t) = E \cos{(\omega t)} \ .
\eeq
In the Euclidean formulation it corresponds to
\beq
\label{oscillatory1}
F_{34} = -i E \cosh{(\omega x_4)} \ .
\eeq
As before for the pulse background, the  Euclidean field  has a minimum at $x_4 = 0$.  A difference with respect to the pulse case is that it does not diverge at any finite value $x_4$.  However the production rate for a given value of electric field is still enhanced.
The solution  is  \cite{Dunne:2005sx} 
\bea
\label{looposcillatory}
x_3(\tau) &=& \frac{1}{\omega} \arcsin \left(\frac{\gamma }{\sqrt{\gamma ^2+1}} \text{cd}\left(4 K\left(\frac{\gamma ^2}{\gamma ^2+1}\right) \tau |\frac{\gamma ^2}{\gamma ^2+1}\right)\right) \nonumber \\
 x_4(\tau) &=& \frac{1}{\omega} {\rm arcsinh} \left(\frac{\gamma}{\sqrt{\gamma ^2+1}}    \text{sd}\left(4 K\left(\frac{\gamma ^2}{\gamma ^2+1}\right) \tau |\frac{\gamma ^2}{\gamma ^2+1}\right)\right)\ .
\eea
Plotting some trajectories for different $E$ and  the maximum of $x_3$ and $x_4$ in Figure \ref{plotmax2} one gets a sense of the change of shape and the transition which happens at $E \simeq \omega  m/q $.  
As for the pulse background,  it becomes an oval shape for small $\gamma \to 0$  but this time $x_3$ goes to a constant while $x_4$ goes to infinity like $\log{(1/E)}$:
\beq
x_3^{\rm max}(\gamma \to \infty) = \frac{\pi}{2 \omega} \qquad x_4^{\rm max}(\gamma \to \infty) = \frac{1}{\omega} \log{\left( 2 \gamma  \right)}\ .
\eeq
Note that the Euclidean field evaluated at the tip of the trajectory goes to a constant value $m \omega / q$.  
The corresponding action is 
\beq
S_E =\frac{4  m^2  \sqrt{1+\gamma^2} }{ q E \gamma^2}  \left(  K\left(\frac{\gamma ^2}{\gamma ^2+1}\right) - E\left(\frac{\gamma ^2}{\gamma ^2+1}\right) \right)
\eeq
The solutions have different characteristics for  small and large values of $\gamma$
\beq
\label{oscaction}
S_E \simeq  \frac{\pi m^2}{q E}  \quad \qquad {\rm for} \qquad \gamma \ll 1 \ , \nonumber \\
S_E \simeq \frac{4 m }{\omega}\log{\gamma} \qquad {\rm for} \qquad \gamma \gg 1 \ . 
\eeq
The pair production probablity, in the large $\gamma$ regime,  is thus
\beq
\Gamma \propto e^{-S_E } \simeq \left( \frac{q E}{m \omega} \right) ^{\frac{4m}{\omega}} \ .
\eeq
This is a pertirbative regime in which $ \Gamma \propto (E^2)^n$ where $n= 2m/\omega$ are the number of photons needed to create the particle pair. The perturbative pair production due to photon absorbtion and the non-perturbative one due to tunneling are continuously connected as $\gamma$ is varied.

We also consider non-homogeneous backgrounds which are space-dependent. For a pulse shape the electric field in Minkowski space is 
\beq
\label{realpulse1space}
E_3(x_3) = \frac{E}{\cosh^2{(k  x_3)}} \ .
\eeq
In the Euclidean formulation this corresponds to a magnetic field in the $34$ plane
\beq
\label{spacepulse1}
F_{34} = \frac{-iE}{\cosh^2{(k x_3)}} \ .
\eeq
Now the Euclidean field has the same dependence on $x_3$ as its Minkowski counterpart and thus it  has actually a maximum at $x_3 =0$.  
An instanton with non zero size thus experiences a field which is in average smaller then the one at $x_4 = 0$.
It is thus natural to expect a suppression in the pair production probability with respect to a locally constant approximation.

\begin{figure}[h!t]
\centerline{
\begin{tabular}{ccc}
\epsfxsize=3.0cm \epsfbox{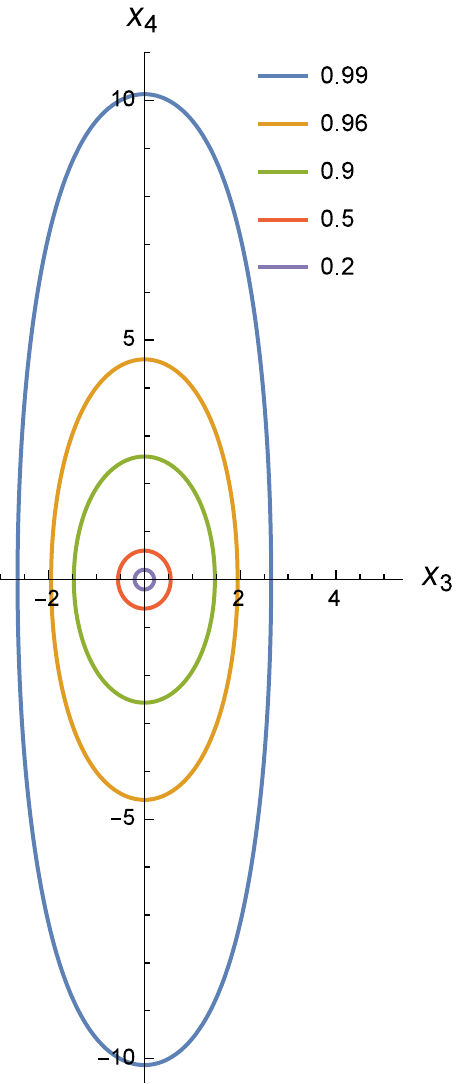} & &
\epsfxsize=8cm \epsfbox{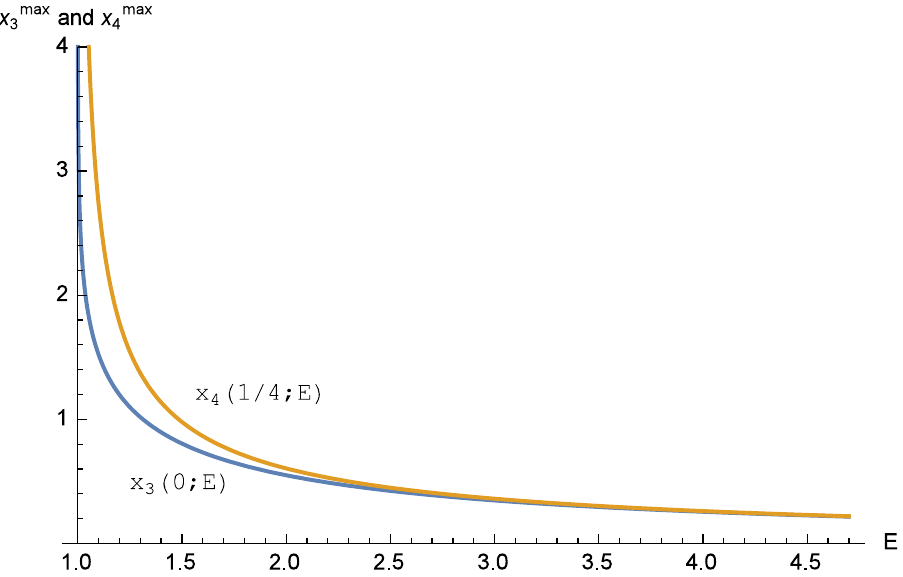}
\end{tabular}}
\caption{{\footnotesize Some trajectories (left) and the $x_3$ and $x_4$ maximal values of the trajectories (right) for the pulse spatial background evaluated for different values of $E$.  We use the normalization $m=1$, $q=1$ and $k=1$.}}
\label{plotspacemax1}
\end{figure}

The explicit analytic solution of equations  \cite{Dunne:2005sx}  is 
\bea
\label{looppulsespatial}
x_3(\tau) &=& \frac{1}{k}\,  {\rm arcsinh}{\left( \frac{\gamma}{ \sqrt{1- \gamma^2}}  \sin{(2\pi\tau)} \right) } \nonumber \\
 x_4(\tau) &=&  \frac{1}{k\sqrt{1- \gamma^2}}  \arcsin{ \left( \gamma \cos{(2\pi \tau)}\right)} \ ,
\eea
where 
\beq
\gamma = \frac{m k}{q E} \ .
\eeq
The corresponding action is 
\beq
S_E =\frac{2\pi m^2}{ q E \left(1 + \sqrt{1-\gamma^2} \right)}  
\eeq
The solutions for small $\gamma$ is well approximated by the trajectory in a constant background. The large $\gamma$ behavior  has  different  characteristics than in the case of the time dependent pulse. First of all there is a lower limit $\gamma \to 1$ where the particle trajectory becomes infinitely large, both in $x_3$ and $x_4$. For $\gamma <1$ there is no closed particle trajectory.   Space dependence is just a tunneling process, 
and it can happen only if $\int_{-\infty}^{+\infty} q E_3(x_3) d x_3 \geq 2m$. This is the reason why no tunneling can happen if $\gamma <1$.\footnote{We thank C. Shubert for an elucidating  comment on this point.}
Some trajectories are plotted in Figure \ref{plotspacemax1}.

\begin{figure}[h!t]
\centerline{
\begin{tabular}{ccc}
\epsfxsize=3.5cm \epsfbox{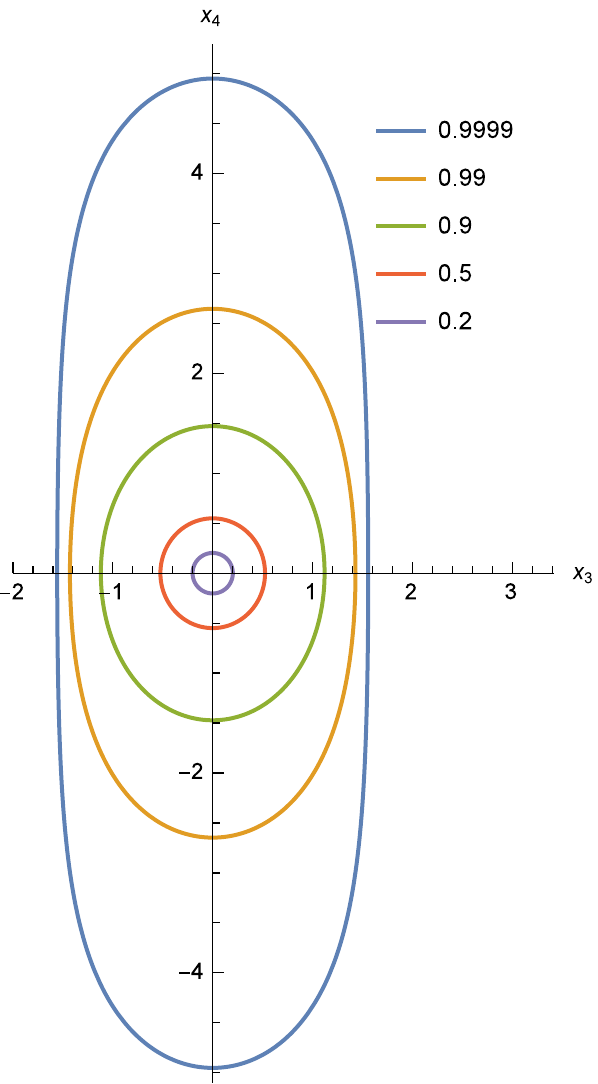} & &
\epsfxsize=8cm \epsfbox{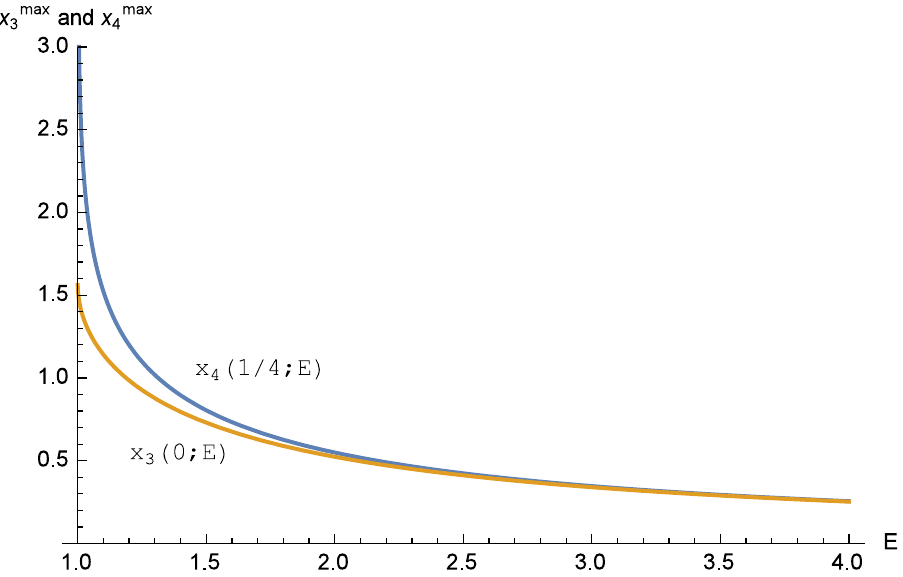}
\end{tabular}}
\caption{{\footnotesize Same as Figure \ref{plotspacemax1} for the  oscillatory spatial background.}}
\label{plotspacemax2}
\end{figure}
Another background which admits an analytic solution is the oscillating field with constant frequency
\beq
\label{realpulse}
E_3(t) = E \cos{(k x_3)} \ .
\eeq
In the Euclidean formulation it corresponds to
\beq
\label{pulse}
F_{34} =  -i E \cos{(k x_3)} \ .
\eeq
As before for the pulse spatial  background, the  Euclidean field has a maximum  at $x_3 = 0$.  
Also in this background field 
we plot some trajectories for different $E$ and  the maximum of $x_3$ and $x_4$ in Figure \ref{plotspacemax2}.   
For  $\gamma \to 1$   $x_3^{\rm max}$ goes to a constant which is exactly the first zero of the electric field $\pi/ 2 k$, while $x_4$ goes to infinity:
\beq
x_3^{\rm max}(\gamma \to 1) = \frac{\pi}{ 2 k} \qquad x_4^{\rm max}(\gamma \to 1) = \frac{1}{2 k} \log{\left( \frac{2 \gamma}{ 1- \gamma }  \right)}\ .
\eeq

\begin{figure}[h!t]
\centerline{
\begin{tabular}{cc}
\epsfxsize=7cm \epsfbox{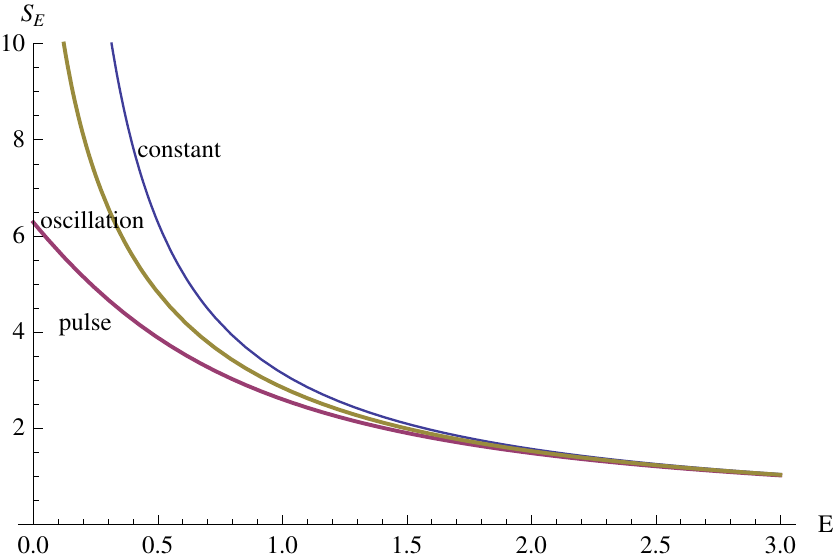}&
\epsfxsize=7cm \epsfbox{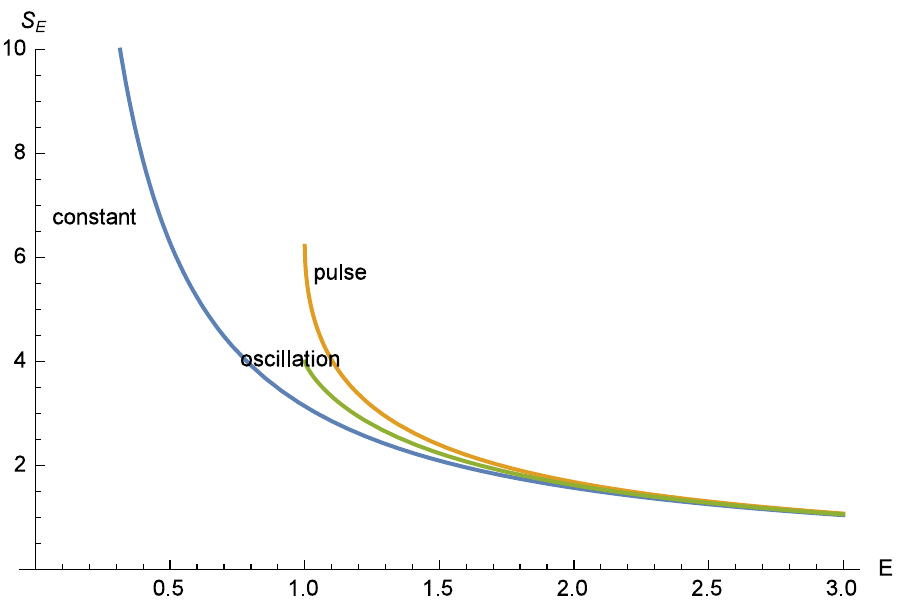}
\end{tabular}}
\caption{{\footnotesize Comparison of pair production probabilities.}}
\label{comparison}
\end{figure}
We now turn to show how these results affects the pair production rate in the different cases. These rates depend on the  actions of the instanton solution. 
A comparison between the actions of the Euclidean solutions as function of $E$ and at fixed $\omega$ or $k$ is given in Figure \ref{comparison}. In the left panel we compare the constant background solution with the pulse and oscillatory temporal backgrounds. Both curves are below the constant background one, indicating an enhancement in the pair production probability $\propto e^{-S_E}$.  In the right panel the pulse and oscillatory spatial backgrounds actions are above the constant background curve, thus indicating suppression in the pair production probability.\footnote{One-dimensional inhomogeneities interpolating between space and time dependence have been discussed in \cite{Ilderton:2015qda}.}
When we say ``enhancement'' or ``suppression'' we are  comparing to a constant background equal to the peak value of the field. 
We may also interpret these results from the ``Minkowskian'' perspective. When there is no time dependence, pair production is just a tunneling event. Since this event is non-local, the pair production is suppressed with respect to a constant field equal to the maximum value. For time dependence instead, we have a mixture between tunneling and energy absorption: the particles can absorb energy from the background field. This is responsible for the enhancement of the pair production.

\section{String pair production: Flat Space}
\label{tre}

\begin{figure}[h!t]
\epsfxsize=6.0cm
\centerline{\epsfbox{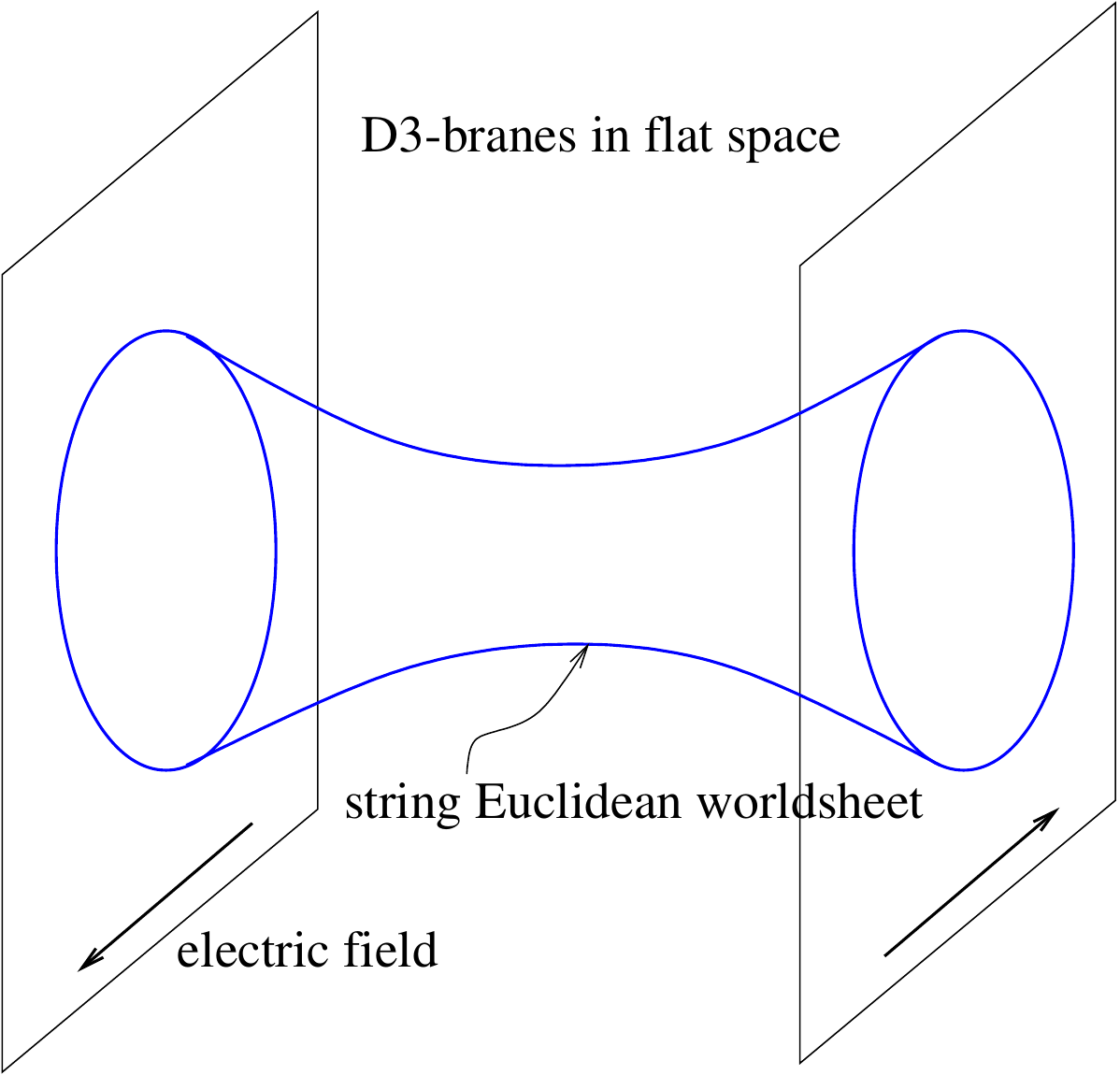}}
\caption{{\footnotesize A string word-sheet instanton stretched between two D-branes.}}
\label{flat}
\end{figure}

String pair production in a constant electric background has been solved in the past thanks to the integrability of the equations of motion. For non homogeneous backgrounds we do not have the same level of analytic control. 
In particular we find it easier to estimate the pair production rate using the instanton method.
We thus want to find cases of string pair production which can treated with the Euclidean instanton technique. For this to be applicable in flat space-time one needs to stretch the string between two branes.
The case we consider in this section is that of a string suspended between two D-branes.
The world-sheet instanton is given in Figure \ref{flat} and has  the topology of a cylinder.
The action in Euclidean formulation is the area spanned by the string, the Euclidean Nambu-Goto action,  plus the boundary interaction with the gauge field
\beq
\label{action}
S = T \int d\sigma d\tau \sqrt{{\rm det} \ g_{2}(\sigma,\tau)}  + i  q \int_{boundary} dX^{\mu} A_{\mu} \ .
\eeq
where $T = 1/l_s^2$ is the string tension.

The two D-branes are at distance $d$ and  the minimal mass for the string state is 
\beq
m = T d \ .
\eeq
On top of this there is a tower of stringy states spaced by $1/l_s$. 
We consider a background field proportional to the generator $(1,-1)$, so that it has opposite field orientation on the two branes. The string also has opposite charges at the two ends so this compensates the change of sign. This configuration is the easiest to discuss because the instanton is symmetric with respect to exchanging the two branes.  We call $q$ the string charge at each of the boundaries so that the total charge of the string state is $2q$.

For a constant electric background field the solution is rotationally symmetric. The minimal surface between two parallel circles is the catenary solution.  We denote by $z$   the coordinate perpendicular to the D-branes  and  by $r(z)$ the radial profile to be determined. The action is
\beq
\label{sphericalcatenaryaction}
S_{E} = T \int_{-d/2}^{d/2} dz 2 \pi r(z) \sqrt{1+r'(z)^2}  - 2 q E \pi R^2 \ .
\eeq
$R$ is the radius of the circle symmetrically traced on each of the D-branes.
The catenary solution, which is the minimal surface for a fixed value of  $R$,  is
\beq
r(z) = \frac{1}{c} \cosh{(c z)}
\eeq
with $c$ determined by the equation
\beq
R = \frac{1}{c} \cosh{(c d/2)}  \ .
\eeq
For  $R$ sufficiently large, this equation gives two solutions for $c$.
These two solutions disappear below a critical radius and after that no solutions can be found. 
One solution is  of a ``thin-neck'' type  and the other is a ``thick-neck'' type.  The thick neck is the one that, in the large $R$ limit,   becomes a  cylinder. In physical terms this corresponds to the particle limit, i.e. a loop of a  particle with mass $m$ and charge $2q$. The thin neck solution is the branch that becomes important close to the critical field.

\begin{figure}[h!t]
\centerline{
\epsfxsize=7.0cm\epsfbox{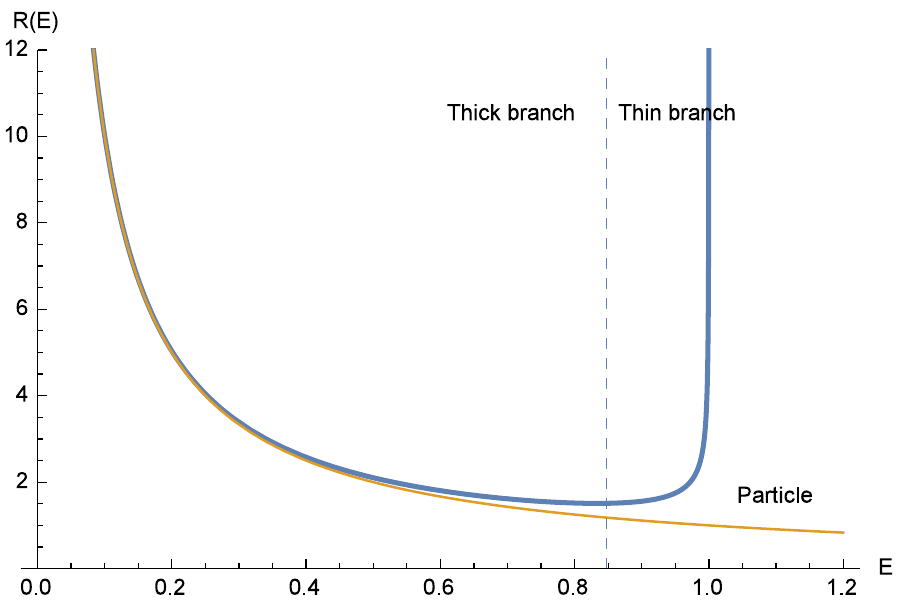} \quad
\epsfxsize=7.0cm\epsfbox{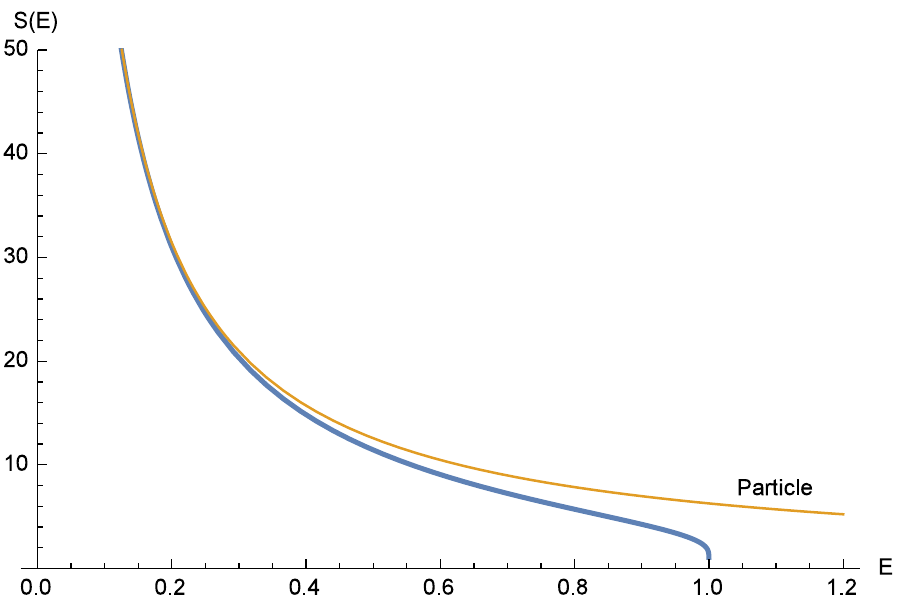}}
\caption{{\footnotesize On the left the radius and on the right the action for the catenary solution that extremises the Euclidean action are compared with the particle limit.
Note that the radius goes to infinity at the critical field and not to zero.
We also plot the corresponding curves for the particle case. Note that the string action is always smaller than the particle action and thus the pair production is always bigger. The normalization of the parameters is $q=T=1$, so that the critical field is $1$, and $d=2$.
}}
\label{cate}
\end{figure}

We can the express everything in terms of the variable $c$. The action evaluated on the catenary solution is 
\beq
S_{E} =\frac{T   \pi  \left(d c+\text{sinh}\left(d c \right)\right)}{ c^2} -\frac{2  \pi  q E \, \text{cosh}^2\left(\frac{d c  }{2}\right)}{c^2 } \ .
\eeq
This must then be extremized as a function of  $c$.  We call $c^*$  the  value of $c$  that extremizes $S_E$. 
The extremization leads  to the following equation
\beq
\label{boundarycircular}
T\, \text{tanh}\left(\frac{d c^*  }{2}\right) = q E
\eeq
which is equivalent to the balance of the two forces at the boundary of the world-sheet: the force due to the magnetic field and the component of the worldsheet tension tangent to the D-brane.
So we have 
\beq
R &=& \frac{q E  d}{2 \epsilon  \sqrt{T^2 -q^2 E^2}} \,\nonumber  \\ 
S_E &=& \frac{d^2 \pi \left( - E q +2 T \epsilon - q E {\rm cosh}\left(2 \epsilon\right)  +T {\rm sinh} \left( 2 \epsilon \right)  \right)}{4 \epsilon^2}
\eeq 
where $\epsilon \equiv {\rm{arctanh}} \left( T/q E \right)$.
We plot the radius $R$ and the instanton action $S_E$ as function of $E$  in Figure \ref{cate} compared with the particle result.

Note that the string action is below the particle one for any value of $E$. This feature will remain true, also in the time and space dependent backgrounds considered below. This implies that the rate of the pair production $\propto e^{-S_E}$ for strings is always bigger than the corresponding particle pair production in the same background. The Euclidean explanation of this effect is the following. The Euclidean action is minimized over all the possible configurations that can ``climb over'' the vacuum and create the pair of particles. The particle action can be reproduced by a string configuration, it is enough to extend the string on an orthogonal surface to the D-branes. The opposite is not true: strings have more degrees of freedom and the minimization can be more effective.

The result in the small field region is well approximated by the particle pair production.  
For larger $E$ we enter in  a stringy regime which has some different features. 
In particular there is a critical field where the action vanishes and has the same value obtained in \cite{Fradkin:1985qd,Burgess:1986dw,Bachas:1992bh}.
Note that the radius $R$ does not go to zero at the critical field, as for example in \cite{Semenoff:2011ng},  but to infinity instead. This is because close to $E_{cr}$, when $R(E)$ reaches its minimum (see the left panel of Figure \ref{cate}), one needs to switch from the thick-neck to  the thin-neck branch  of solutions. The switch between the two branches of solutions happens at $E \simeq 0.848\,  T/q$ and $ R \simeq 0.755  \, d $; this is the minimal radius for which a minimal surface solution exists.

We now move to the case of a non-homogeneous electric background.
The strategy to solve the problem for a non-homogeneous field background is the following. 
We  have to extremize the original action (\ref{action}). 
The minimal area surface provides a solution to the bulk equation. 
For any given shape of the boundary curve, a discrete number of solutions exist. 
A boundary equation then  fixes uniquely  the shape of the boundary curve.

We work in cylindrical coordinates $(z,r,\theta)$ 
\beq
\label{polar}
r \cos \theta = x_3 \qquad \qquad  r \sin \theta = x_4 \ .
\eeq
A suitable gauge choice for the world-sheet coordinates  is
\beq
\sigma = z \qquad \qquad  2 \pi \tau = \theta \ .
\eeq
The string world-sheet is then parametrized by a single function of two variables $r(z,\theta)$ which we shall denote by $r$ for simplicity. 
The Euclidean action is 
\beq
S_{E} = T \int_{-d/2}^{d/2} dz \int_{0}^{2 \pi} d\theta  \sqrt{r^2(1+ (\partial_z r)^2) +(\partial_{\theta} r)^2 } -i q  \int_{0}^{2 \pi} d\theta  (A_{\theta} + A_r\partial_{\theta} r ) \ ,
\nonumber \\ 
\eeq
where the last boundary term takes into account both boundaries at $z = \pm d/2$.
For the constant background one has
\beq
r=r(z) \qquad A_{\theta} = -i E r^2 \sin^2{\theta} \qquad A_{r} =  i E r \sin{\theta} \cos{\theta}
\eeq
and we recover exactly the action (\ref{sphericalcatenaryaction}).
For a generic time dependent background we have
\beq
 A_{\theta} = -i E f(r  \sin{\theta}) r \sin{\theta} \qquad \qquad A_{r} = i E f(r \sin{\theta}) \cos{\theta} \ .
\eeq
For example the cases of pulse and oscillatory fields are given by
\beq
&&f(x_4) = \frac{\tan{(\omega x_4)}}{\omega} \qquad {\rm pulse} \nonumber  \\
&&f(x_4) =  \frac{\sinh{(\omega x_4)}}{\omega} \qquad {\rm oscillating}
\eeq

The bulk equation is the Euler-Lagrange equation for the minimization of the area which is
\beq
\label{pdeuno}
&&r \, \partial_z^2 r \, (r^2 + (\partial_{\theta}r)^2)
+ r \, \partial_{\theta}^2 r \, (1+ (\partial_{z}r)^2) + \nonumber \\
&& \quad -2\,  r \, \partial_z r \, \partial_{\theta} r\,  \partial_z \partial_{\theta} r
-2 \, (\partial_{\theta} r)^2
- r^2 \, (\partial_{z} r)^2
-r^2 = 0 \ .
\eeq
The boundary term is 
\beq
\label{boundarycondition}
T \frac{r^2 \partial_z r}{ \sqrt{r^2(1+ (\partial_z r)^2) +(\partial_{\theta} r)^2 }} = q E \left(  r - \partial_{\theta} r \sin{\theta} \cos{\theta} \right) f'(r \sin{\theta})  \ .
\eeq
When there is no $\theta$ dependence (\ref{pdeuno}) reduces to the catenary equation
\beq
r \partial_z^2 r -  (\partial_z r)^2 -1 =0 \ ,
\eeq
and for the constant background (\ref{boundarycondition}) reduces to
\beq
T \frac{ \partial_z  r}{ \sqrt{1+ (\partial_z r)^2}} = q E     
\eeq
which is equivalent to (\ref{boundarycircular}).

So one is led to solve the PDE (\ref{pdeuno}) with one function of two variables  $r(z, \theta)$ defined on the cylinder $-d/2 \leq z \leq d/2$, $0 < \theta \leq 2\pi$ and with the boundary condition (\ref{boundarycondition}) at two boundaries $z = \pm d/2$. 
We solve this numerically. The numerical solver involves a pseudo-spectral method nested in a Newton relaxation procedure. The function $r(z,\theta)$ is expanded in a double basis using a standard Fourier series in the $\theta$ coordinate and Chebyshev polynomials in the $z$ direction. The solver iteratively finds the solutions of the linearized equation until convergence. Throughout our numerical procedure we use $T=1$ and $q=1$. All solutions are tested for convergence with increasing number of basis modes. 

\begin{figure}[h!]
\centerline{
\begin{tabular}{ccc}
\epsfxsize=5cm\epsfbox{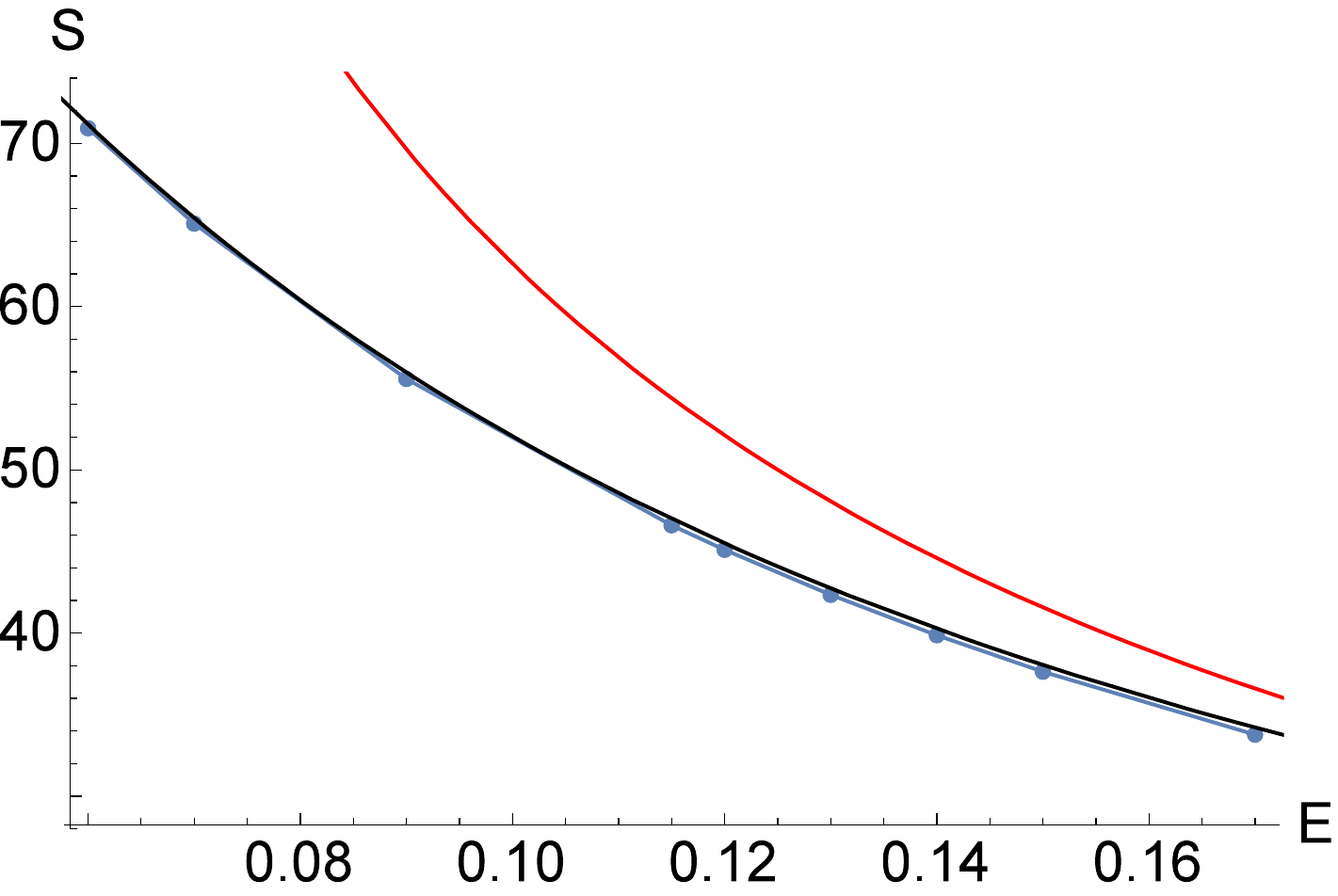}&
\epsfxsize=5cm\epsfbox{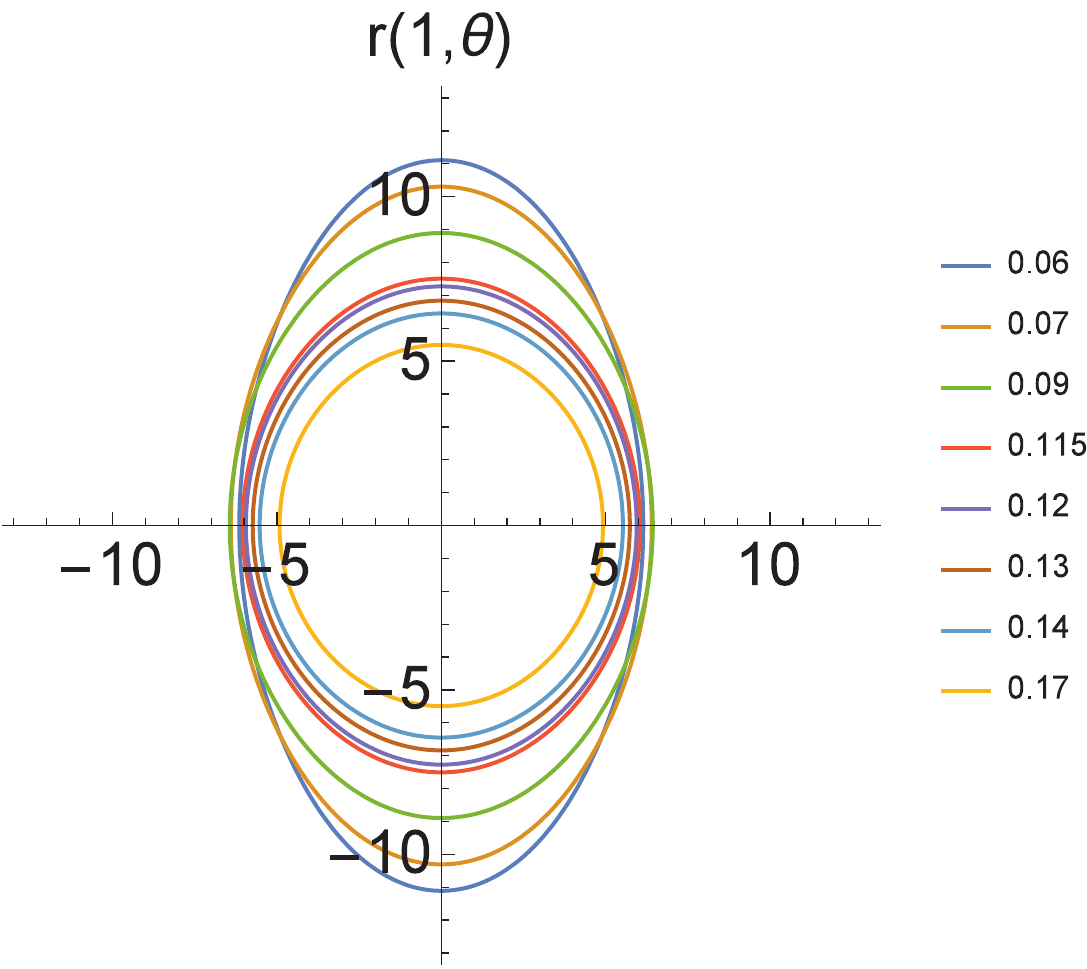}&
\includegraphics[width=36mm]{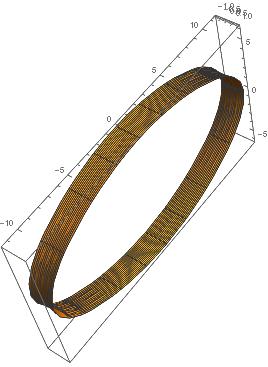}\\
\epsfxsize=5cm\epsfbox{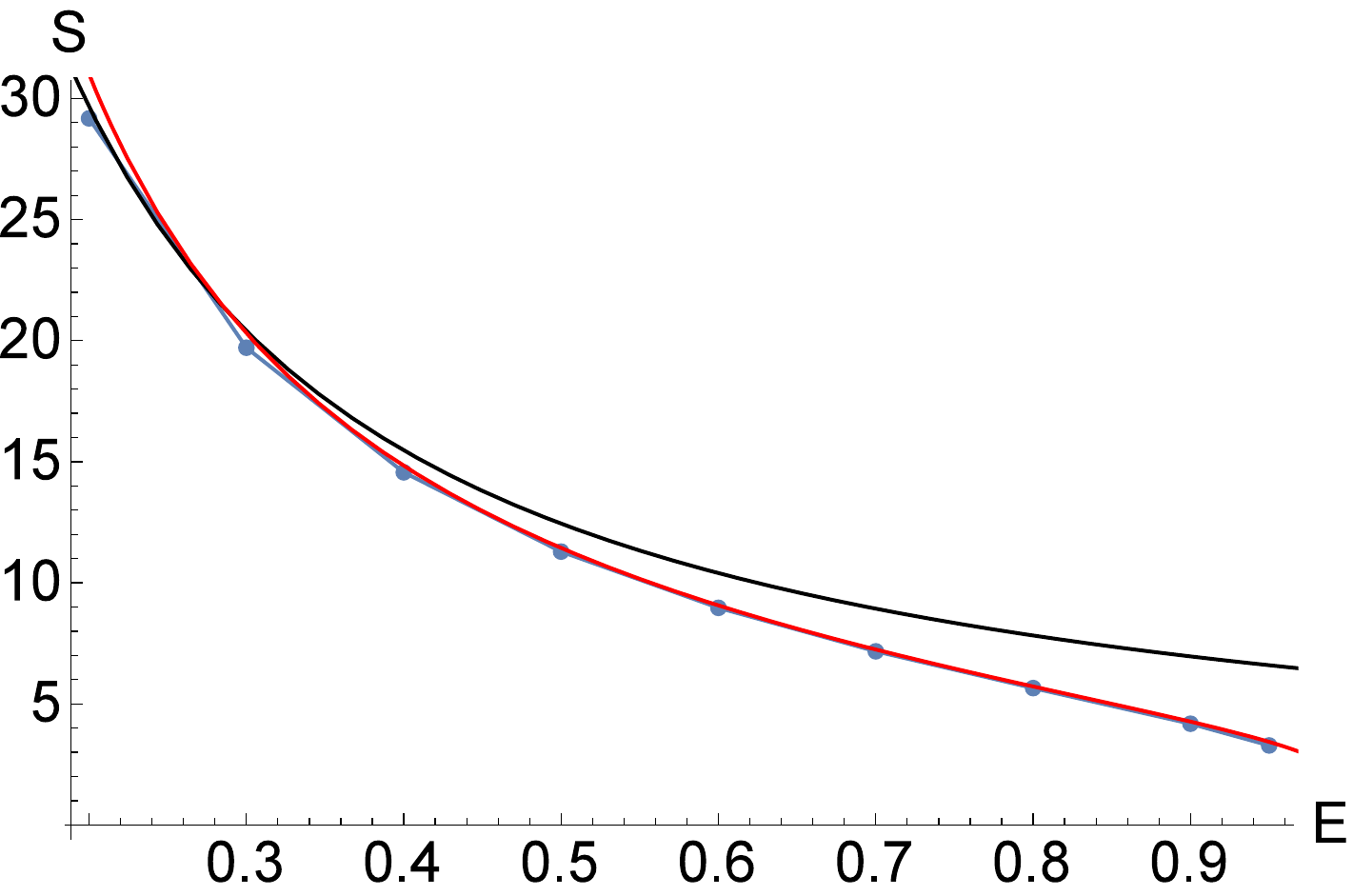}&
\epsfxsize=5cm\epsfbox{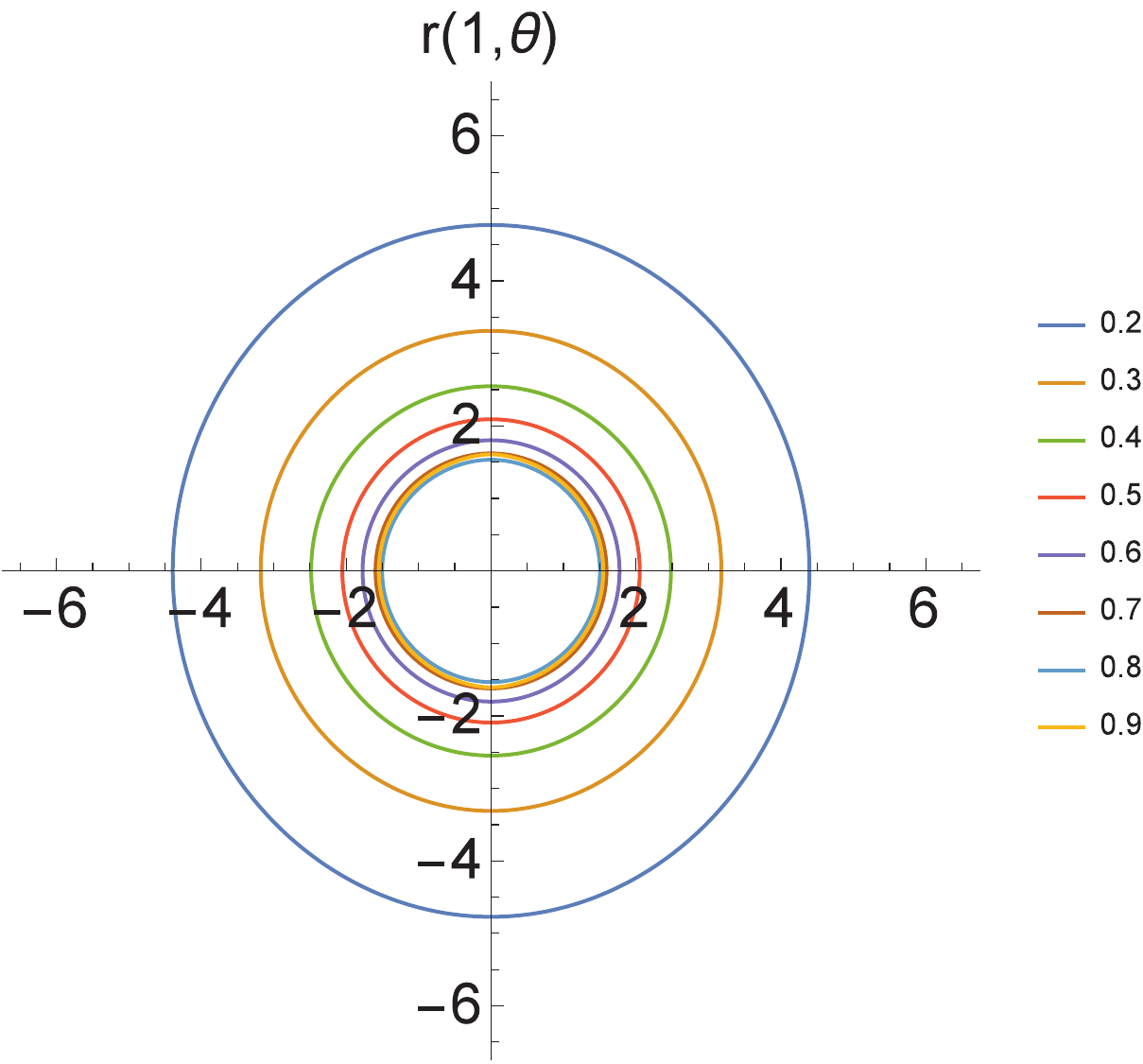}& 
\includegraphics[width=43mm]{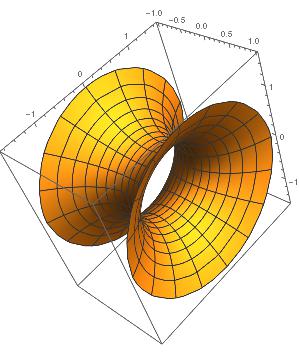}
\end{tabular}}
\caption{{\footnotesize On the left column the action is plotted  for a  pulse background as a function of $E$ at $\omega=0.1$. For this value of $\omega$ the action is almost a perfect interpolation between the two analytic approximations: the action for the catenary solution (red curve) and the pulse solution for particles (black curve). 
On the center column we display  the corresponding boundary on the world-sheet in terms of  a polar plot of $r(z,\theta)$ at $z=1.$ The legend indicates the value of $E$ for the corresponding solution. 
On the top right column represent a revolution plot at $E=0.06$ and on the  bottom at $E=0.9$.}}
\label{cate1}
\end{figure}
\begin{figure}[h!]
\centerline{
\begin{tabular}{ccc}
\epsfxsize=5cm\epsfbox{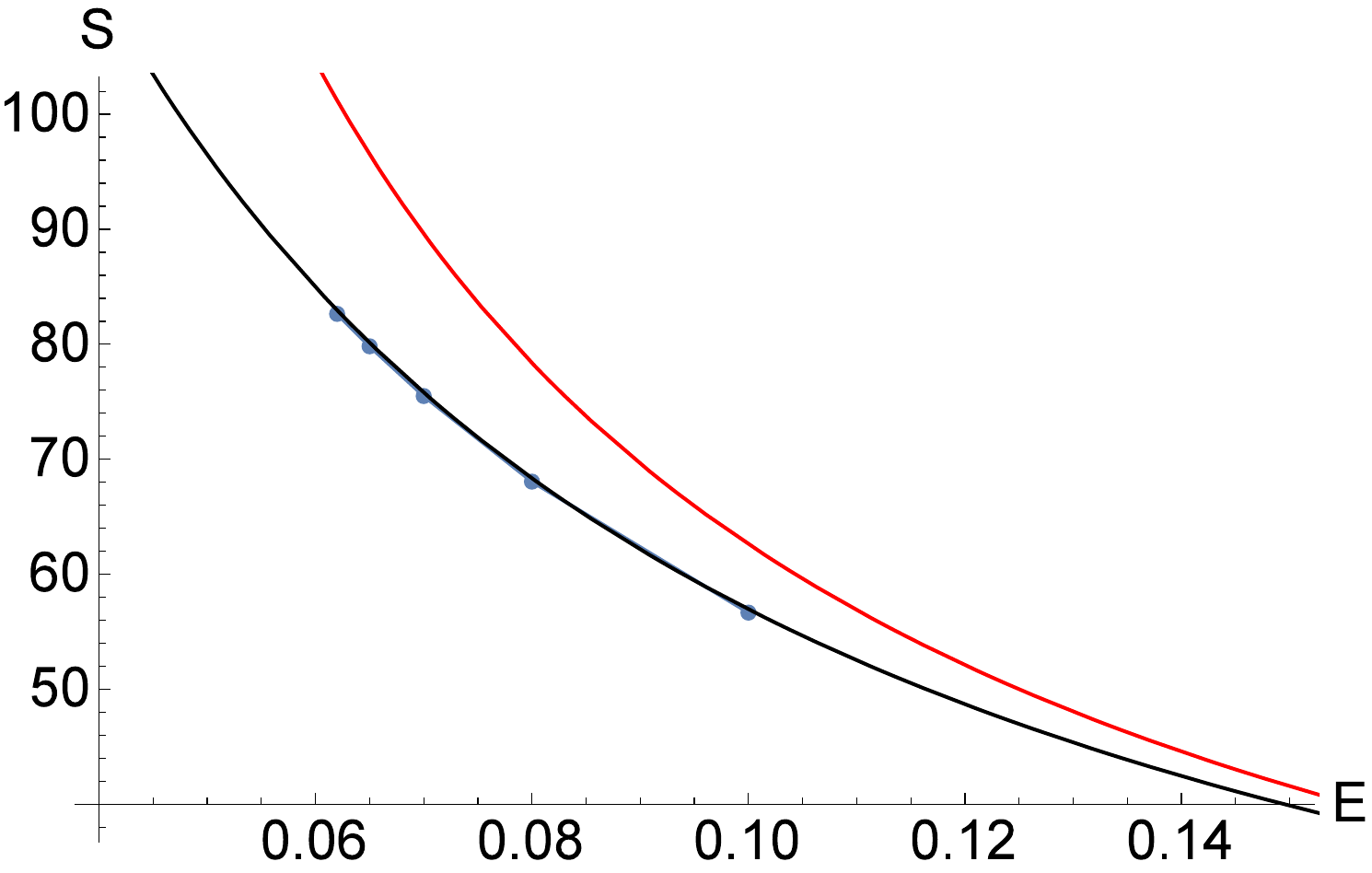}&
\epsfxsize=5cm\epsfbox{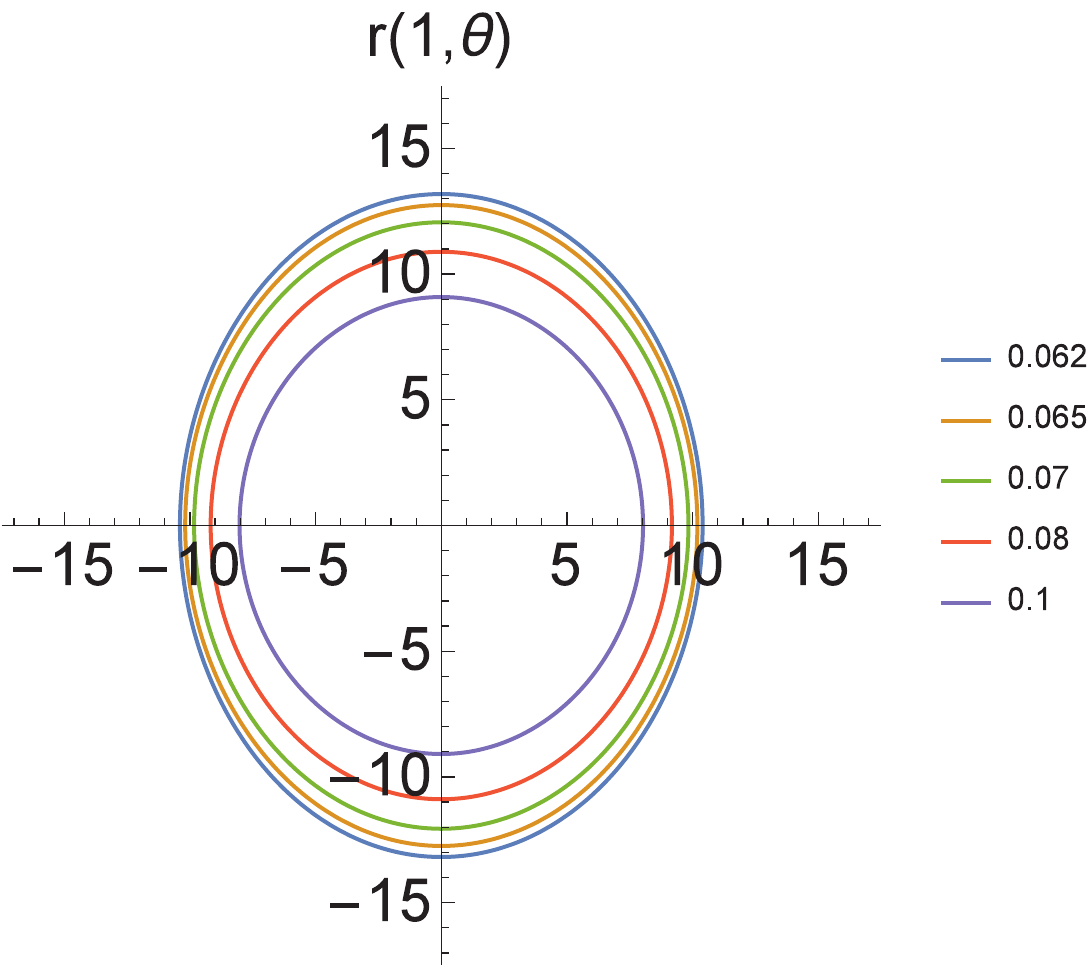}&
\includegraphics[width=33mm]{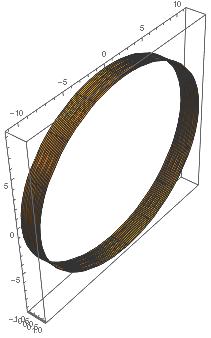}\\
\epsfxsize=5cm\epsfbox{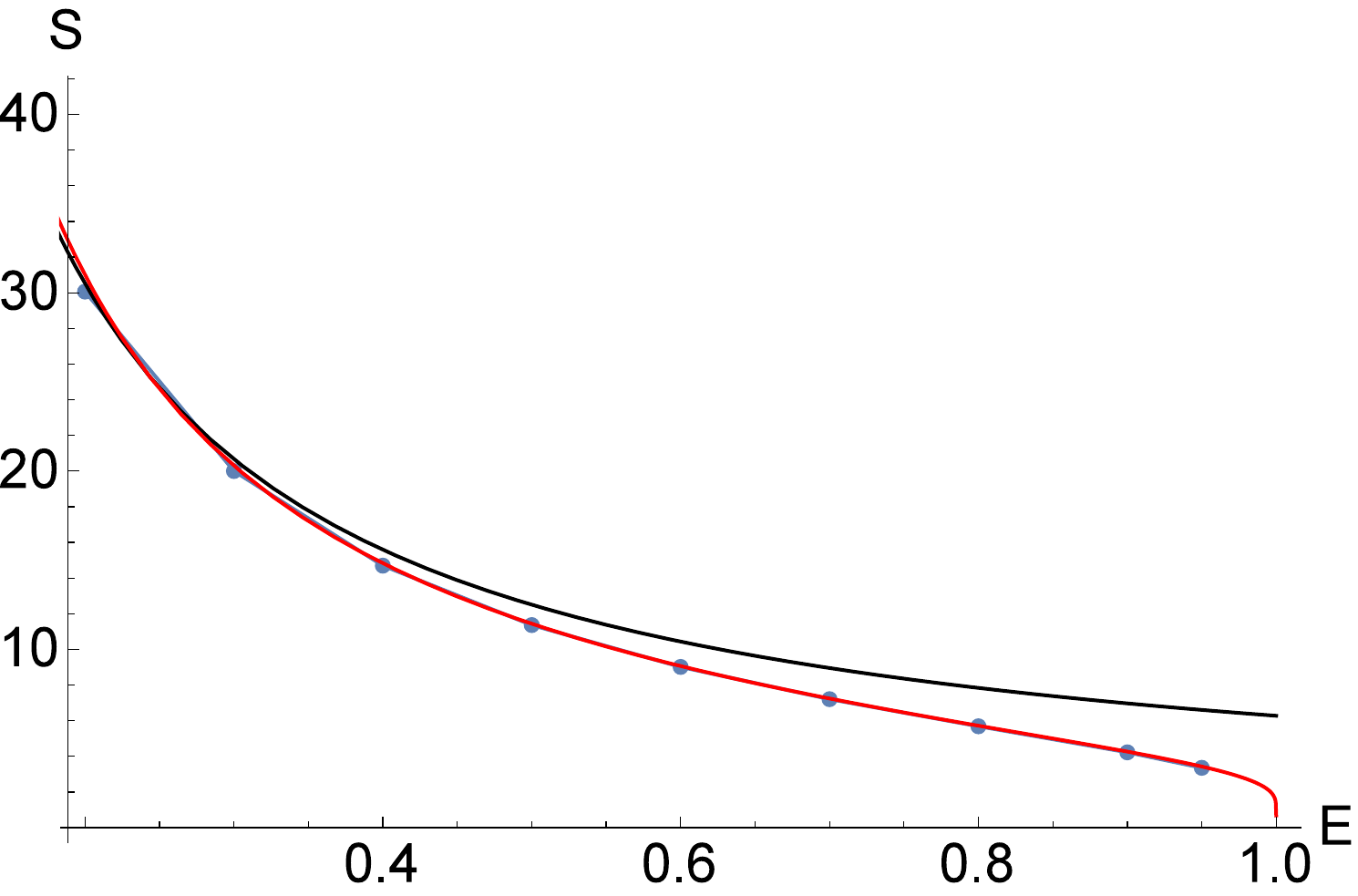}&
\epsfxsize=5cm\epsfbox{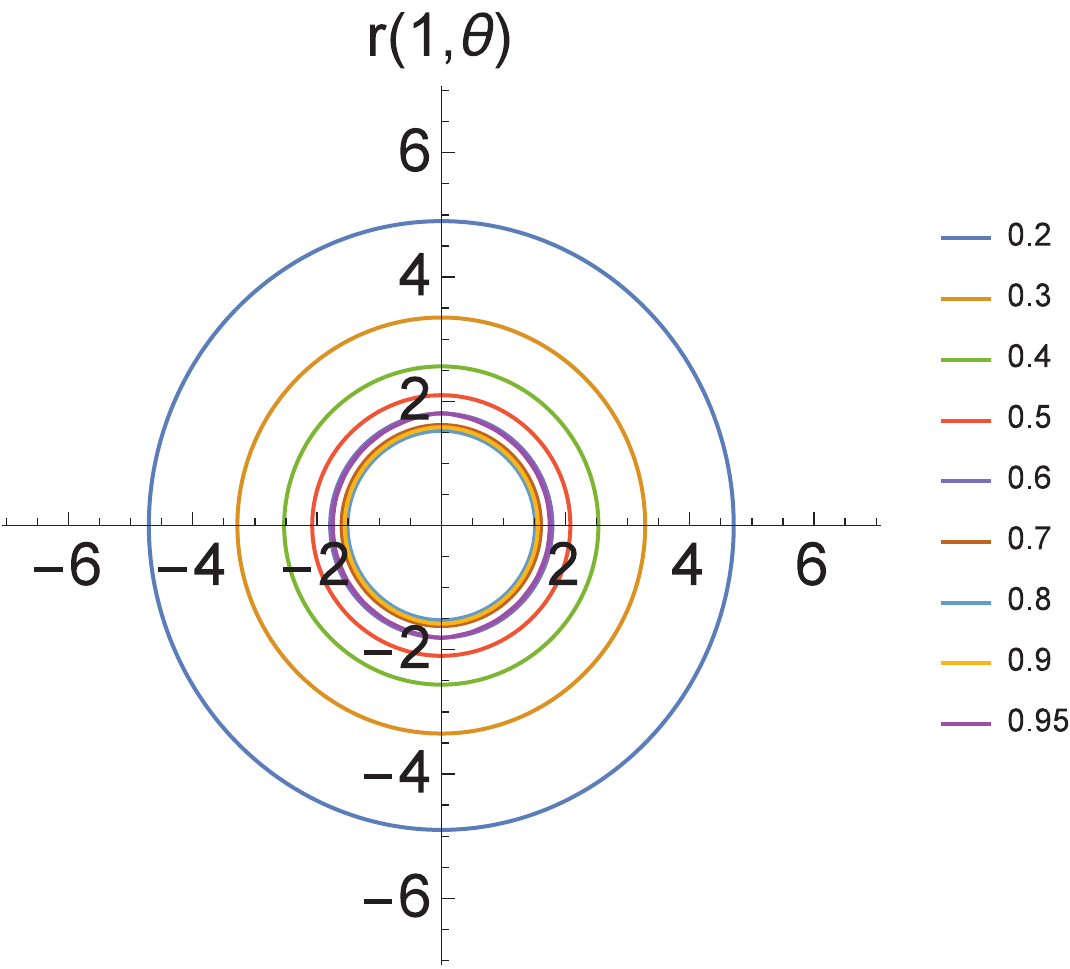}&
\includegraphics[width=42mm]{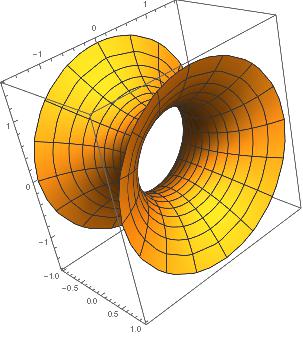}
\end{tabular}}
\caption{{\footnotesize Same as in Figure \ref{cate1} but this time for the oscillatory background.  On the right column on top revolution plot at $E=0.08$ and bottom at $E=0.9$.}}
\label{cate2}
\end{figure}
We start with small $\omega=0.1$ and we find that solutions  interpolate between the analytic solutions representing catenary solutions in the constant background approximation and particle trajectories in the respective non-homogeneous background fields. These solutions, along with their action values, are shown in Figure \ref{cate1} for a pulse background  and Figure \ref{cate2} for an oscillatory background. 
A distinction between the ``high" and ``low" $E$ regimes is made as, by the explanation above, we expect to see the two analytic approximations to be valid in this regime. 
We find convergence for $\omega=0.1$ down to values of $E\approx 0.06$ 
with no remarkable variation on the particle trajectory. Similar behavior is seen 
and up to $E\approx0.9$ for both backgrounds (pulse and oscillatory) 
with no remarkable variation from the catenary solutions. 
Above and below these values the solver loses convergence, a high number of modes in the basis expansions may be needed. 
The particle behavior at small electric field is quite visible also from the revolution pots that show a strip-like shape of the world-sheet instanton. For large $E$ instead the instanton becomes highly pinched in the middle and thus shows a typical string deviation from the particle approximation.

\begin{figure}[t!]
\centerline{
\begin{tabular}{ccc}
\epsfxsize=5cm\epsfbox{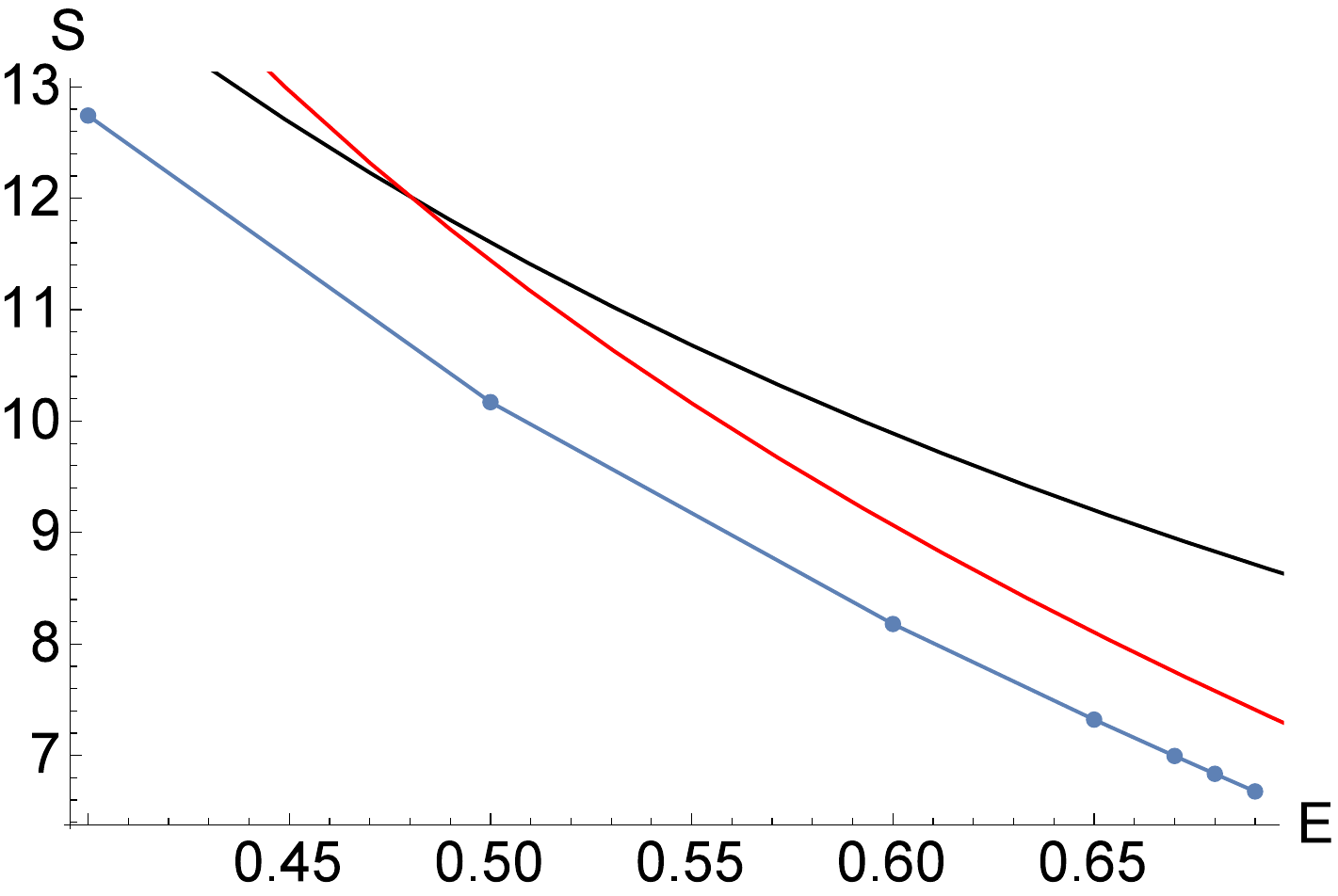}& 
\epsfxsize=5cm\epsfbox{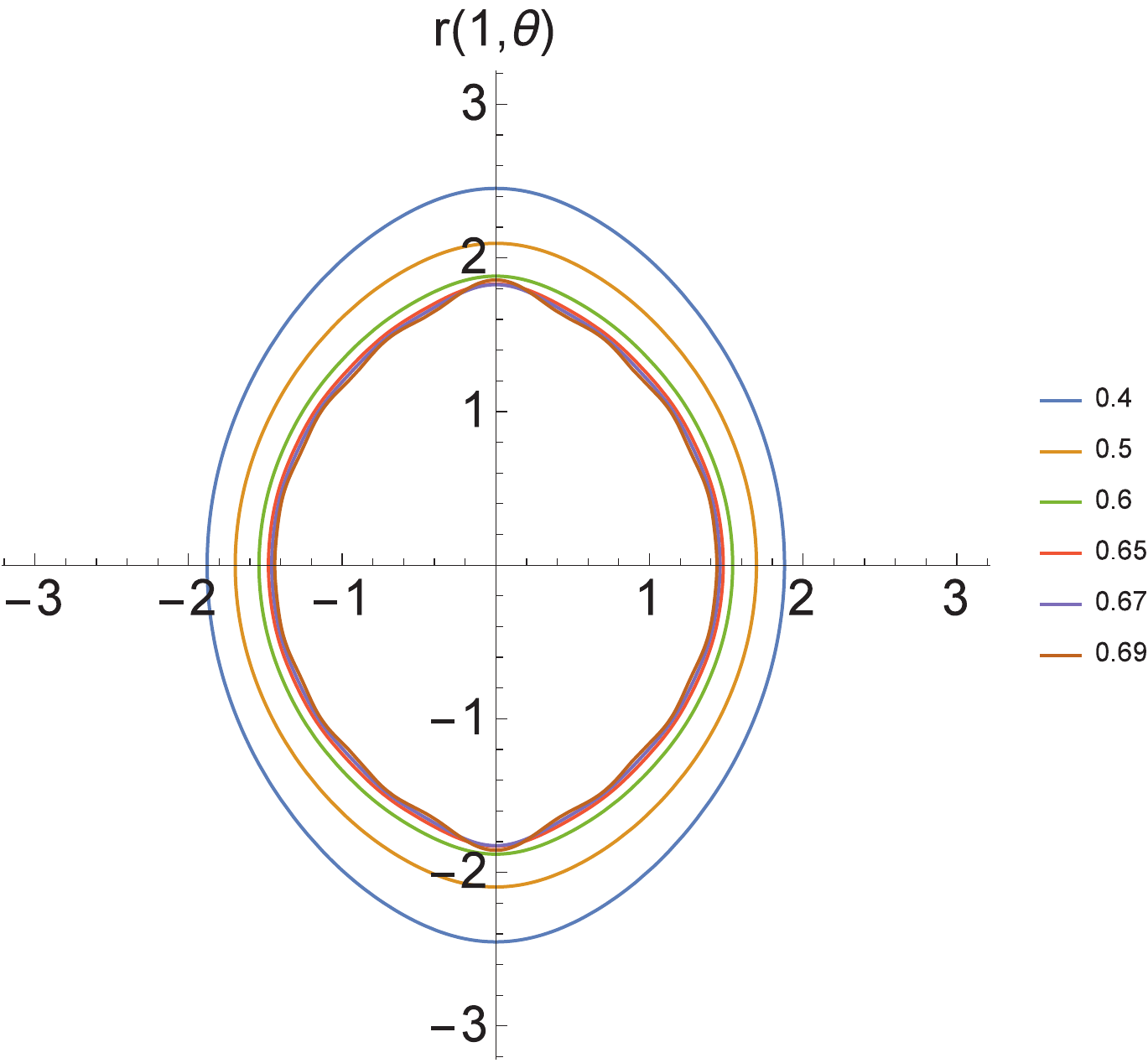}& 
 \includegraphics[width=45mm]{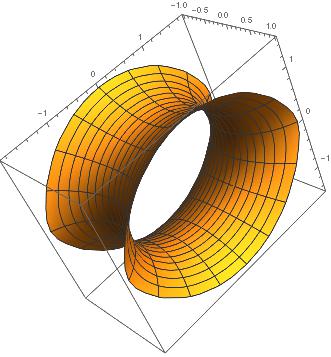}
\end{tabular}}
\caption{{\footnotesize On the left is plotted the action for a pulse background at $\omega=0.3$. At higher $\omega$ the action values are further from the analytic solutions.  On the right we display polar plots of the profiles $r(z,\theta)$ at $z=1$  for the pulse background at $\omega=0.3$. The legend indicates the value of $E$ for the corresponding solution. Revolution plot at $E=0.65$.}}
\label{cate3}
\end{figure}

\begin{figure}[h!]
\centerline{
\begin{tabular}{ccc}
\epsfxsize=5cm\epsfbox{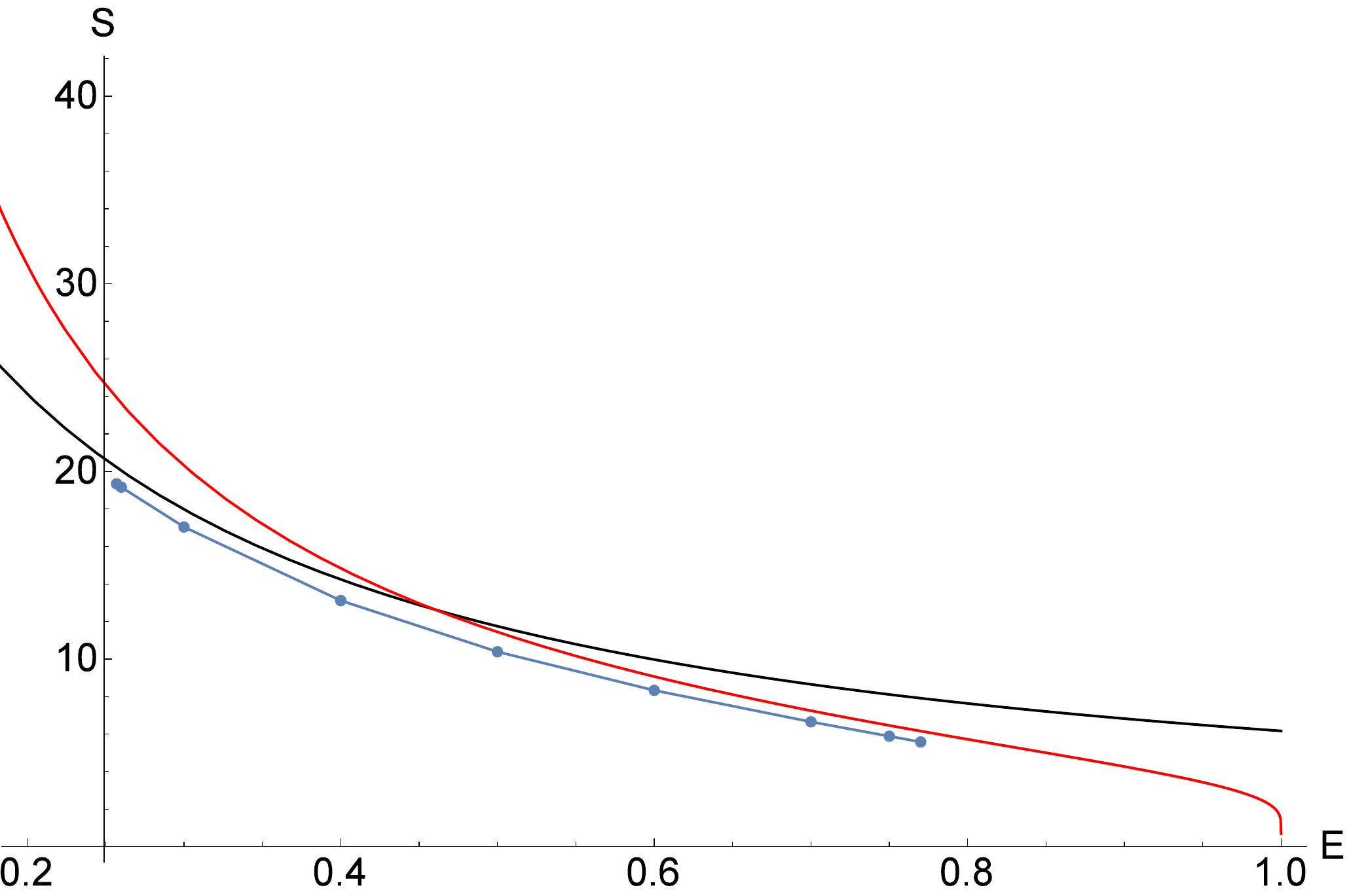}&
\epsfxsize=5cm\epsfbox{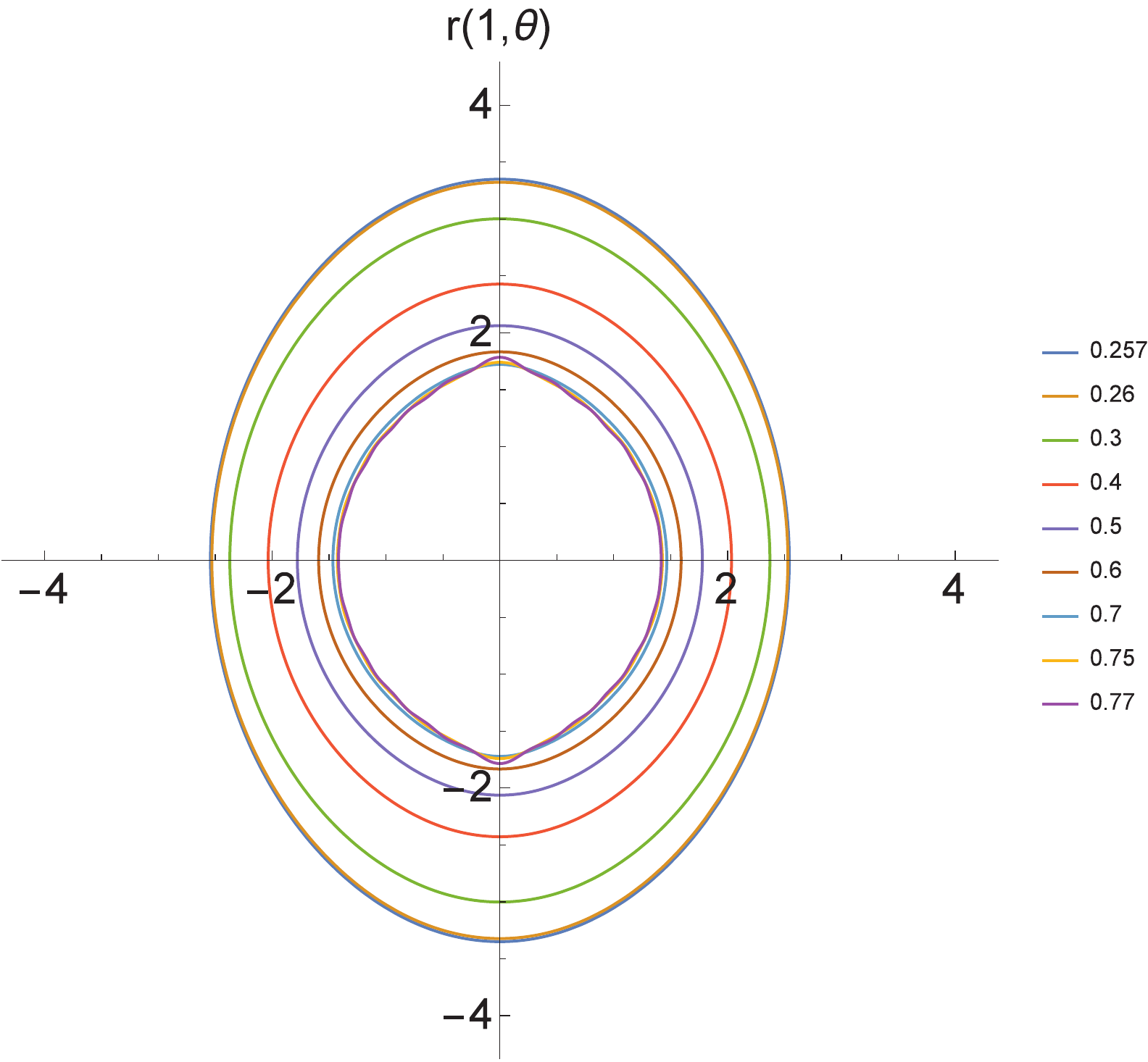}& 
   \includegraphics[width=54mm]{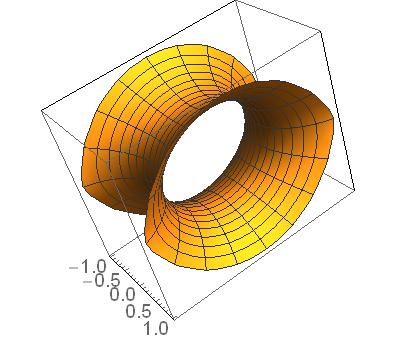}
\end{tabular}}
\caption{{\footnotesize Same of Figure \ref{cate3} for  oscillatory  background at $\omega=0.4$.  On the right a revolution plot the case $E=0.75$.
}}
\label{cate4}
\end{figure}

The previous examples are analyzed for relatively low frequency ($\omega=0.1$) and show the transition from the particle to the string regime. Nevertheless they do not show any effects which are both stringy and related to non-homogeneity.
String effects are expected to be important  when the field is large enough,  of order  the string tension $q E \simeq T$. On the other hand non-homogeneous effects are expected to be important at low-field $ E < m \omega / q$. So if we want to see some effect which is at the same time stringy and related to non-homogeneity the most natural assumption would be to take the frequency large enough so that $T \leq  m \omega$, that is :
\beq
\omega \geq \frac{1}{d} \ .
\eeq
At larger $\omega$ indeed a more significant deviation from the analytic solutions (the particle in the non-homogeneous background and the string the locally constant approximation) is seen.   
In Figures \ref{cate3} and \ref{cate4} we present a plot of the numerical actions for the solutions at $\omega = 0.3$. 
We see from the shape of the revolution plots that the world-sheet instanton has features that are both stringy (the surface is pinched in the middle) and related to the inhomogeneity of the background (the surface has no longer a circular shape).
We see that for both cases the numerical solution for the action is significantly lower than the analytic actions. We thus see that time dependence and string nature work together to enhance the pair production.\footnote{The perturbative regime of particle pair production is valid, according to (\ref{oscaction}),  when $S_E \simeq \frac{4 m }{\omega}\log{\gamma}$. This happens, up to $10\%$ error, when $E \simeq \omega/2$. Exploring lower $E$ regimes exploting  the numerical analysis which led to   Figure \ref{cate4} would allow us to test the perturbative string regime. At the moment, 
with the values of E availiable, it is,  is not possible to disentangle the perturbative and non-perturbative behavior in the  string pair production. }


We find that for high values of $E$ the solutions close to the D-branes develop protuberances around $\theta =\pi/2$. These are visible in the plots of Figures \ref{cate3} and  \ref{cate4} for the highest values of $E$ we are able to reach.
 This effect is due to the non-homogeneity of the background. The Euclidean field for both pulse (\ref{pulse1}) and oscillatory (\ref{oscillatory1}) backgrounds is increasing with the modulus of $x_4$. The electric field on the other hand sets the angle by which the worldsheet surface terminates on the D-brane. So the surface has to be flatter when $|x_4|$ is bigger.

A rather surprising effect emerges in the $E \to 0$ limit for  the pulse background. 
In the $E \to 0$ limit the area enclosed in the particle loop (\ref{looppulse}) goes to zero. 
We estimate the area inside the loop roughly as 
\beq
A_{L} \simeq 4 x_3^{\rm max} x_4^{\rm max} = \frac{2 \pi q E}{m \omega^3 } \log{\left(\frac{2 m \omega}{ q E}\right)} \ .
\eeq
The particle approximation is valid if  the string world-sheet instanton looks like a strip with the boundary exactly equal to the particle trajectory (\ref{looppulse}). This has been confirmed to be the case at least up to the smaller values of $E$ for which we have convergence in our numerical method. But for smaller values of $E$ some deviation is expected otherwise, at a certain point,  
the surface area of the strip would be in general bigger than the area inside the loop. 
In this limit, assuming the particle picture is still a good approximation, the area of the world-sheet instanton would be 
\beq
A_{I} \simeq 4   x_4^{\rm max}  d = \frac{2 \pi m}{ T \omega}
\eeq
So, at a certain low value  $E_{\rm low}$ given by  
\beq
 \frac{ T q E_{\rm low}}{\pi m^2 \omega^2 } \log{\left(\frac{2 m \omega}{q E_{\rm low}}\right)} \simeq 1
\eeq
the area inside the loop $A_{L}$ becomes smaller than the ``supposed'' word-sheet instanton area $A_{I}$.
So we expect that for  $E < E_{\rm low}$ the real solution will not be well approximated by the particle loop. 
We can give a physical explanation for this. The time dependent pulse with a fixed $\omega$, when decomposed in Fourier components, contains all sort of frequencies, also the ones near the string scale. For small enough $E$ the string scale photons are the ones who dominate the pair production process.  
If we consider the specific example of Figure \ref{cate1} this small field would be $E_{\rm low} \simeq 0.0026$ which is smaller than the values we were able to test numerically. We do not have yet numerical evidence for this new effect at low electric field.

We expect a modification close to  critical electric field of the locally constant approximation. 
For any value of $E$ and $\omega \neq 0$ the Euclidean field $F_{34}$ reaches the string scale at a certain value of $x_4$. This is true both for the pulse (\ref{pulse1}) and oscillatory (\ref{oscillatory1}) backgrounds. The value $\bar{x}_4$ at which $q F_{34}(\bar{x}_4) = T$ is
\beq
\bar{x}_4 = \left\{ 
\begin{array}{cc}
  \frac{1}{\omega} \arccos{ \sqrt{\frac{q E}{T}} } & {\rm pulse} \\
 \frac{1}{\omega} {\rm arccosh} \left( \frac{T}{qE} \right)    & {\rm oscillatory}
\end{array}
\right.
\eeq 
The Euclidean world-sheet never exceeds this limit. 
At low frequency $\omega$, in the high-$E$, the locally constant approximation is thus expected to be valid as long as $R <  \bar{x}_4 $ 
\beq
\frac{q E  d}{2 {\rm arctanh}\left( T/q E\right) \sqrt{T^2 -q^2 E^2}}  <  \bar{x}_4 
\eeq
This condition is certainly violated in the proximity of the critical field $E_{\rm cr} = T/q$ due to the fact that $R$, switching to the thin-neck branch of solutions, becomes infinity in the limit  $E  \to E_{\rm cr}$. For $E_{\rm cr}-E \ll E_{\rm cr}$ the condition above becomes
\beq
\frac{d \sqrt{T}}{ \sqrt{E_{\rm cr}-E} \log{\left( E_{\rm cr}-E \right)} }
< 
\left\{ 
\begin{array}{cc}
  \frac{\sqrt{2}\sqrt[4]{E_{\rm cr}-E} }{\omega \sqrt[4]{E_{\rm cr}}} & {\rm pulse} \\
  \frac{\sqrt{2}\sqrt{E_{\rm cr}-E} }{\omega \sqrt{E_{\rm cr}} }   & {\rm oscillatory}
\end{array}   
\right.
\eeq
This condition, both for the pulse and oscillatory backgrounds, is violated when $E$ is sufficiently close to $E_{\rm cr}$. This means that the locally constant approximation close to the critical field is certainly violated, even when $\omega$ is very small. For example in the case studied in the Figure \ref{cate1}, which has relatively small $\omega$ and a big range at large $E$ where the locally constant approximation is valid, this expected deviation would be seen at $E_{\rm cr}-E \ll E_{\rm cr} \simeq 0.003 E_{\rm cr}$ which is outside the region of parameter we are able to test numerically.

\begin{figure}[h!t]
\centerline{
\begin{tabular}{ccc}
\epsfxsize=5cm\epsfbox{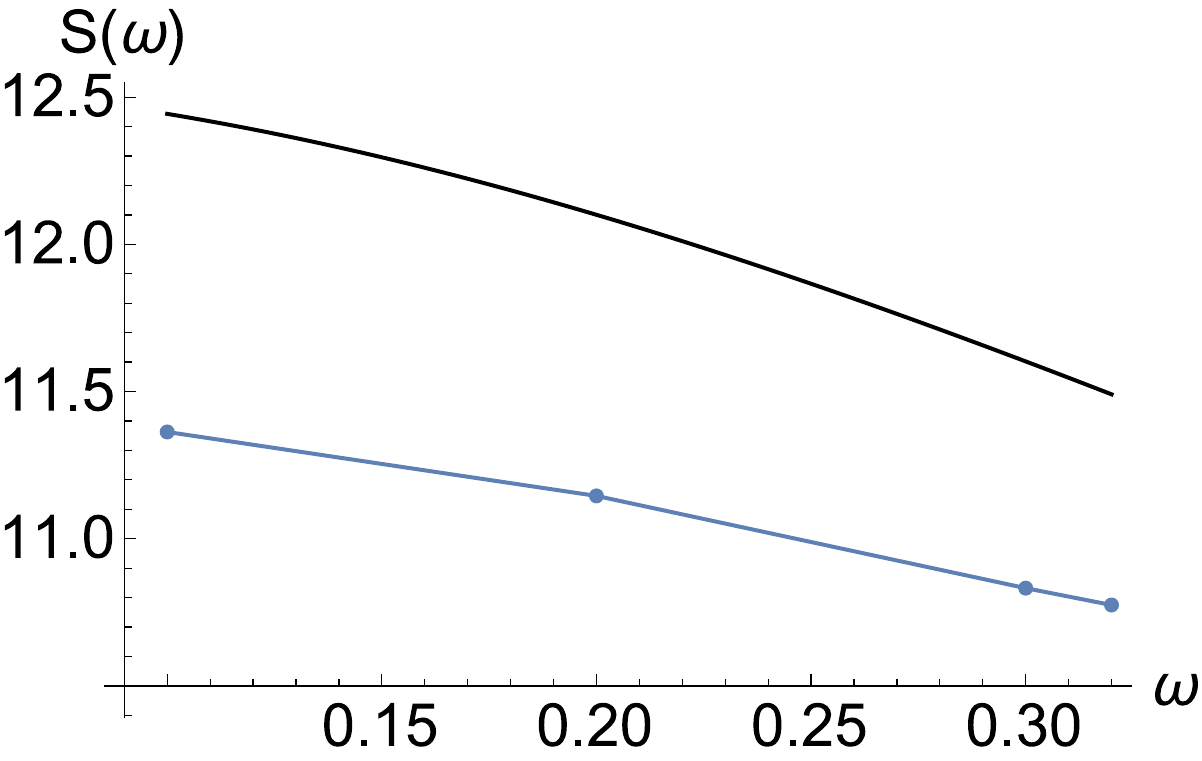}&
\epsfxsize=5cm\epsfbox{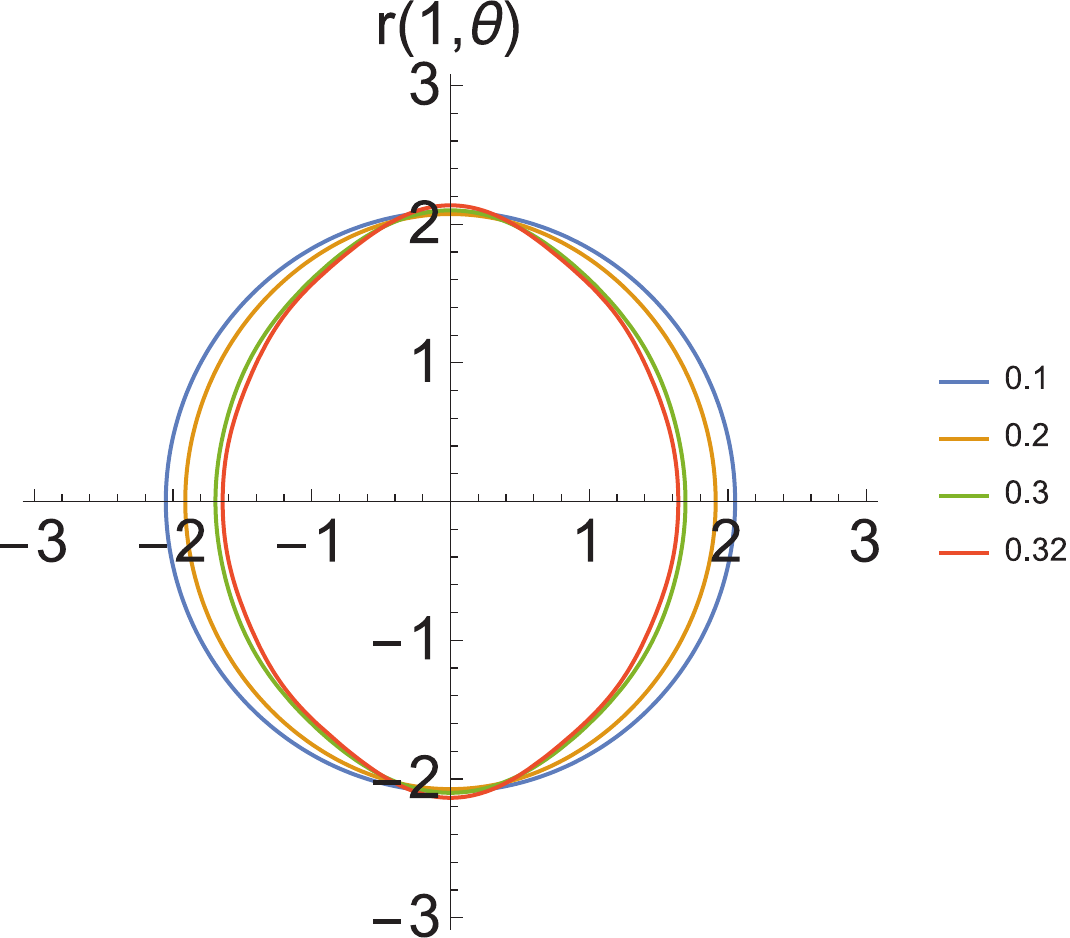}&
\includegraphics[width=43mm]{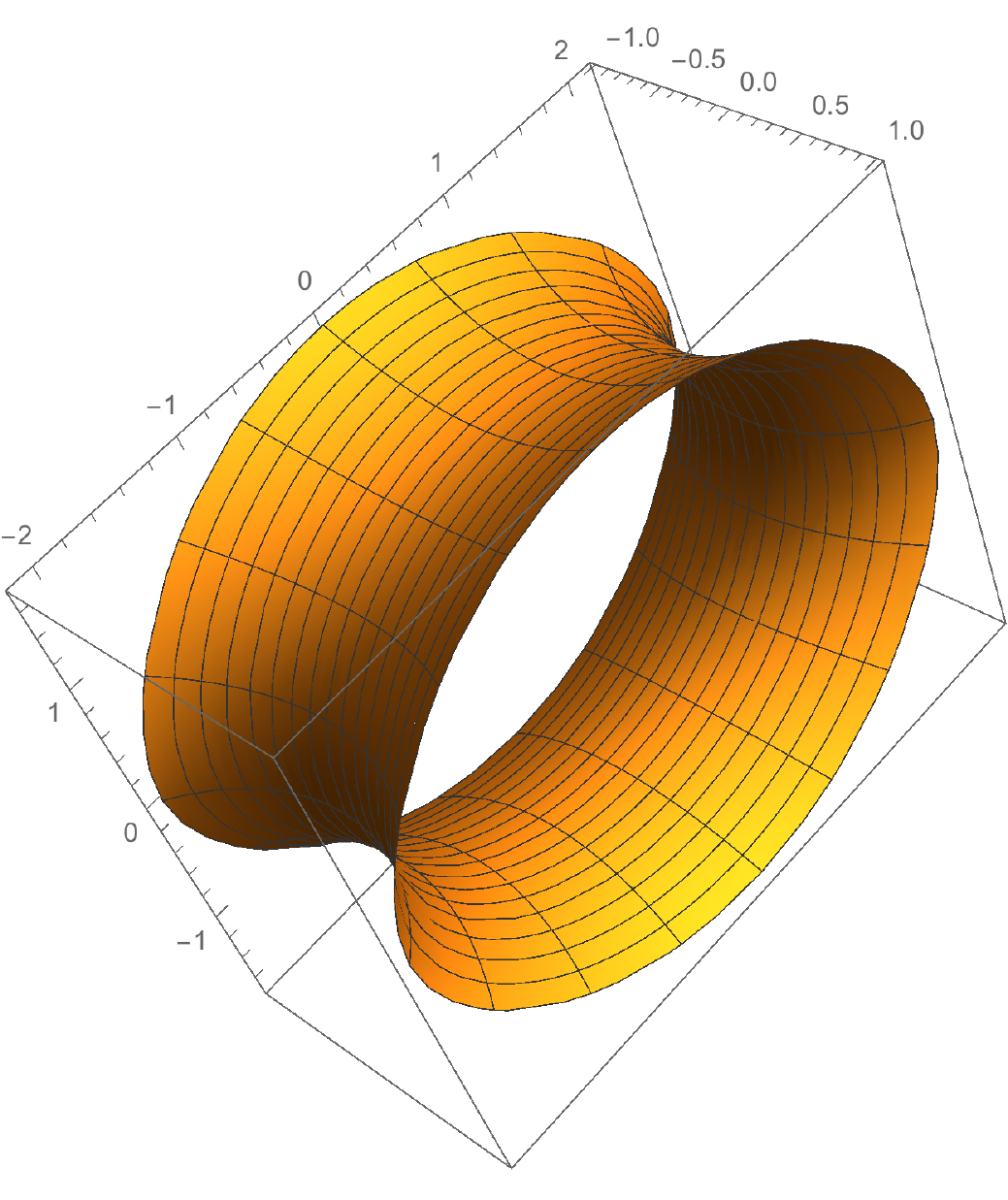}
\end{tabular}}
\caption{{\footnotesize Action and solutions at fixed $E=0.5$ for the pulse background as a function of $\omega$.}}
\label{cate5}
\end{figure}
\begin{figure}[h!t]
\centerline{
\begin{tabular}{ccc}
\epsfxsize=5cm\epsfbox{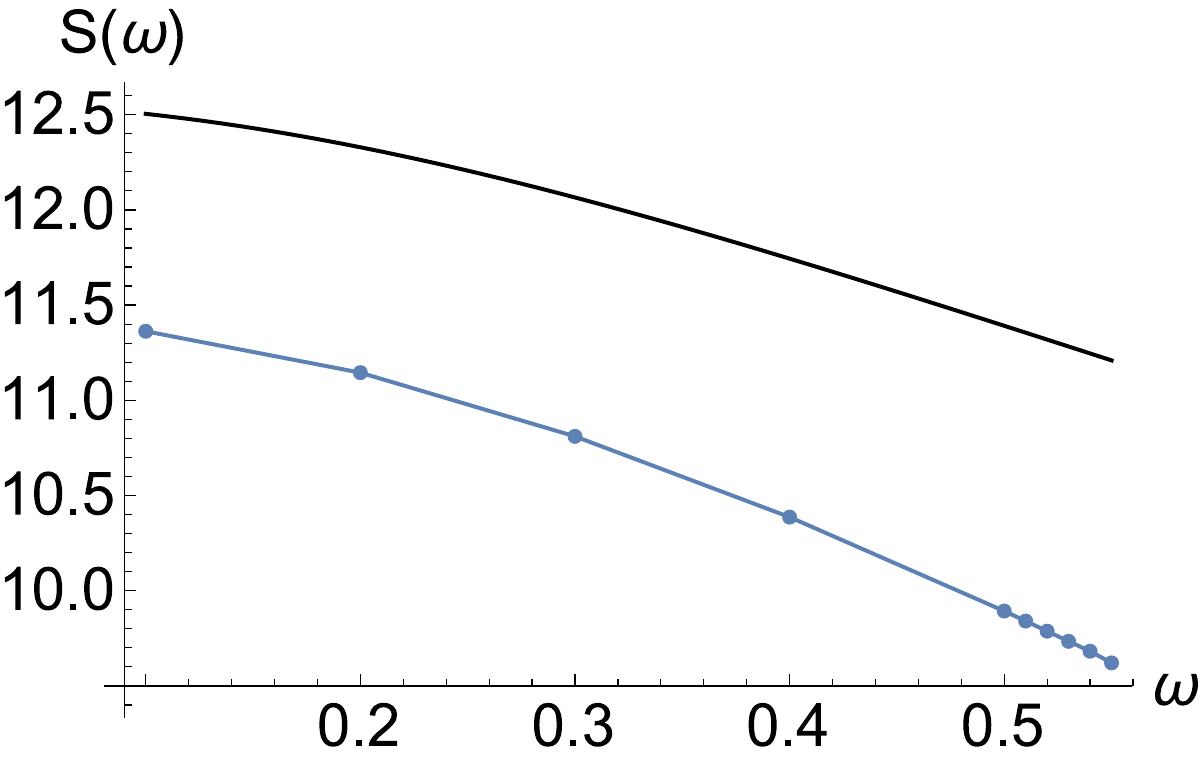}&
\epsfxsize=5cm\epsfbox{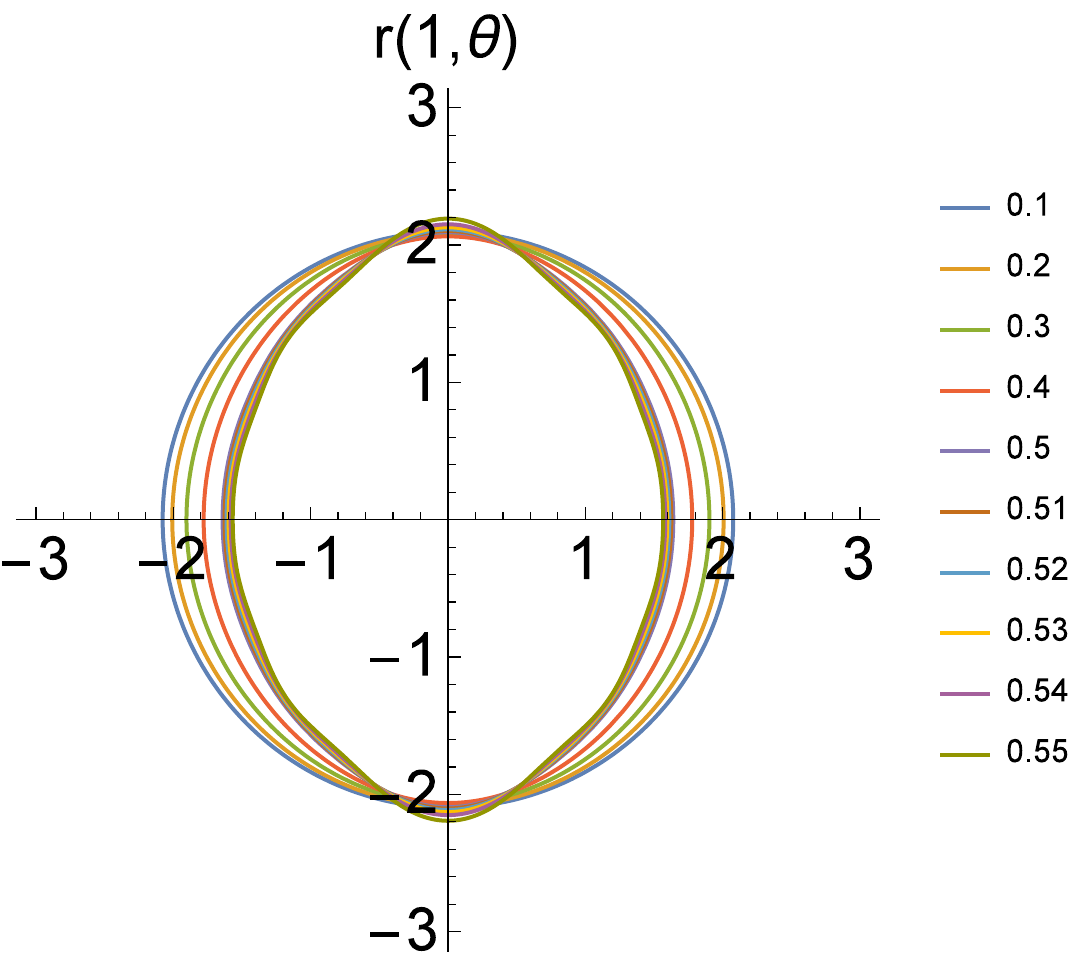}&
\includegraphics[width=43mm]{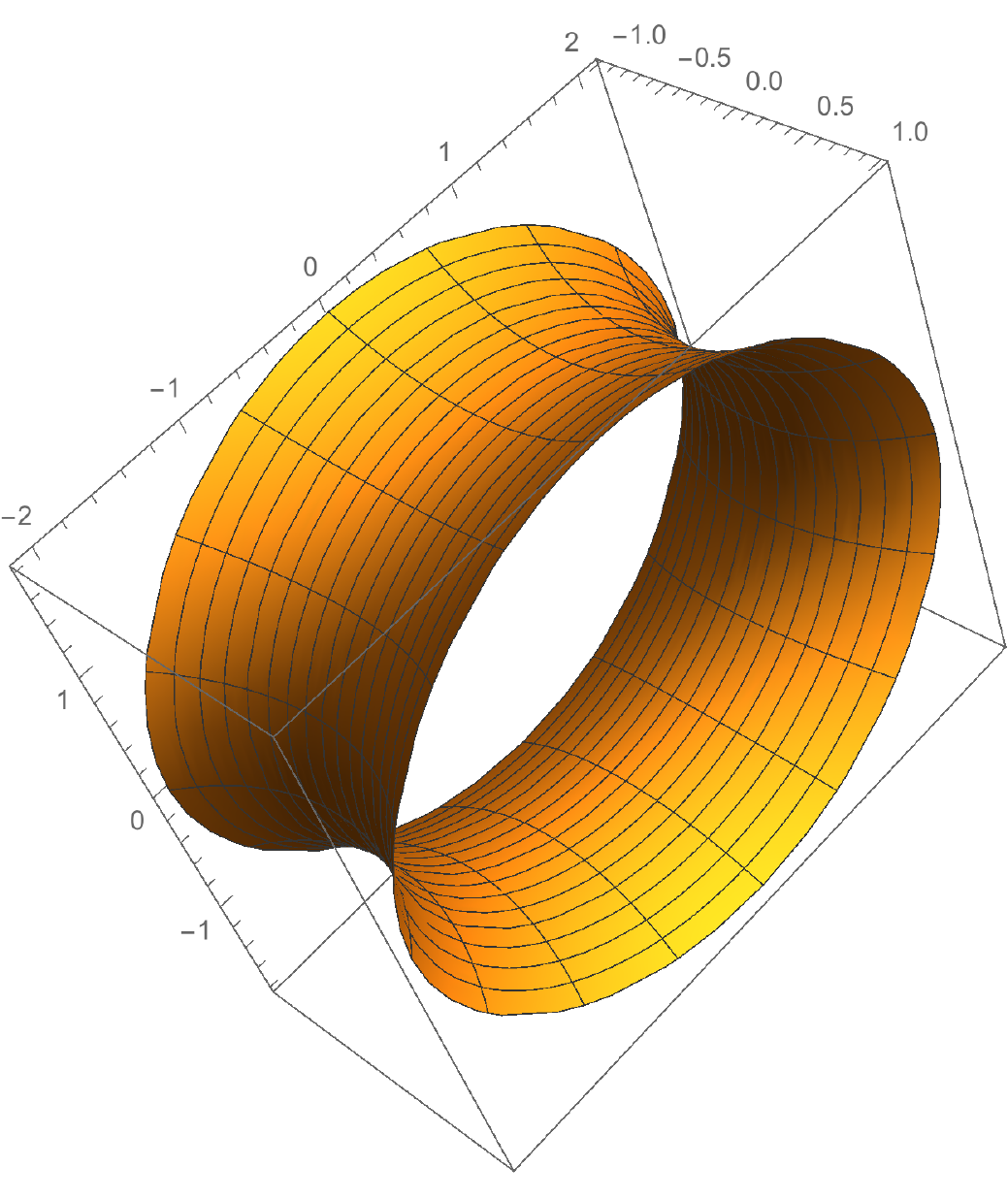}
\end{tabular}}
\caption{{\footnotesize  Same as Figure \ref{cate5} for  oscillatory  background.}}
\label{cate6}
\end{figure}
We also show the effect of variations of $\omega$ at fixed $E$. We set $E=0.5$ (this value is the most stable in the relaxation procedure) and found similar behavior for both field backgrounds, this is shown in Figures \ref{cate5} and \ref{cate6} which show the solutions and the behavior of the action as a function of $\omega$. The action decreases as $\omega$ increases, signaling the enhancement of the pair production. Moreover the action is always smaller than one obtained in the particle limit. At large $\omega$ we see a similar behavior to the large $E$ case discussed above, namely the solutions at the D-brane junction begin to develop protuberances at the maxima in $|x_4|$.

Now we consider the space dependent electric field backgrounds. In polar coordinates, the same coordinates used before (\ref{polar}),  we have
\beq
 A_{\theta} = -i E f(r  \cos{\theta}) r \cos{\theta} \qquad \qquad A_{r} = - i E f(r \cos{\theta}) \sin{\theta} \ .
\eeq
For the case of pulse and oscillatory fields the function $f$ is given by
\beq
&&f(x_3) = \frac{\tanh{(k x_3)}}{k} \qquad {\rm pulse} \nonumber  \\
&&f(x_3) =  \frac{\sin{(k x_3)}}{k} \qquad {\rm oscillating}
\eeq
The bulk equation is the Euler-Lagrange equation for the minimization of the area and is the same as before (\ref{pdeuno}).
The boundary terms are instead
\beq
\label{boundaryconditionspace}
T \frac{r^2 \partial_z r}{ \sqrt{r^2(1+ (\partial_z r)^2) +(\partial_{\theta} r)^2 }} = q E  \left(  r + \partial_{\theta} r \sin{\theta} \cos{\theta} \right) f'(r \cos{\theta})\ .
\eeq

\begin{figure}[h!]
\centerline{
\begin{tabular}{ccc}
\epsfxsize=5cm\epsfbox{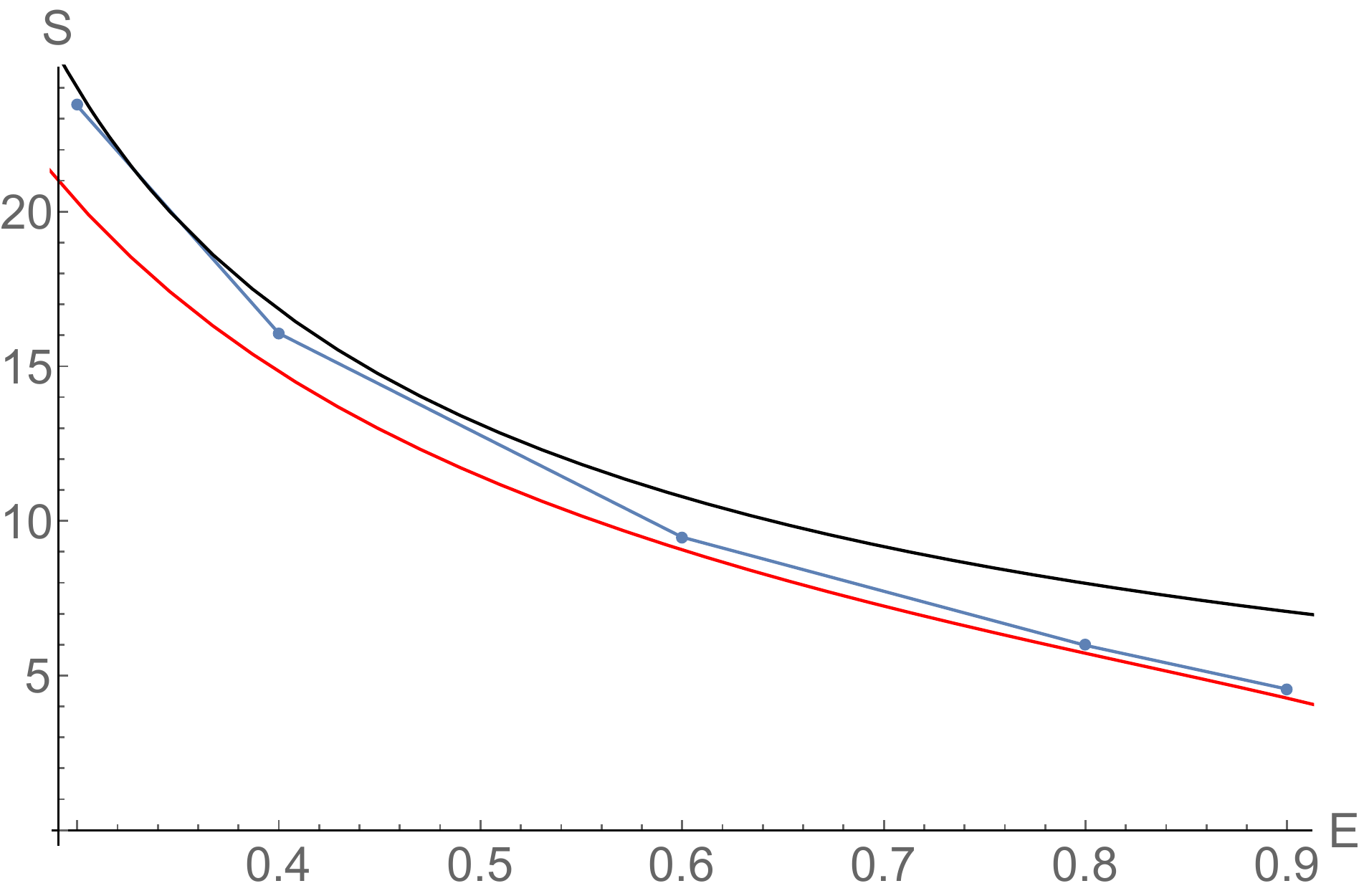}&
\epsfxsize=5cm\epsfbox{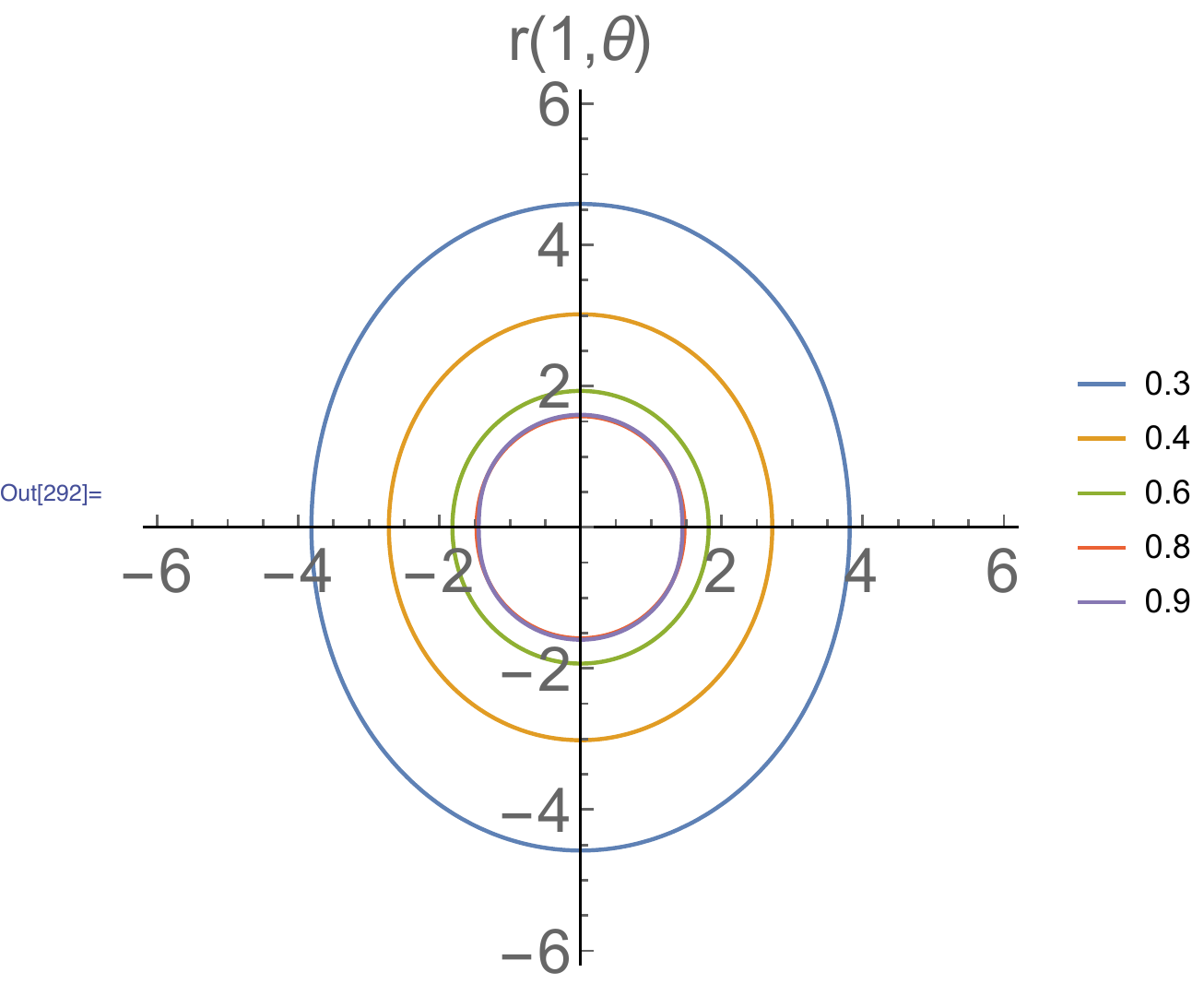}
\end{tabular}
\begin{tabular}{c}
\includegraphics[width=28mm]{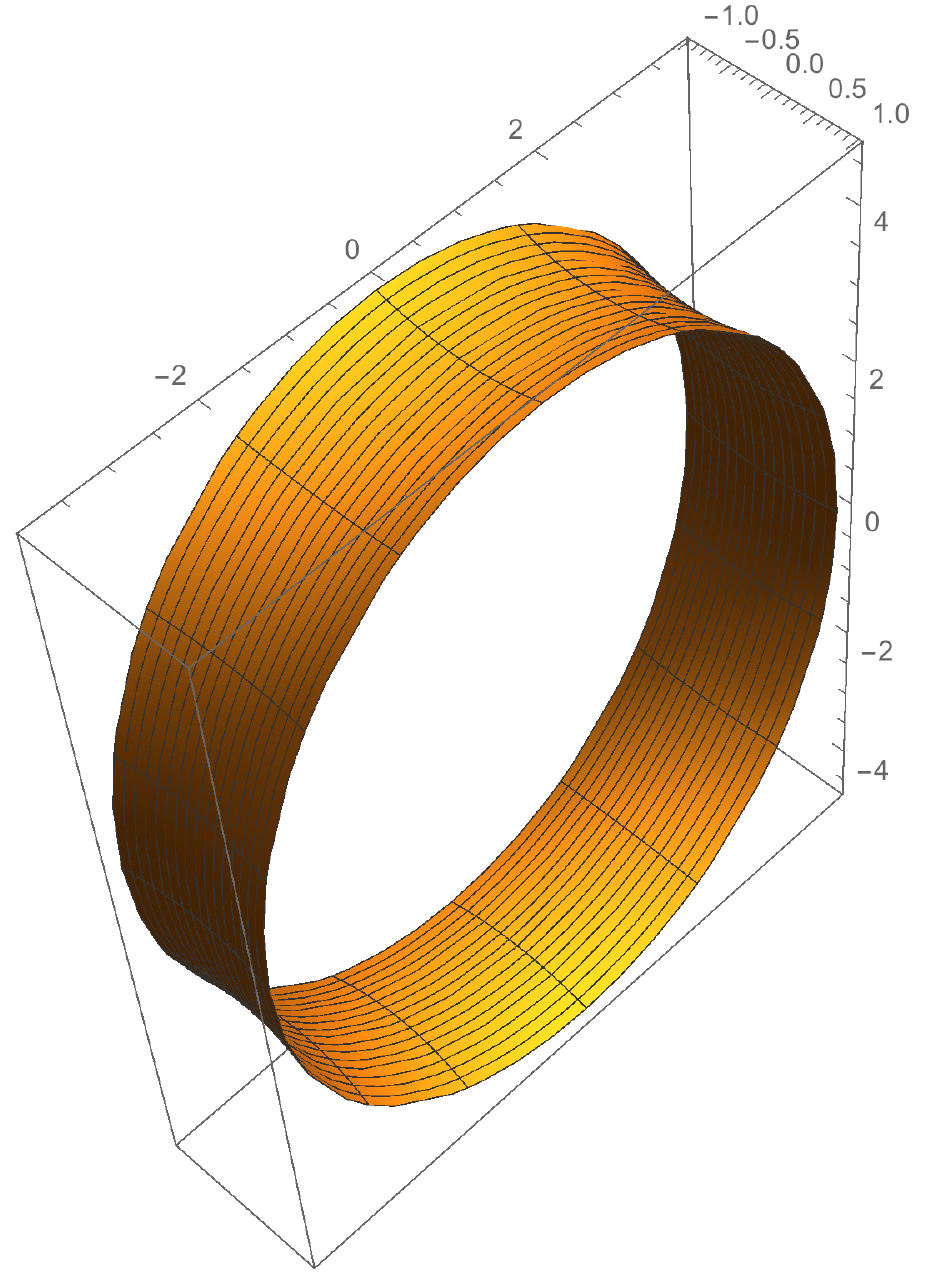} \\
\includegraphics[width=36mm]{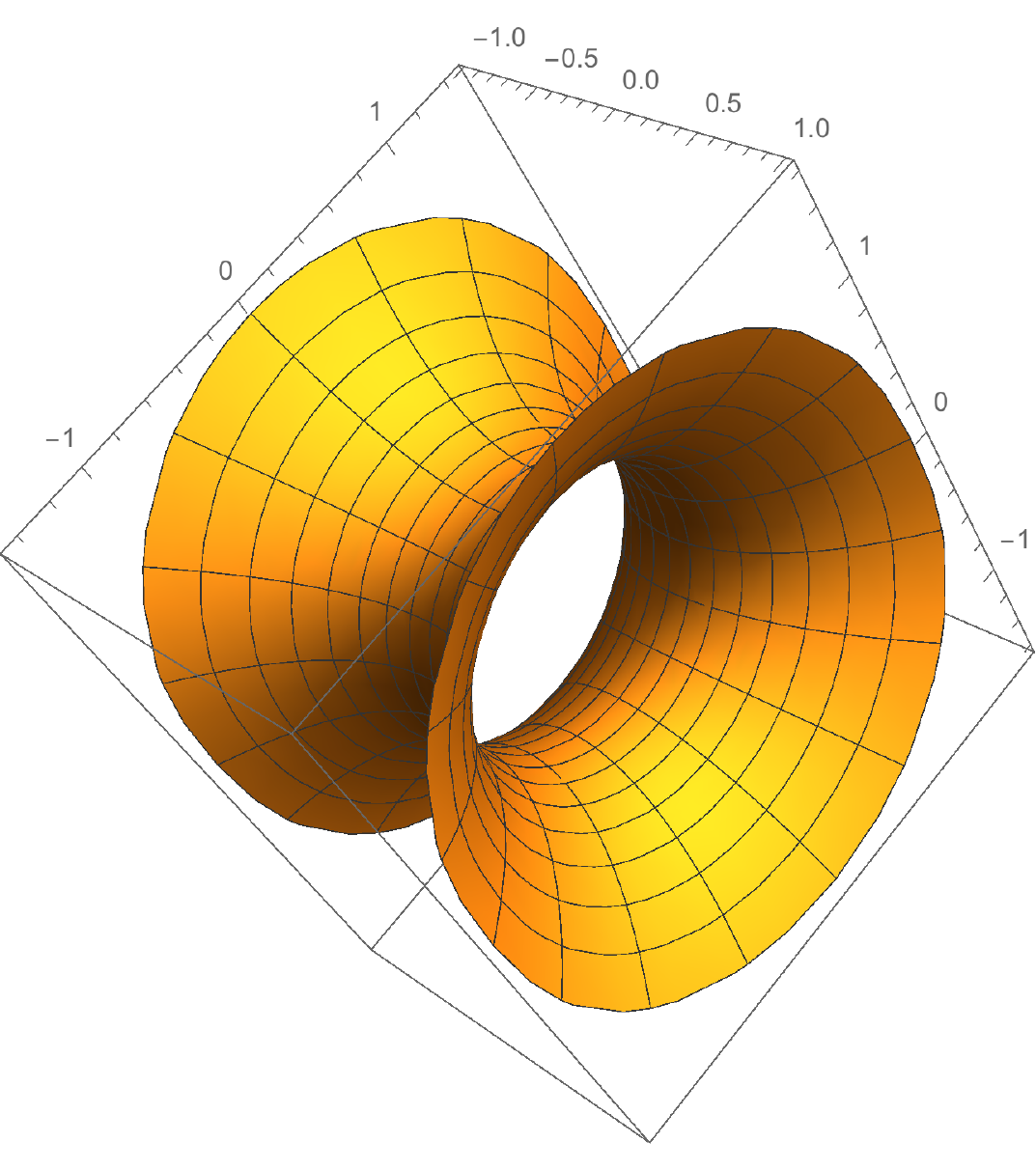}
\end{tabular}}
\caption{{\footnotesize The top graphs show the action and solutions for the spatially varying pulse background at various values of $E$ (shown in the legend) at $k=0.2$. The bottom graphs show the revolution plots  for $E=0.3$  and $E=0.9$ at the same $k$.}}
\label{sc1}
\end{figure}
\begin{figure}[h!]
\centerline{
\begin{tabular}{ccc}
\epsfxsize=5cm\epsfbox{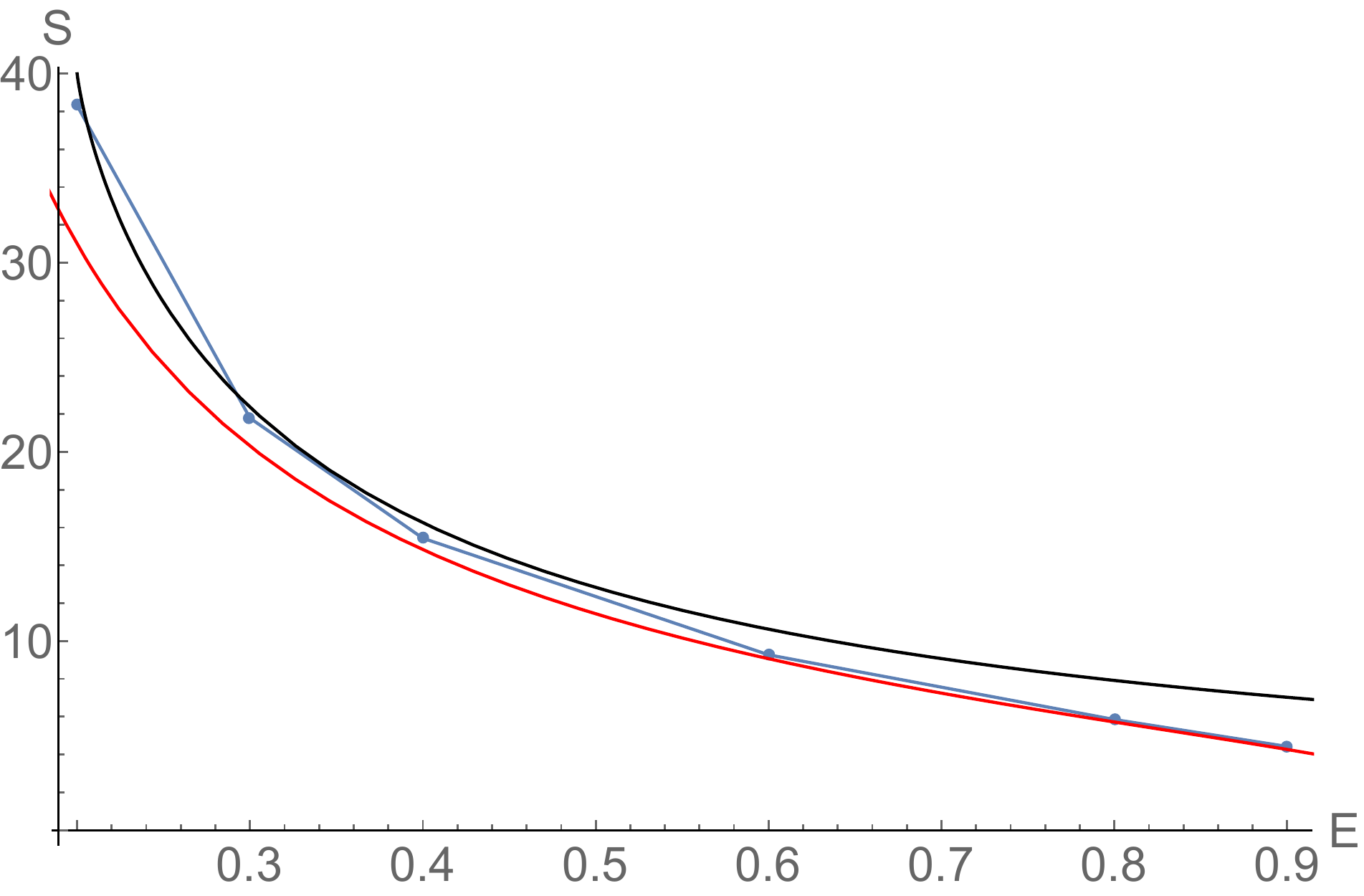}&
\epsfxsize=5cm\epsfbox{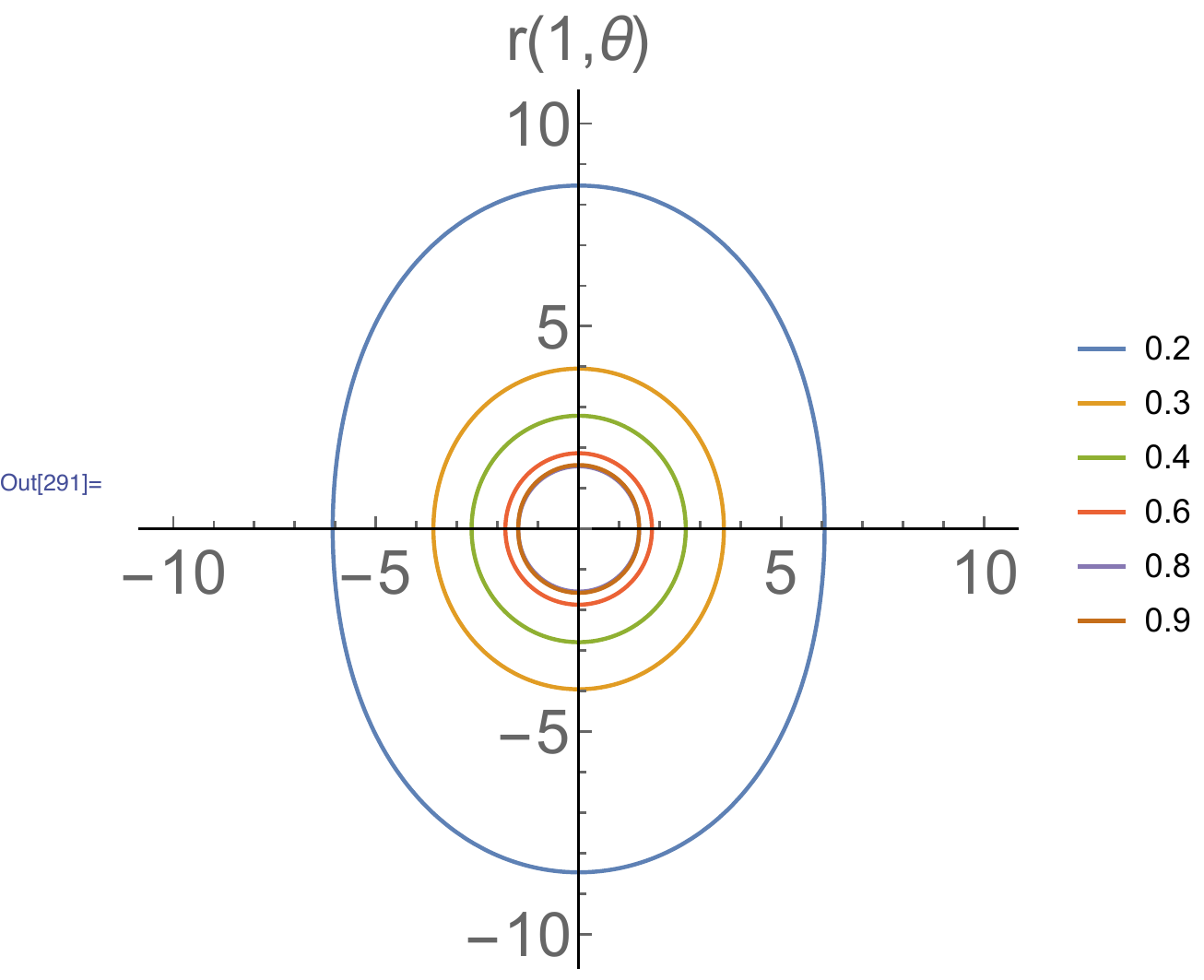}
\end{tabular}
\begin{tabular}{c}
\includegraphics[width=28mm]{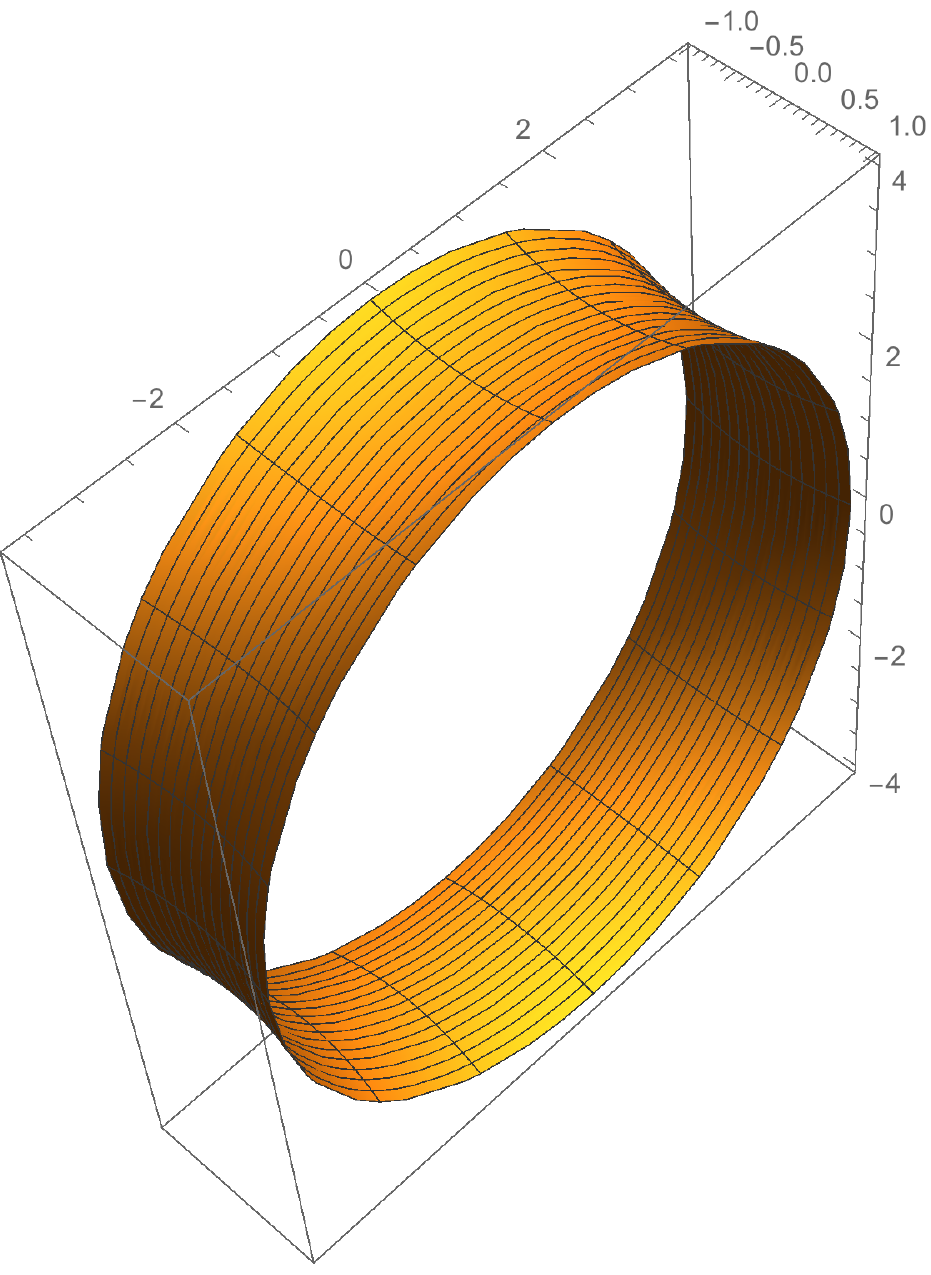}\\
\includegraphics[width=32mm]{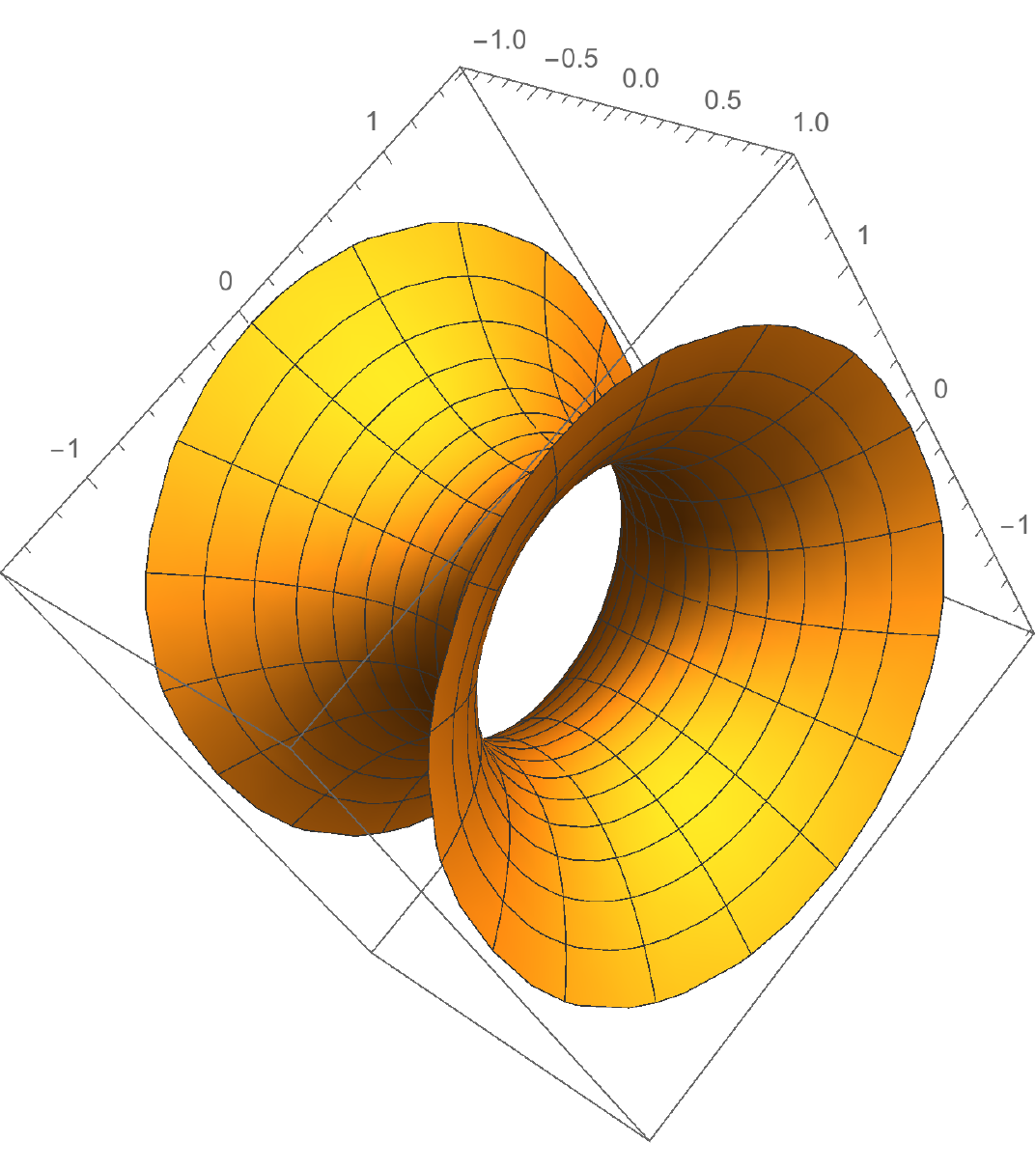}
\end{tabular}
}
\caption{{\footnotesize Same as Figure \ref{sc1} but this time for the oscillatory spatial background at $k=0.2$.}}
\label{sc2}
\end{figure}
\begin{figure}[h!]
\centerline{
\begin{tabular}{ccc}
\epsfxsize=5cm\epsfbox{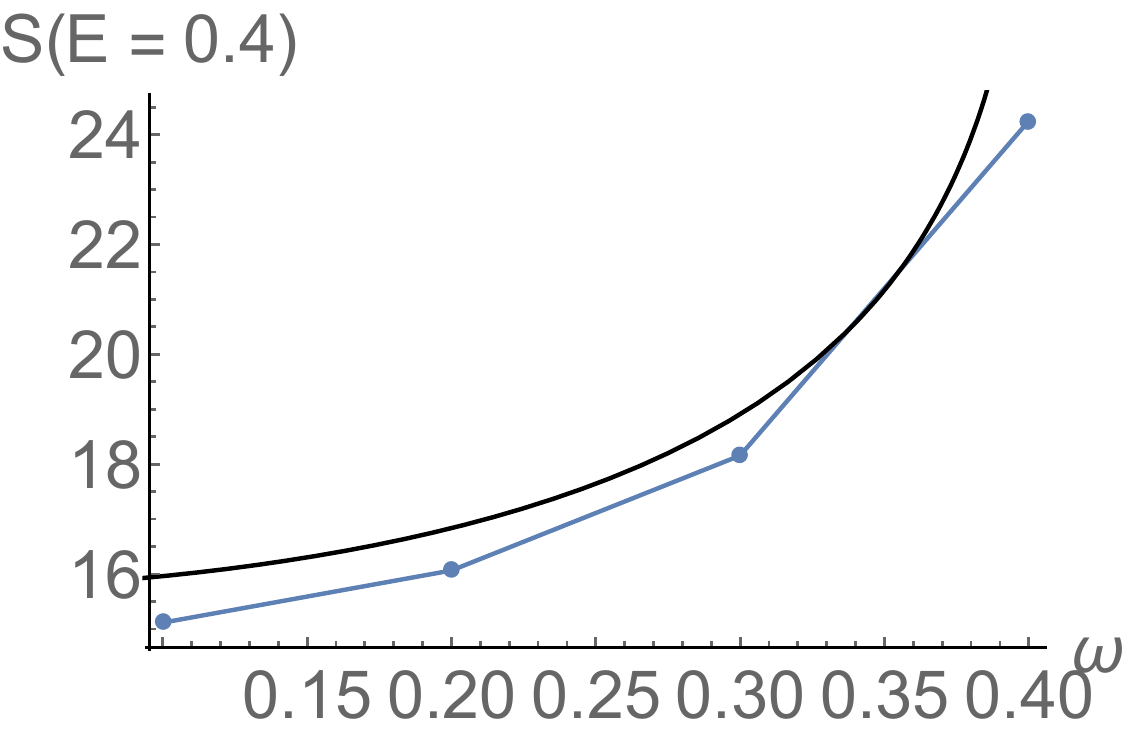}&
\epsfxsize=5cm\epsfbox{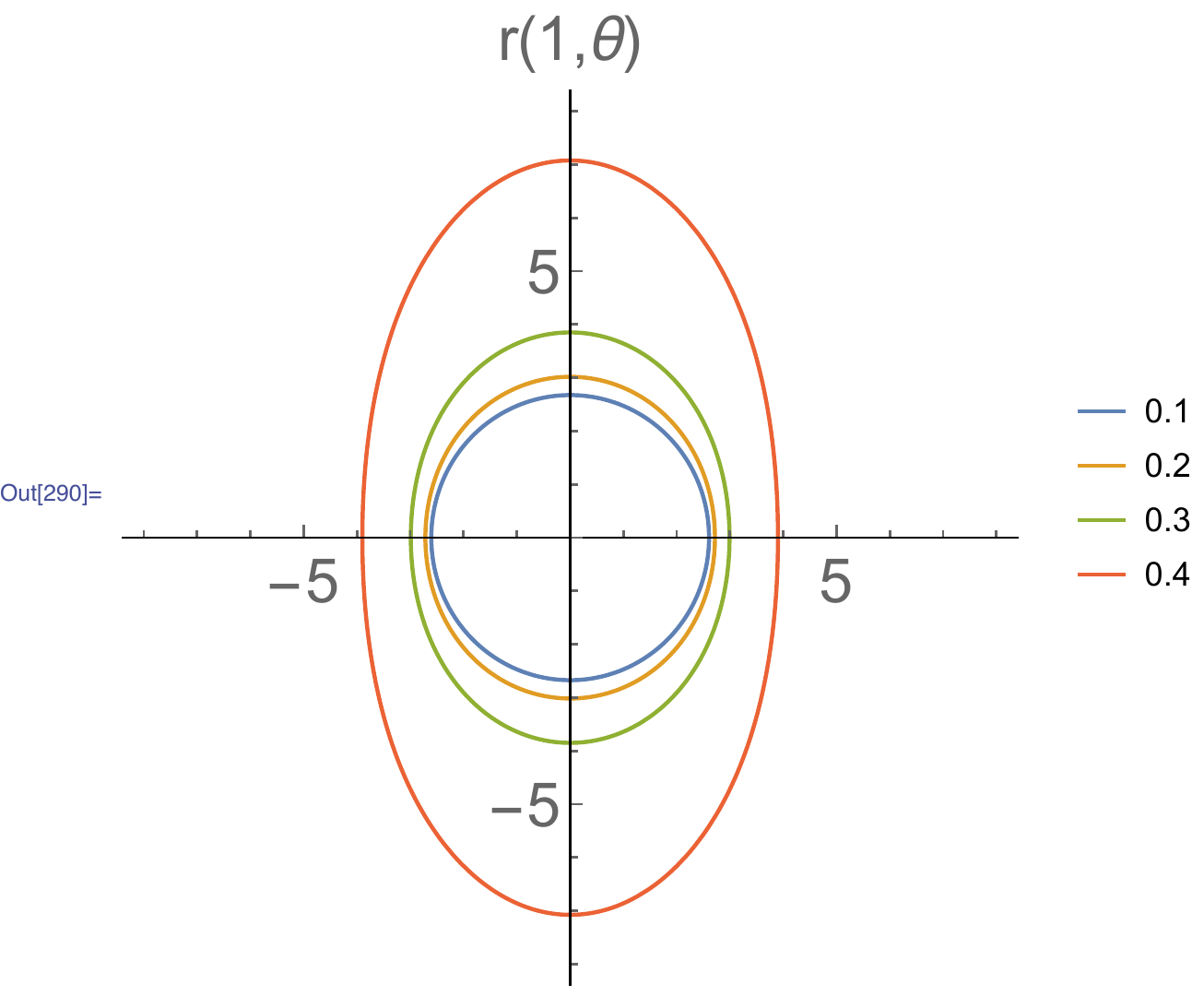}&
\includegraphics[width=33mm]{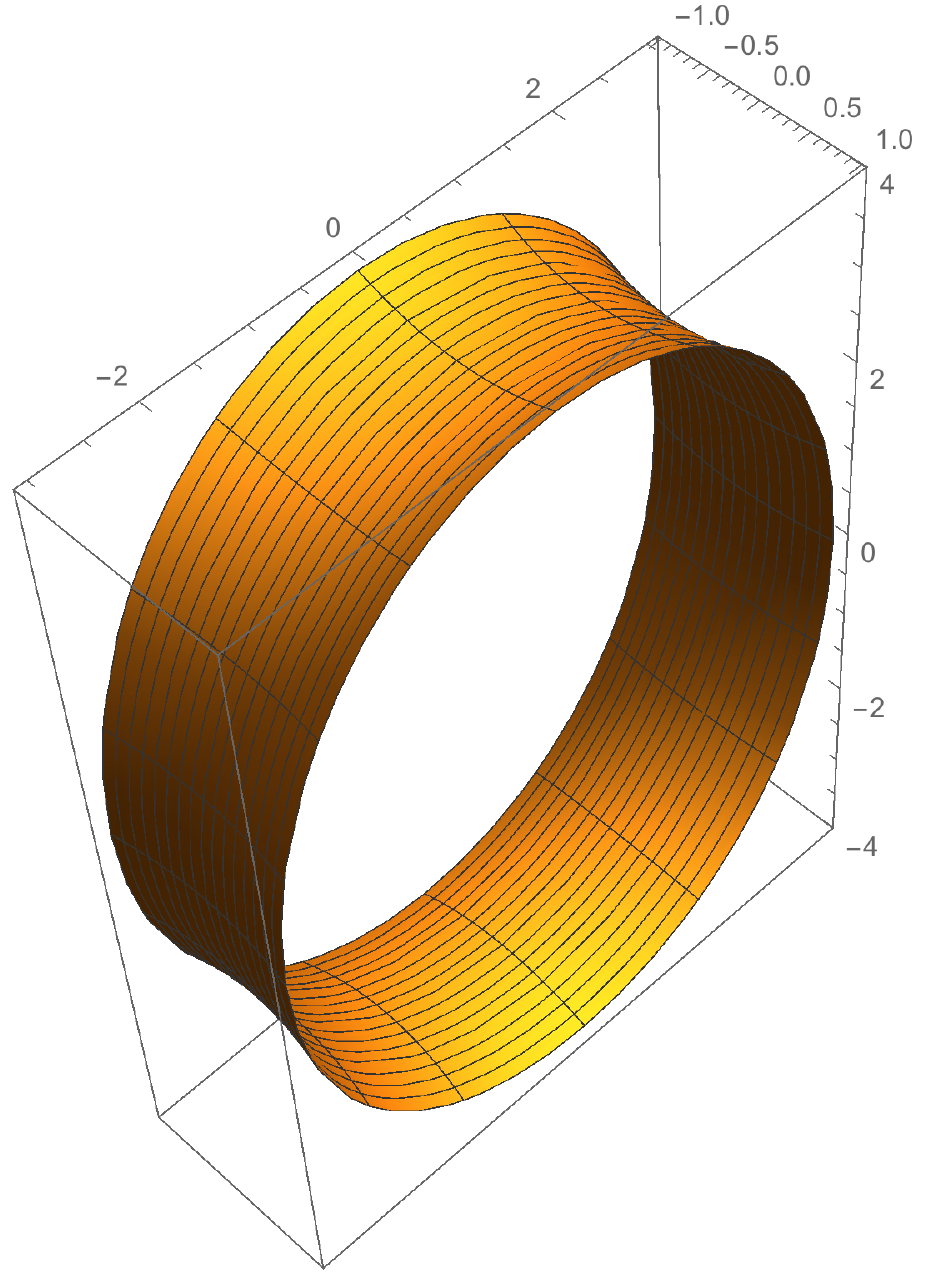}
\end{tabular}}
\caption{{\footnotesize Action and solutions at fixed $E=0.4$ for the spatial pulse background as a function of $k$.}}
\label{sc3}
\end{figure}
\begin{figure}[h!]
\centerline{
\begin{tabular}{ccc}
\epsfxsize=5cm\epsfbox{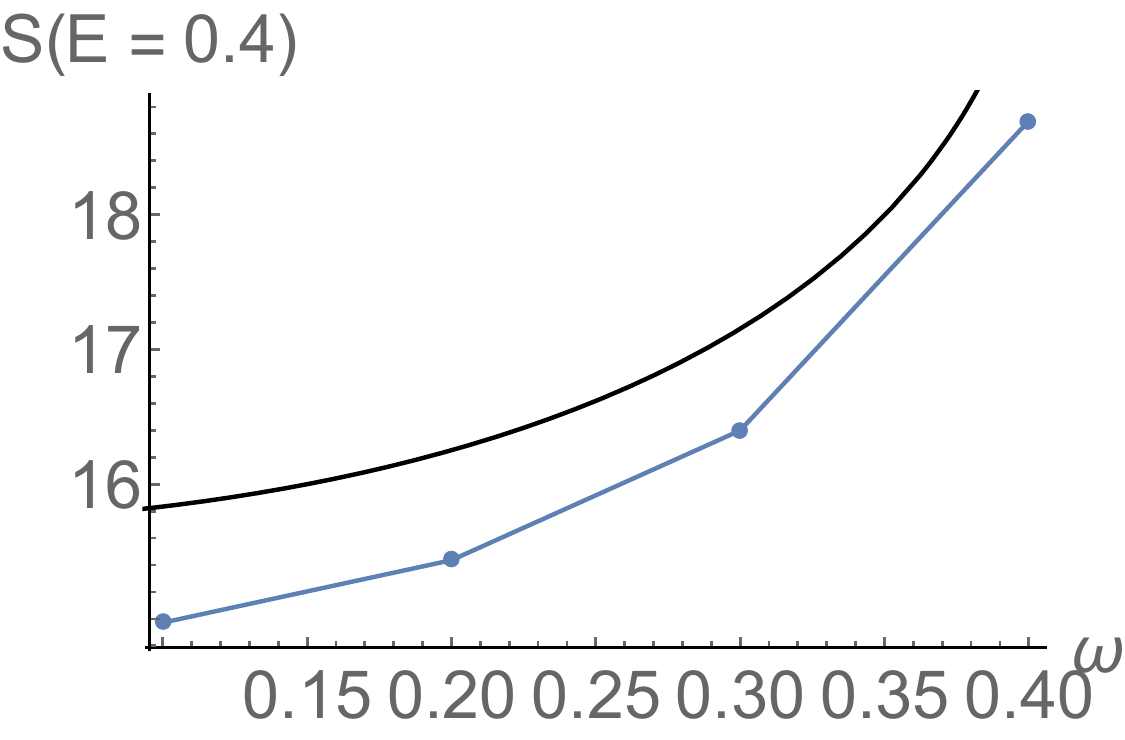}&
\epsfxsize=5cm\epsfbox{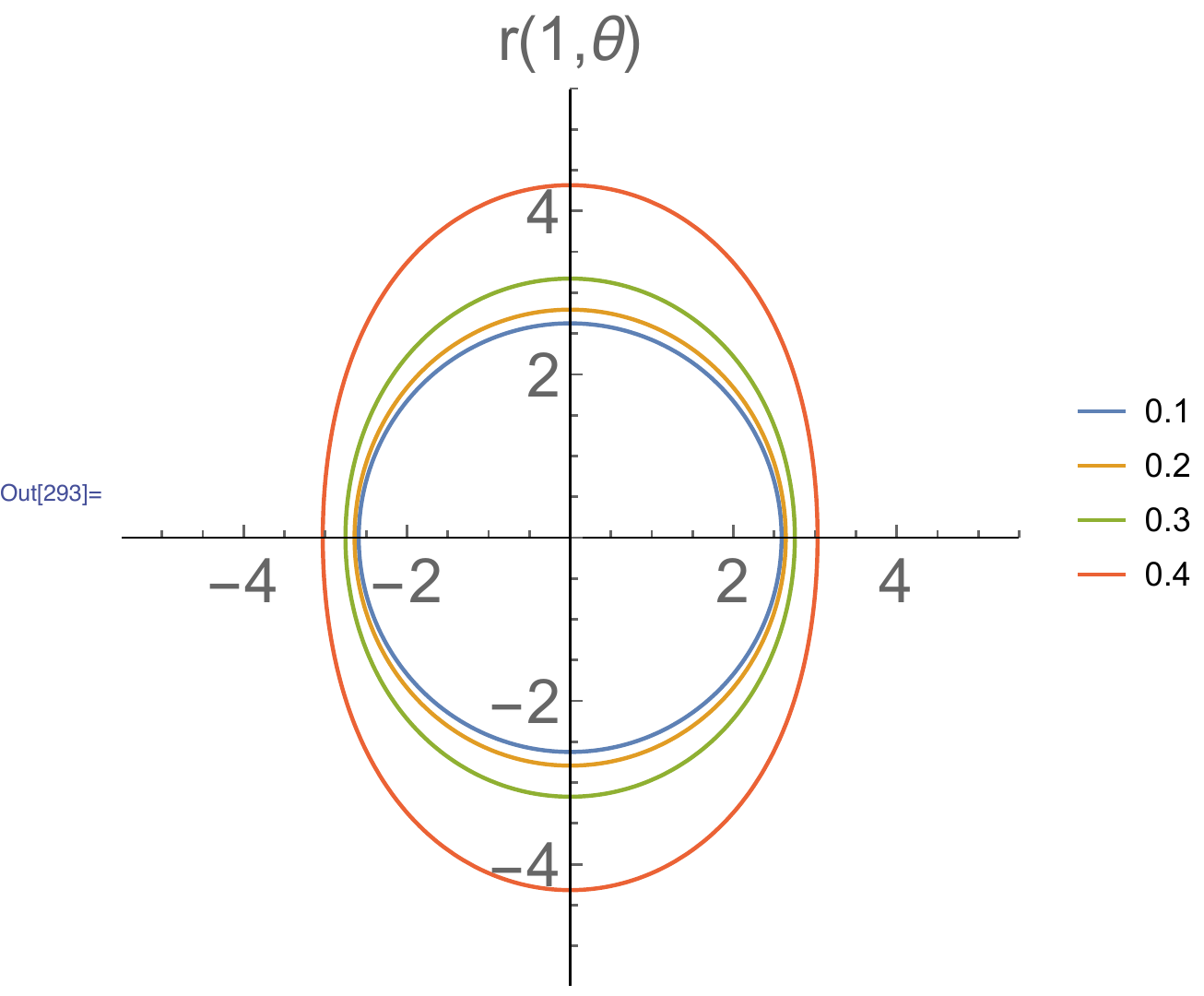}&
\includegraphics[width=33mm]{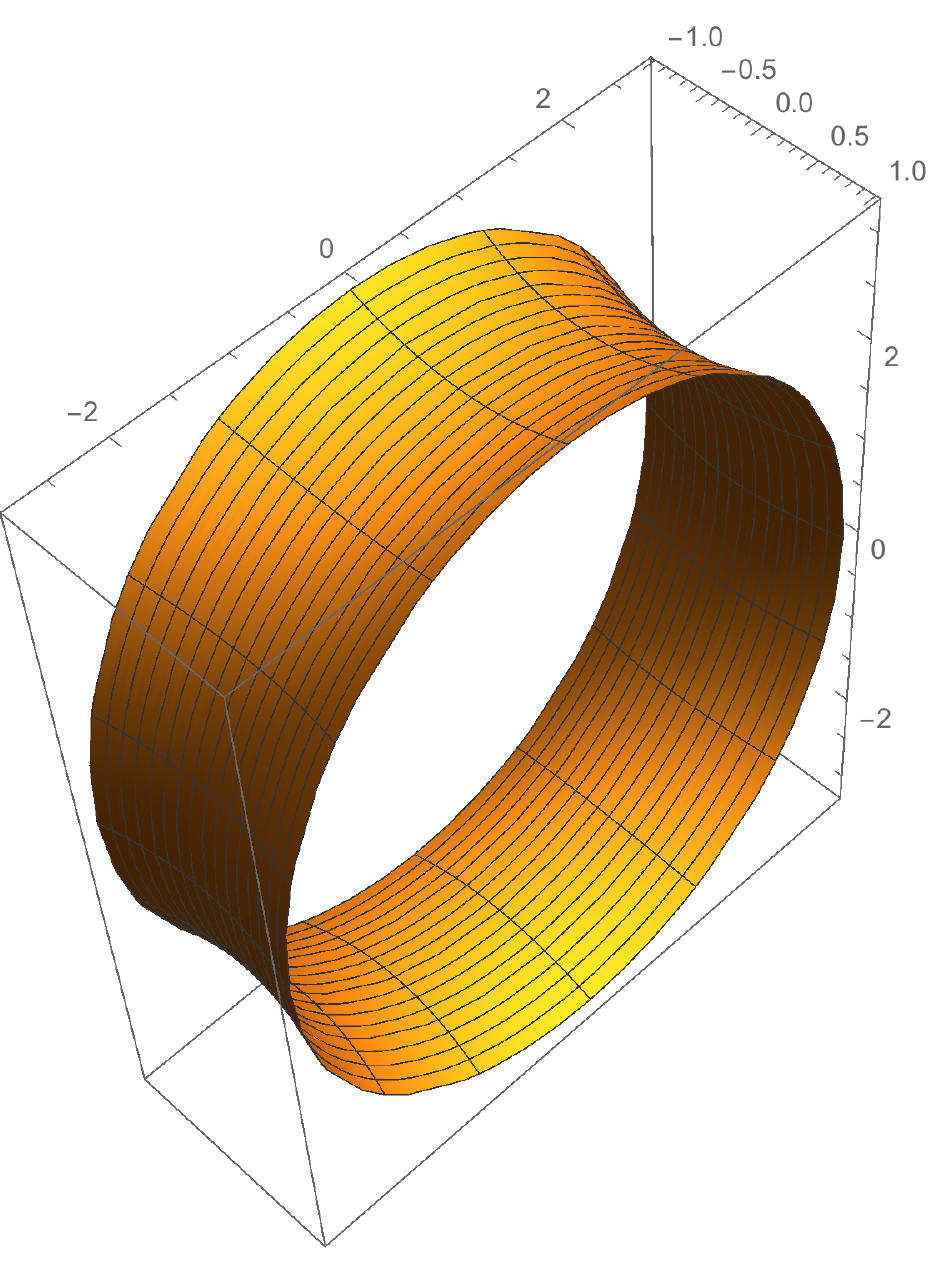}
\end{tabular}}
\caption{{\footnotesize Action and solutions at fixed $E=0.4$ for the spatial oscillating background as a function of $k$.}}
\label{sc4}
\end{figure}

Results for $k=0.2$ are shown in Figure \ref{sc1} for a pulse background  and Figure \ref{sc2} for an oscillatory background. As for the time dependent backgrounds, we see a distinction between the ``high" and ``low" $E$ regimes where we find the two analytic approximations to be valid. In between we find deviation from the two analytical curves. 
We find convergence  down to values of $E\approx 0.03$ 
with no remarkable variation on the particle trajectory.
We also show the effect of variations of $k$ at fixed $E$. We set $E=0.4$  in Figures \ref{sc3} and \ref{sc4} which show the solutions and the behavior of the action as a function of $k$. The action increases as $k$ increases, signaling the suppression of the pair production. Moreover the action is always smaller than one obtained in the particle limit: this is the enhancement due to the string nature.

\section{String pair production: Holography}
\label{quattro}

\begin{figure}[h!t]
\centerline{
\begin{tabular}{cc}
\epsfxsize=6cm
\epsfbox{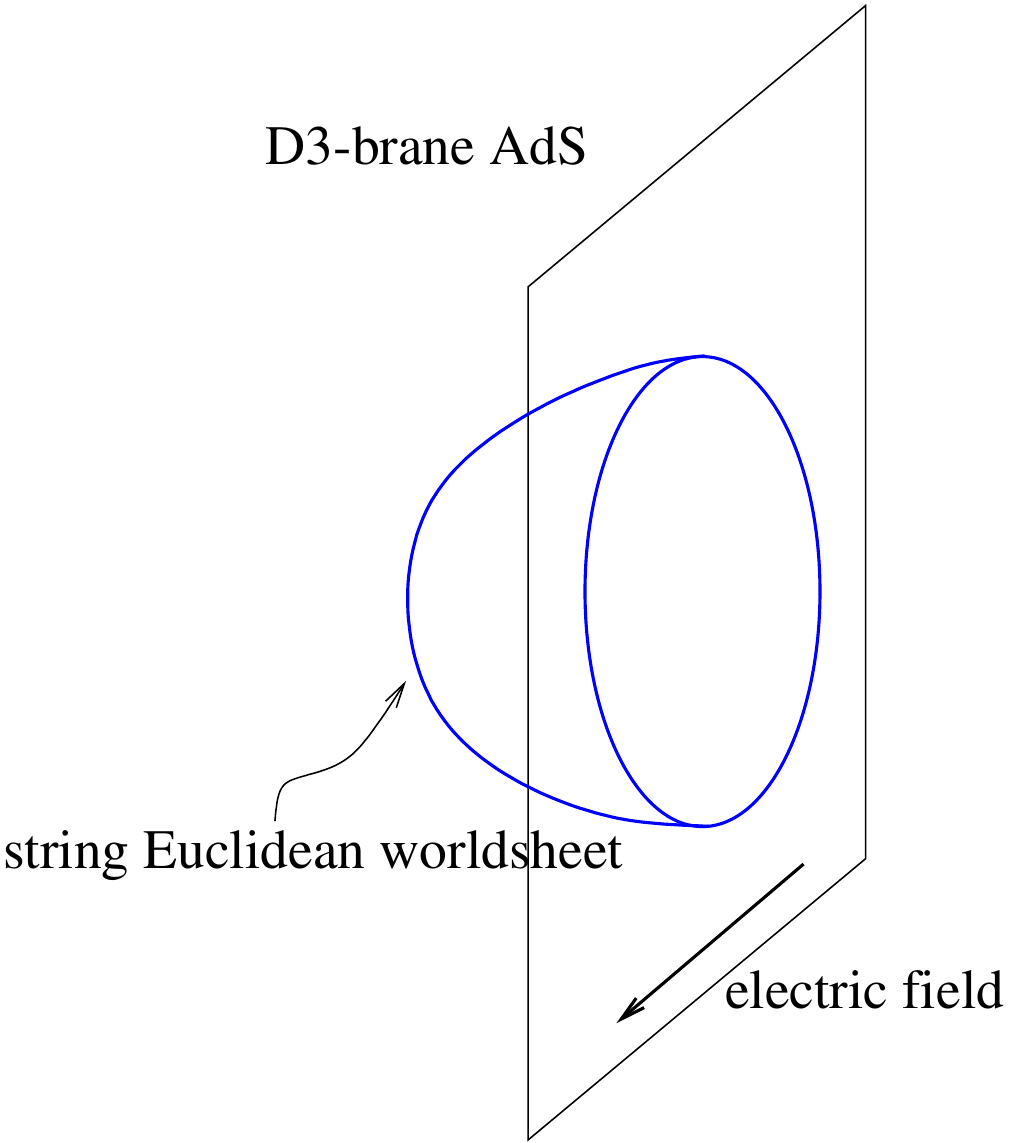} \ \ \ \ \qquad  &
\epsfxsize=5cm\epsfbox{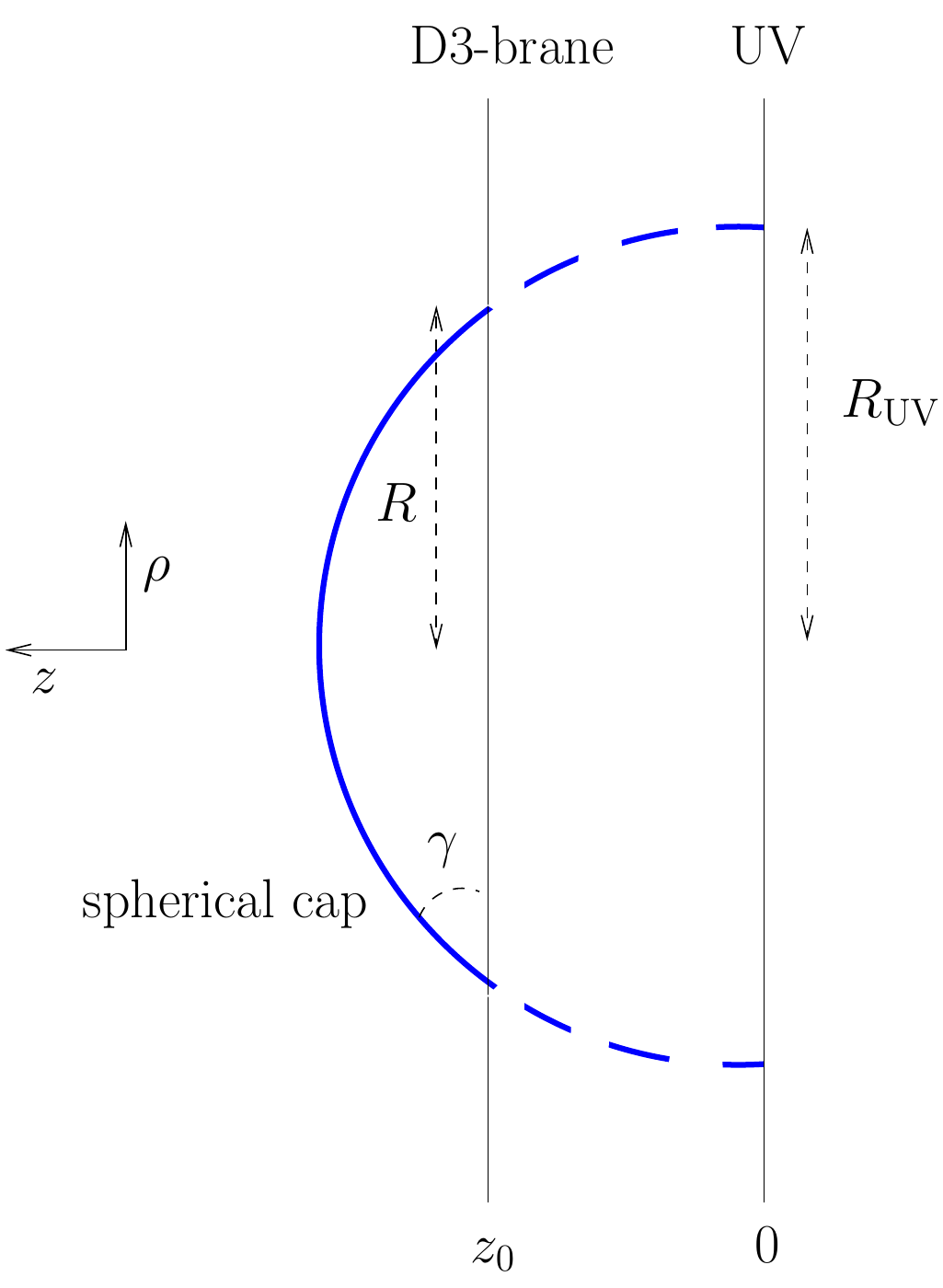} \qquad
\end{tabular}}
\caption{{\footnotesize String pair production in AdS.  Geometry of the world-sheet bounce. The dashed part is not physical, just a continuation of a minimal surface solution in AdS to the UV boundary.}}
\label{ads}
\end{figure}

Next we turn to consider the holographic Schwinger effect \cite{Semenoff:2011ng}. The string is suspended on a D-brane probe which lives in AdS space. 
The world-sheet instanton is given in Figure \ref{ads}. It is  a world-sheet with a disk topology with the boundary located inside the D-brane.

It is convenient to use coordinates  where the metric is manifestly conformally flat
\beq
ds^2 = \frac{L^2}{z^2} \left( dz^2 + dx_{\mu}dx^{\mu} \right)
\eeq
For constant background the  instanton is circular and  the Euclidean action is then
\beq
\label{actiond1string}
S_{E} = T \int_0^{R} d\rho 2\pi \rho  \left(\frac{L}{z(\rho)}\right)^2  \sqrt{1+ z'(\rho)^2 } -q E \pi R^2
\eeq
A  minimal surface in hyperbolic space is  given by a half sphere 
\beq
\label{circlesolution}
z(\rho) = \sqrt{ R_{\rm UV}^2 - \rho^2 }
\eeq
These are the stationary solutions to the first part of the action (\ref{actiond1string}).
This curve should be truncated at $z=z_0$ where the string ends on the D3-brane.
The radius $R$ is measured at $z=z_0$ and is given by 
\beq
\label{rtilde}
R^2 + z_0^2 =R_{\rm UV}^2  \ .
\eeq 
The action as a function of $R$ has the following expression:
\beq
\label{actionER}
S_E=2\pi T L^2 \left( \sqrt{1+ \frac{R^2}{z_0^2}}-1 \right) -q E \pi R^2
\eeq
The maximization with respect to $R$ is equivalent to  the  balance of forces at the boundary
\beq
T \cos(\gamma) = q  F_{\rm loc} 
\eeq
The radius at the stationary point is given by
\beq 
\label{reuclidholo}
R = z_0  \sqrt{\left(\frac{T L^2 }{q E z_0^2}\right)^2 - 1} \ ,
\eeq
and it leads to the action
\beq
\label{seuclidholo}
S_{E} = \frac{\pi}{ q E z_0^2 }\left(\left(\frac{T L^2 }{q E z_0^2}\right)   - 1 \right)^2 \ .
\eeq
A critical point is reached when the radius $R$ and the classical action vanish. This happens at the following critical value for the field 
\beq
\label{criticalF}
q E_{\rm cr} = \frac{T L^2}{ z_0^2}
\eeq
This is when the sphere in Figure \ref{ads} becomes exactly tangent to the D3 brane and nothing is left for the physical cap.

The weak field limit $E \ll E_{cr}$ corresponds to the field theory limit. In this case the action becomes the particle action with the mass
\beq
\label{mass}
m = \frac{T L^2}{z_0}
\eeq
which is the mass of a string with tension $T$ stretched from $z_0$ to the horizon at $z \to \infty$.

We can now look at the non-homogeneous background case, for which the world-sheet has a generic, non circular shape. 
The world sheet topology is a disk of radius one parametrized by coordinates $\rho, \theta$.  We use the following gauge
\beq
(z_0 - Z_0) \rho + Z_0 = z
\eeq
where $z=z_0$ is the position of the brane and $z=Z_0$ the tip of the world-sheet. $Z_0$ is an unknown parameter to be determined by the equations. 
The action is 
\beq
S_{E} =T \int_{z_0}^{Z_0} dz \int_{0}^{2 \pi} d\theta  \frac{L^2}{z^2}\sqrt{r^2(1 + (\partial_z r)^2) + (\partial_{\theta} r)^2  }  -i q  \int_{0}^{2 \pi} d\theta  (A_{\theta} + A_r\partial_{\theta} r )
\eeq
The bulk equation is that of a minimal surface in AdS 
\beq
\label{pdedue}
&&-\frac{2 r \partial_{z}r}{z}\left( r^2\, (1+ (\partial_{z}r)^2) + (\partial_{\theta}r)^2 \right) \nonumber \\
&&r \, \partial_z^2 r \, (r^2 + (\partial_{\theta}r)^2)
+ r \, \partial_{\theta}^2 r \, (1+ (\partial_{z}r)^2)  \nonumber \\
&& \quad -2\,  r \, \partial_z r \, \partial_{\theta} r\,  \partial_z \partial_{\theta} r
-2 \, (\partial_{\theta} r)^2
- r^2 \, (\partial_{z} r)^2
-r^2 = 0 \ .
\eeq
With no $\theta$ dependence this reduces to 
\beq
-\frac{2 r^3 \partial_{z}r}{z} (1+ (\partial_{z}r)^2)+r \partial_z^2 r -  (\partial_z r)^2 -1 = 0
\eeq
and we recover the previous result.
The boundary term at the D-brane is 
\beq
\label{boundaryconditiondue}
T\frac{L^2}{z_0^2} \frac{r^2 \partial_z r}{ \sqrt{r^2(1+ (\partial_z r)^2) +(\partial_{\theta} r)^2 }} = q E f' \left(  r - \partial_{\theta} r \sin{\theta} \cos{\theta} \right) \ .
\eeq
which, apart from a redshift rescaling, is  the same as we had before for the flat space-time case  (\ref{boundarycondition}).
The other boundary condition at the tip of the world-sheet is 
\beq
\label{bctip}
r(Z_0,\theta) = 0, \qquad r'(Z_0,\theta) = \infty. 
\eeq
So at the end we need to solve the PDE (\ref{pdedue}) with one variable $r(z, \theta)$ defined on $z_0 \leq z \leq  Z_0$, $0 < \theta \leq 2\pi$ and the boundary condition (\ref{boundaryconditiondue}) at $z = z_0$ and (\ref{bctip}) at $z= Z_0$.

Our numerical solver mimics that used in the previous section, that is a spectral procedure in a Fourier and Chebyshev basis. However when solving the system in $AdS$ one faces a  restriction with our gauge choice: the minimal surfaces of the strings in $AdS$ are hemispheres defined up to  $Z_0$ which is not known in advance but must be an outcome of the solution process. 
Our strategy is the following. 
We consider a generic $Z$ and  map the interval $z_0 \leq z \leq Z$ to the standard Chebyshev interval $-1\leq \tilde{z}\leq 1$ with the following linear transformation
\beq
\tilde{z} = \frac{2 z - Z -z_0}{Z- z_0} \ .
\eeq
If $ Z $ is smaller than $Z_0$, the solver (the same we used in Section \ref{tre}) will  be able to find the solution truncated to $Z$.
We then manually increase $Z$ until we can cover the entire solution with the Chebyshev interval and thus we find also the correct value of $Z_0$.

\begin{figure}[h!]
\centerline{
\begin{tabular}{ccc}
\epsfxsize=5cm\epsfbox{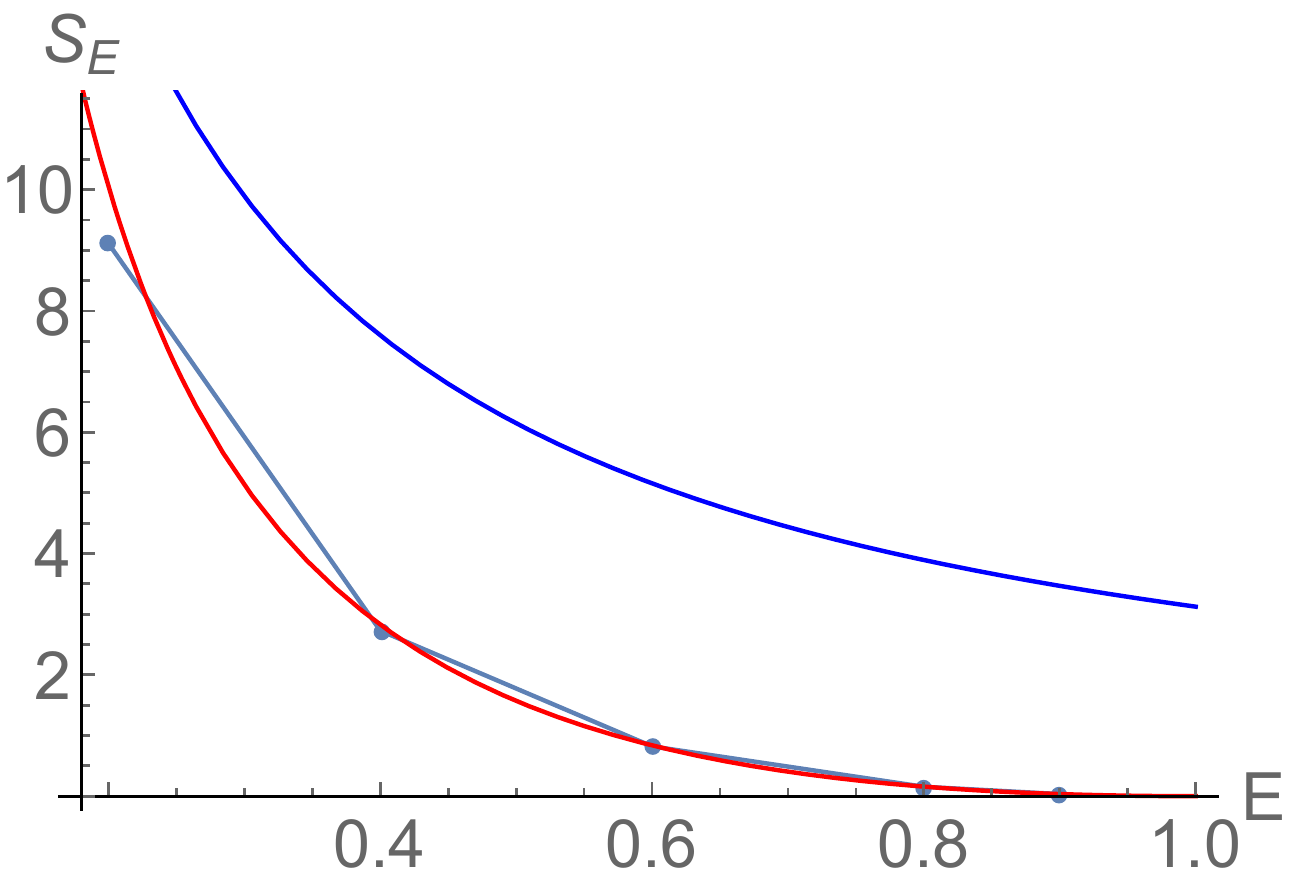}&
\epsfxsize=5cm\epsfbox{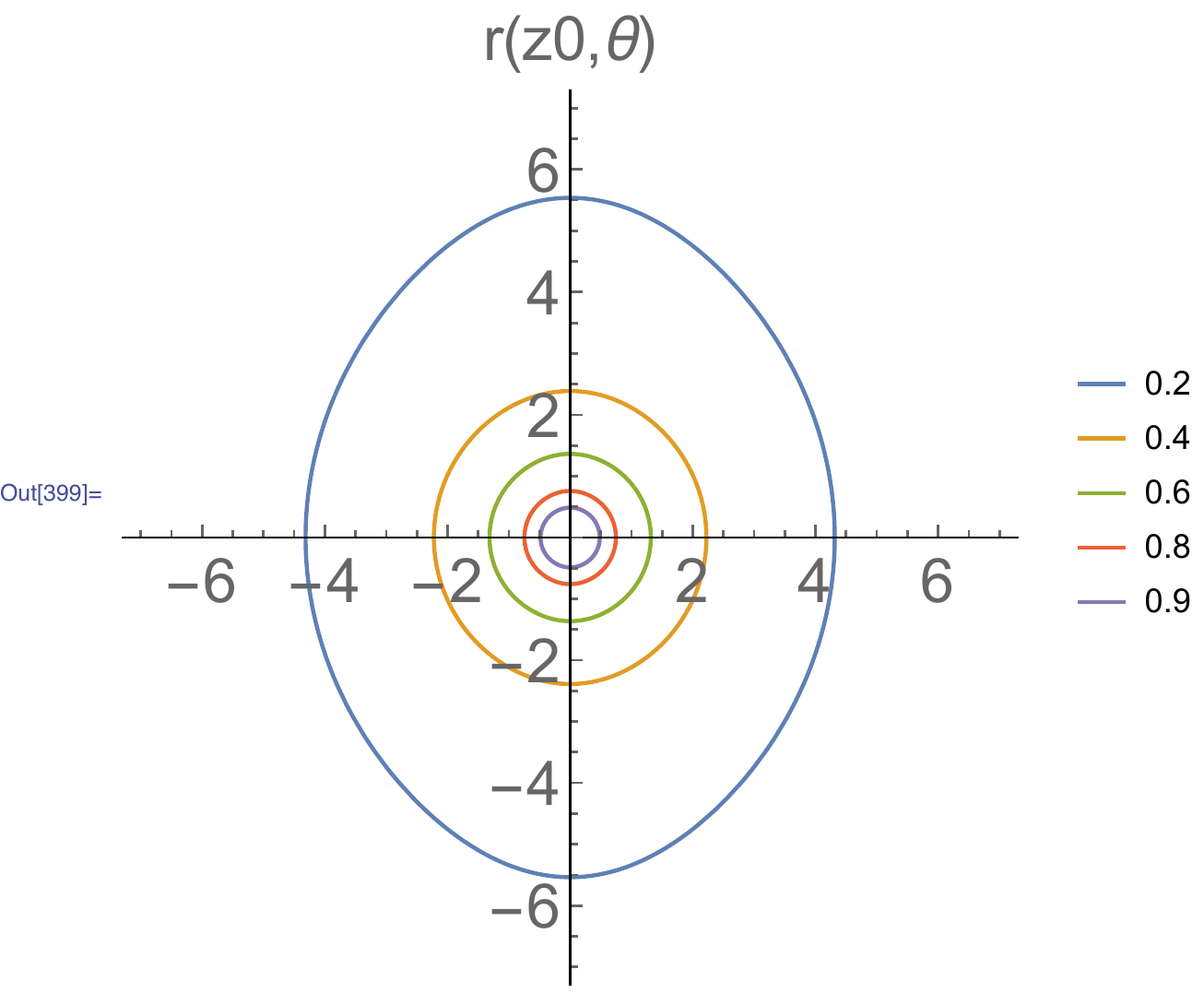}
\end{tabular}
\begin{tabular}{c}
\includegraphics[width=45mm]{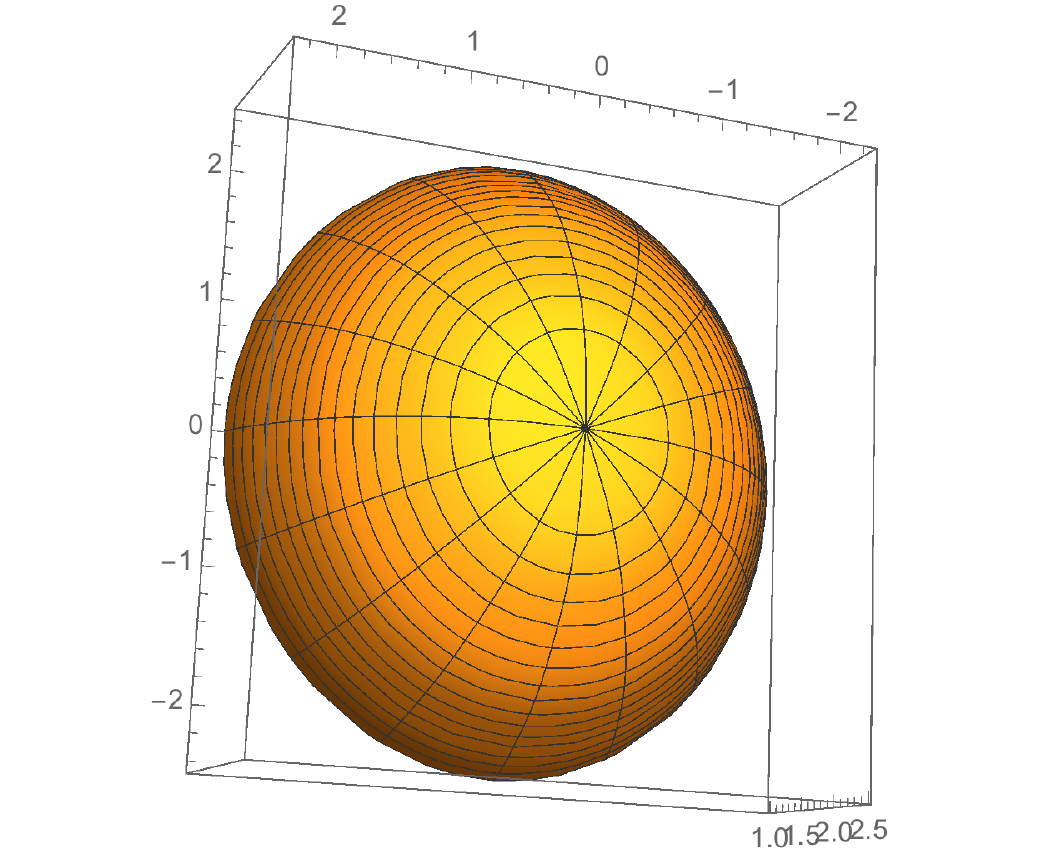} \\
\includegraphics[width=45mm]{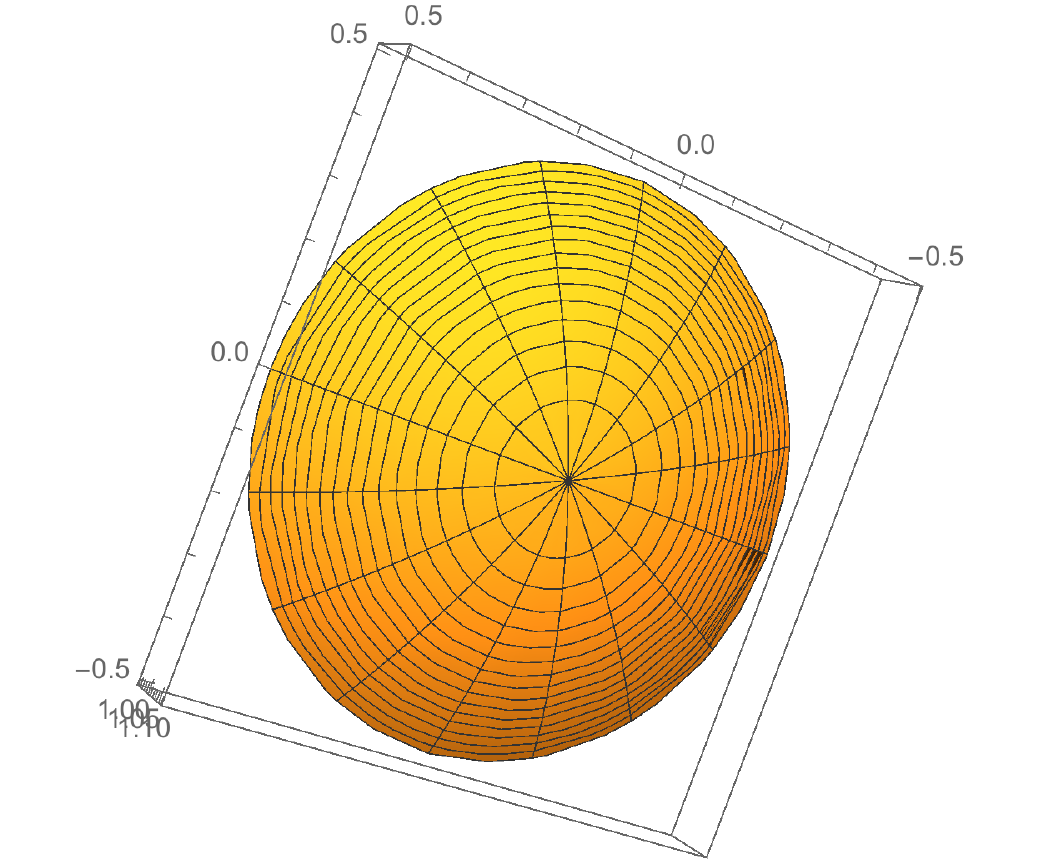}
\end{tabular}}
\caption{{\footnotesize Plots for $AdS$ pulse solutions at $\omega=0.15$. The red line denotes the action integral for $\theta$ independent solution. The black solid line denotes the oscillatory particle action, whilst blue solid is pulse particle action. Oscillatory revolution plots on the bottom line are at $E=0.4$ (top) and $E=0.9$ (bottom).}}
\label{holo1}
\end{figure}
\begin{figure}[h!]
\centerline{
\begin{tabular}{ccc}
\epsfxsize=5cm\epsfbox{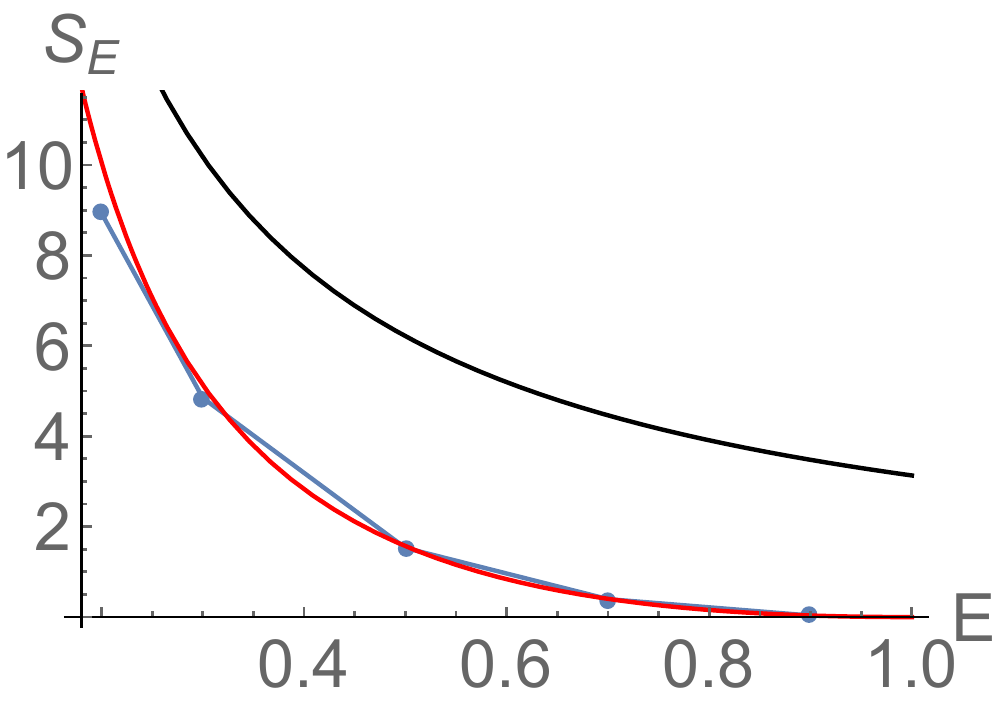}&
\epsfxsize=5cm\epsfbox{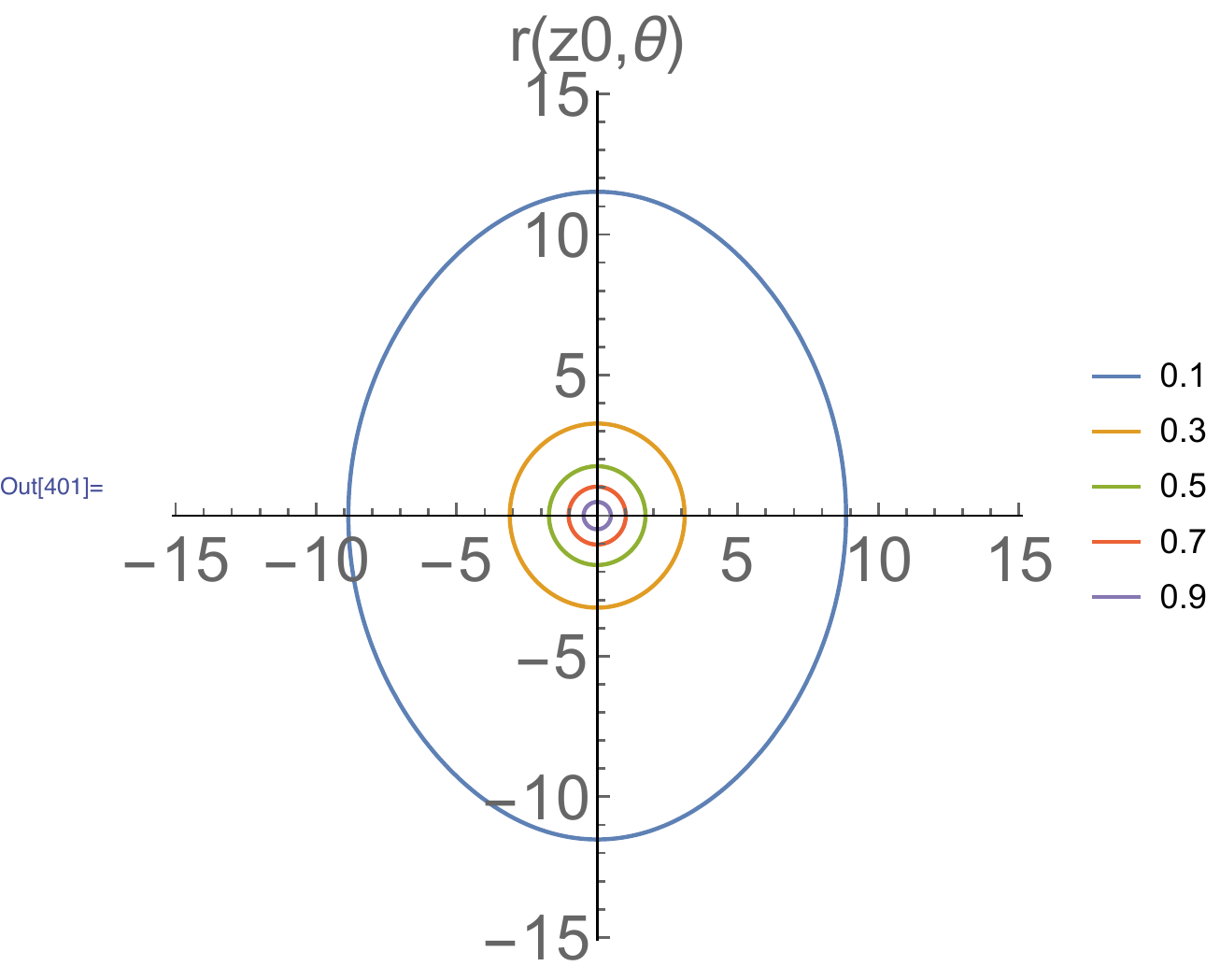}
\end{tabular}
\begin{tabular}{c}
\includegraphics[width=44mm]{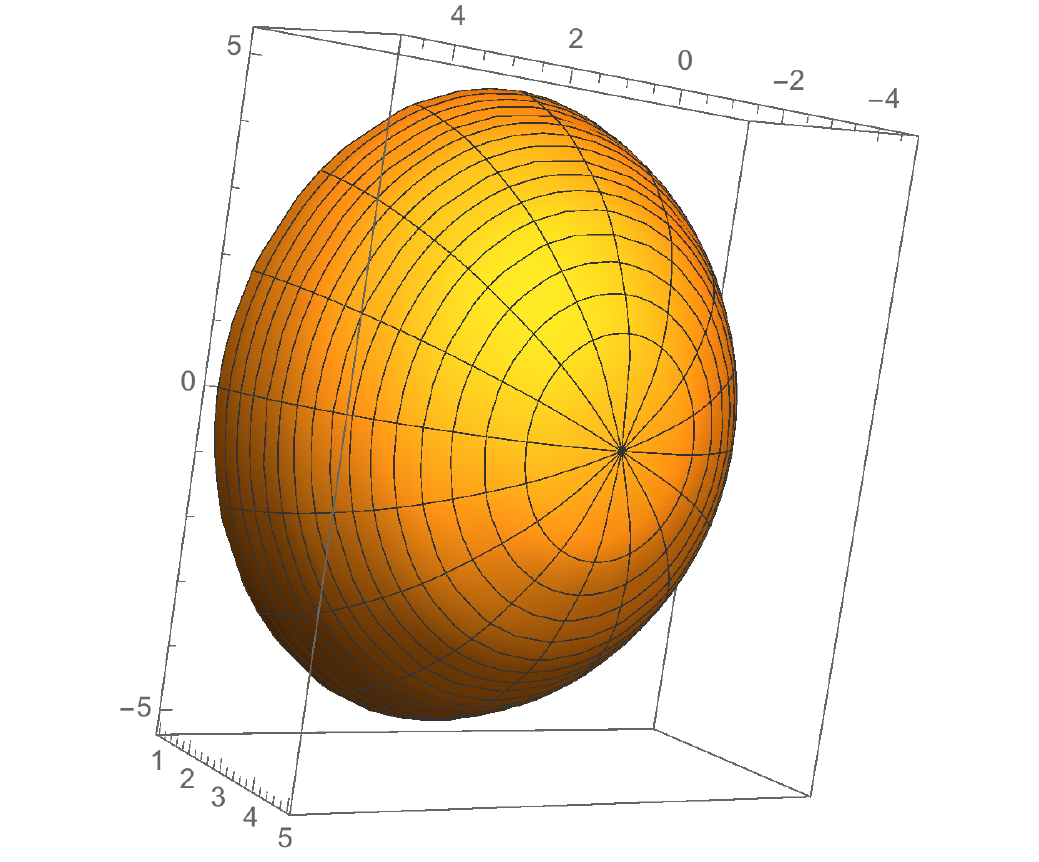}\\
\includegraphics[width=45mm]{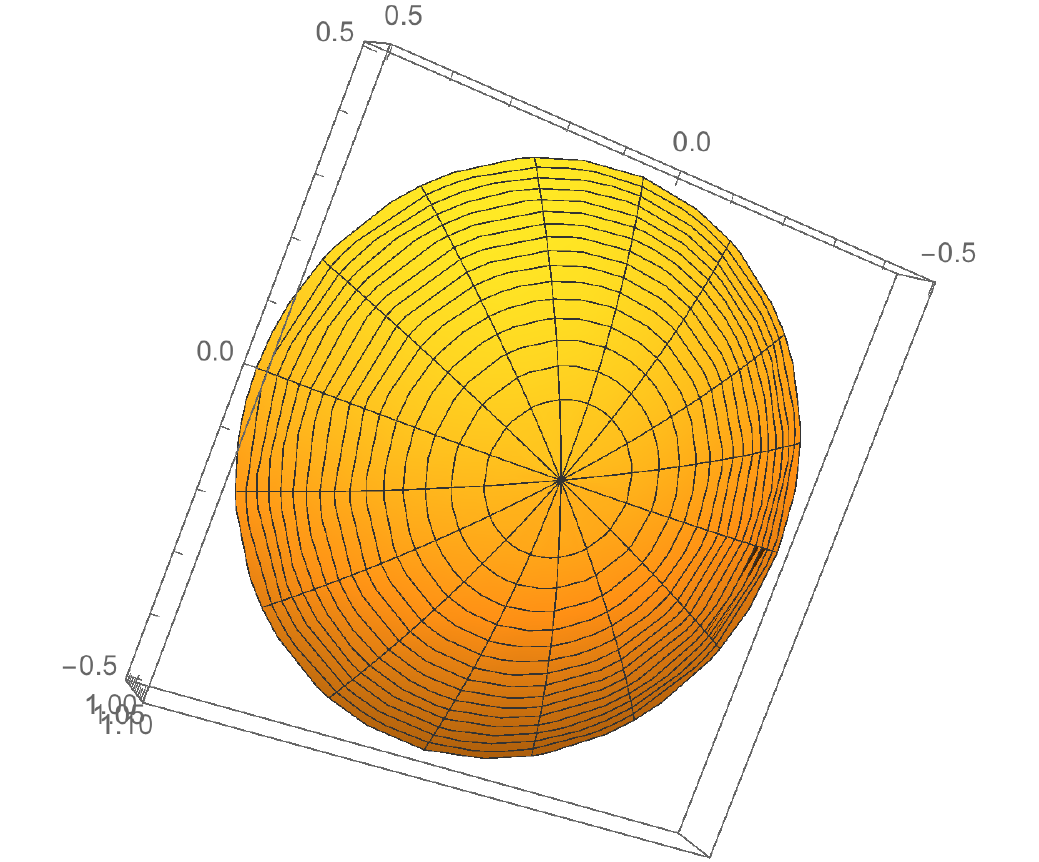}
\end{tabular}}
\caption{{\footnotesize  Same as Figure \ref{holo1} for  oscillatory  background.}}
\label{holo2}
\end{figure}

\begin{figure}[h!]
\centerline{
\begin{tabular}{ccc}
\epsfxsize=5cm\epsfbox{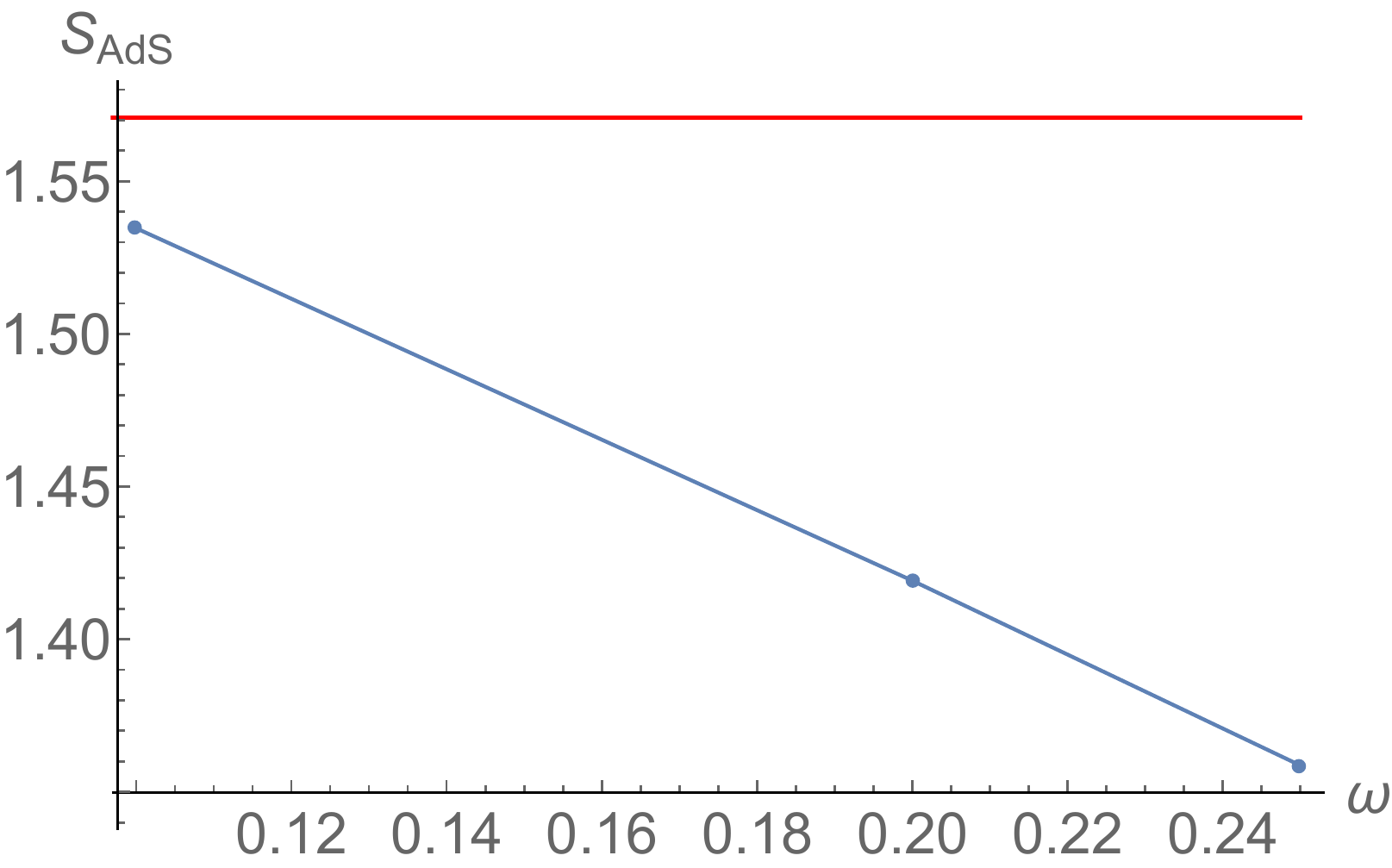}&
\epsfxsize=4.5cm\epsfbox{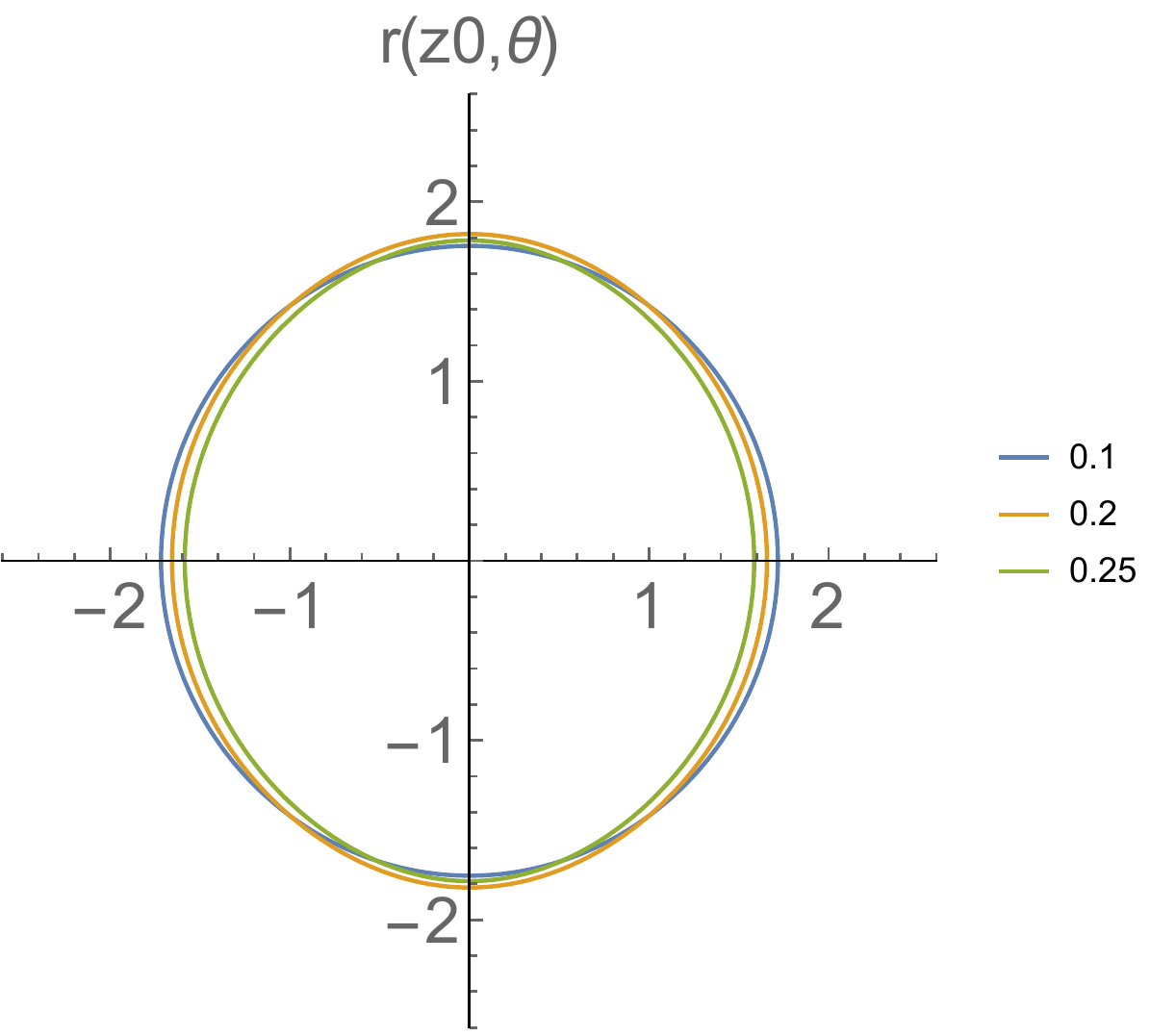}& 
\includegraphics[width=42mm]{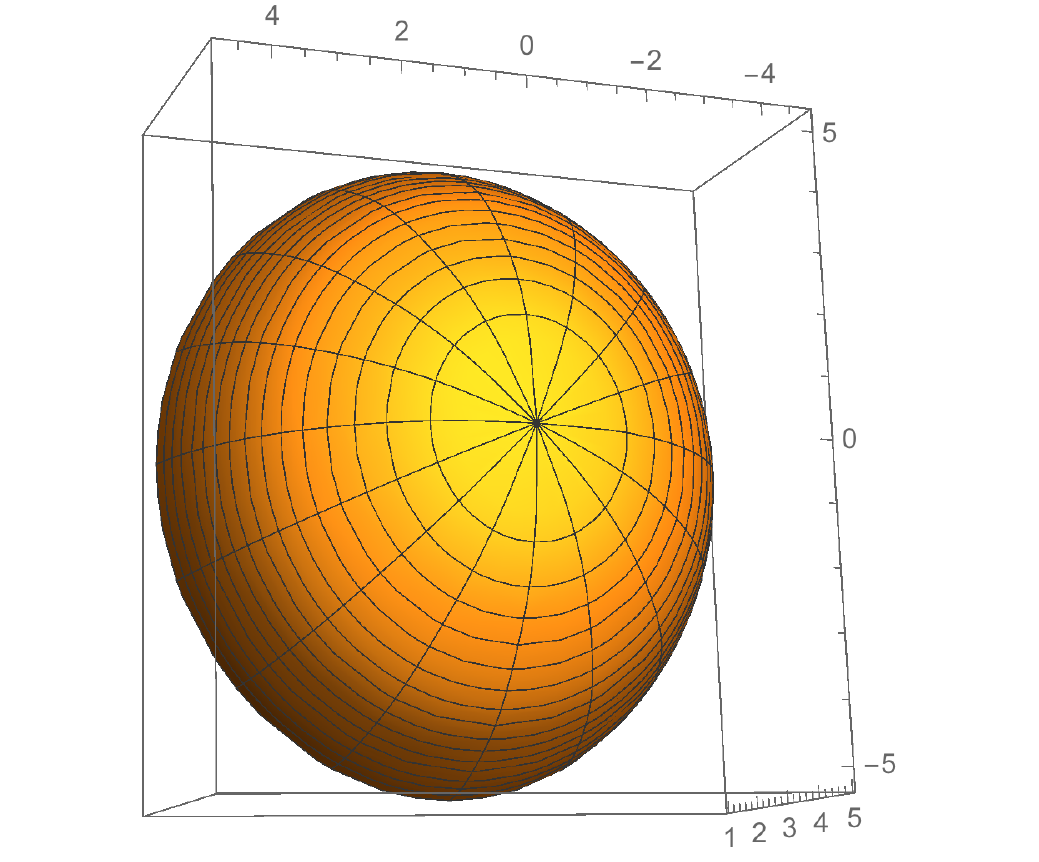}
\end{tabular}}
\caption{{\footnotesize Action and solutions at fixed $E=0.2$ for the pulse background in $AdS$ as a function of $\omega$  whose values are shown in the legend. Revolution plot is for $w=0.1$.}}
\label{holo3}
\end{figure}

\begin{figure}[h]
\centerline{
\begin{tabular}{ccc}
\epsfxsize=5cm\epsfbox{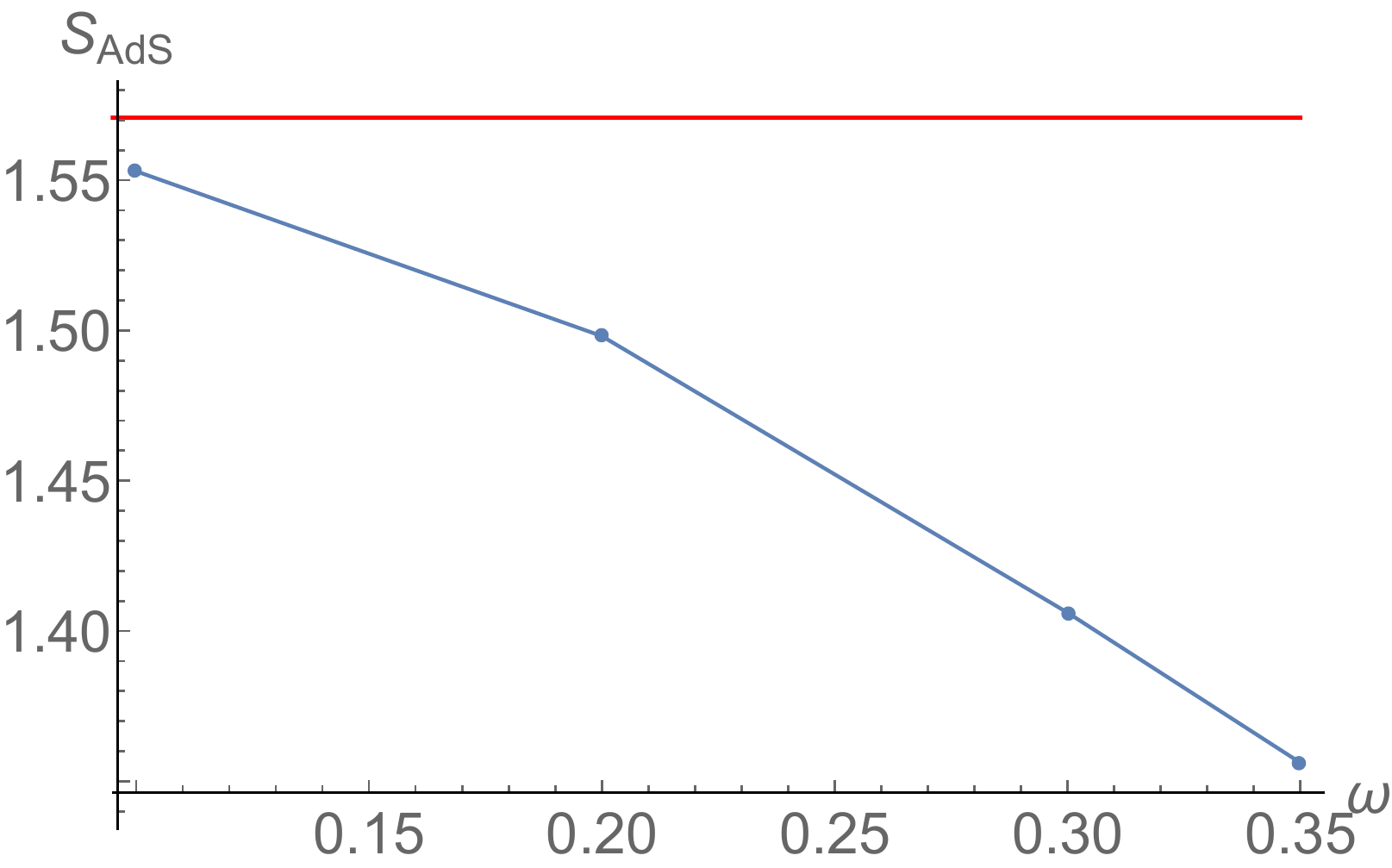}&
\epsfxsize=5cm\epsfbox{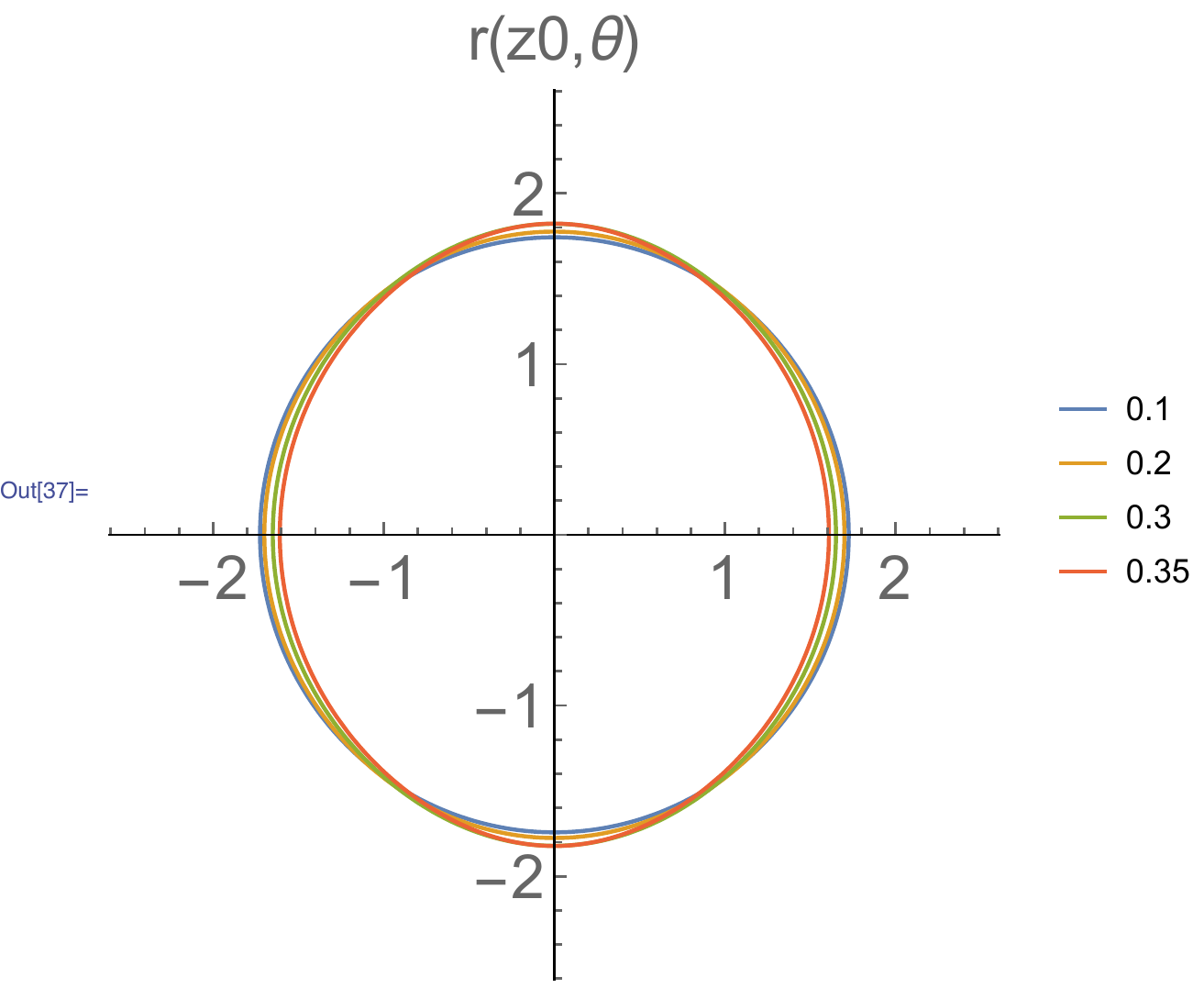}& 
\includegraphics[width=33mm]{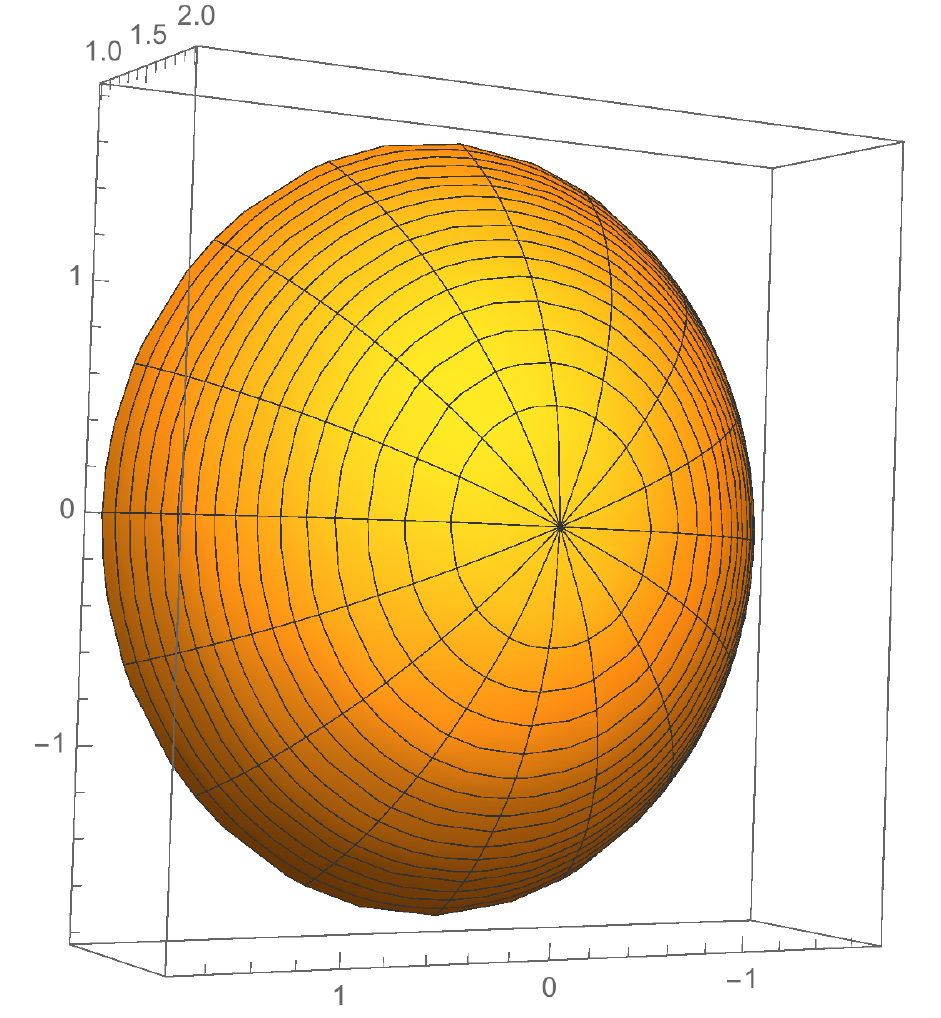}
\end{tabular}}
\caption{{\footnotesize Action and solutions at fixed $E=0.5$ for the oscillatory background in $AdS$ as a function of $\omega$  whose values are shown in the legend. Revolution plot is for $w=0.1$.}}
\label{holo4}
\end{figure}

\begin{figure}[h!]
\centerline{
\begin{tabular}{ccc}
\epsfxsize=5cm\epsfbox{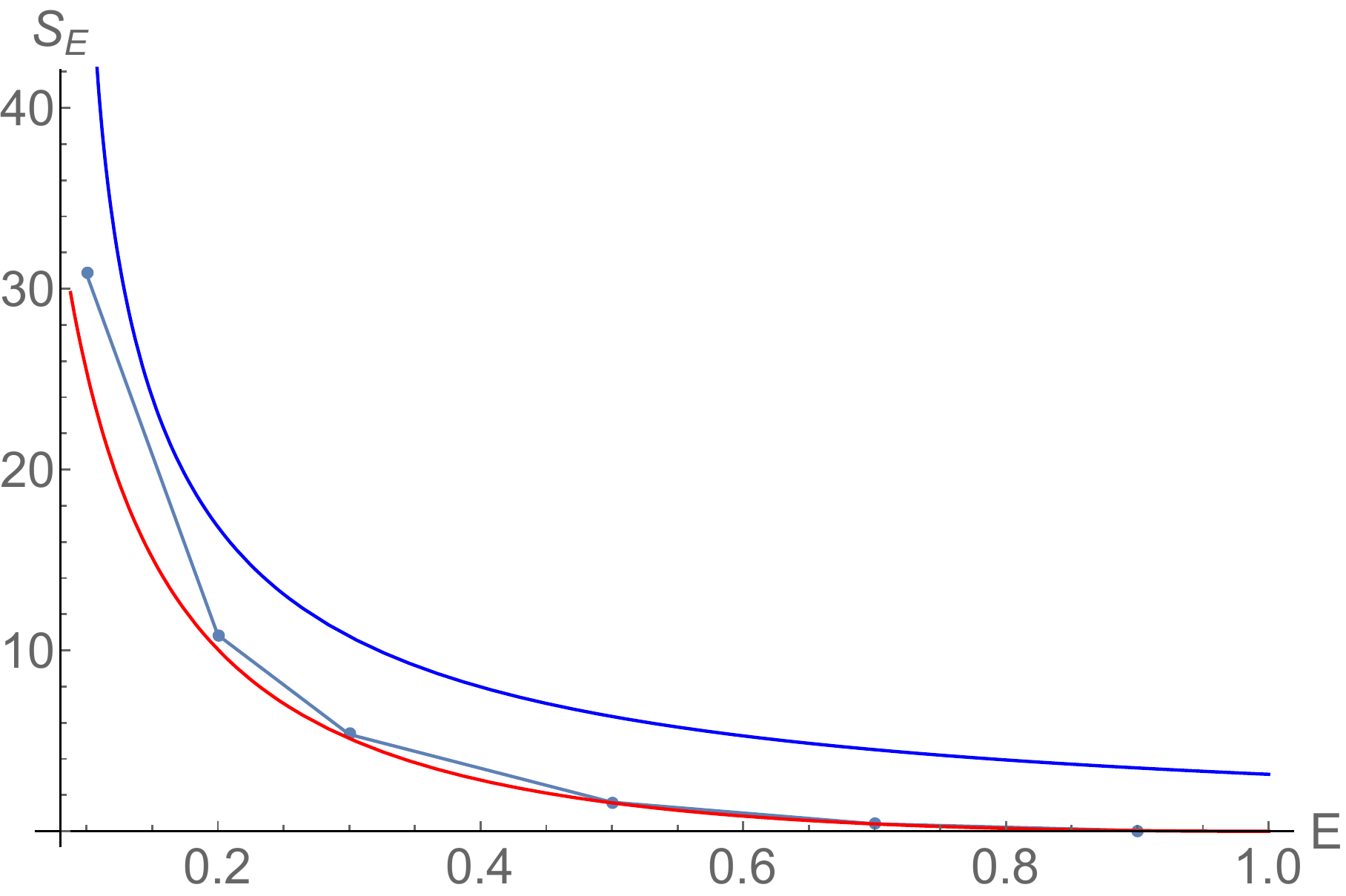}&
\epsfxsize=5cm\epsfbox{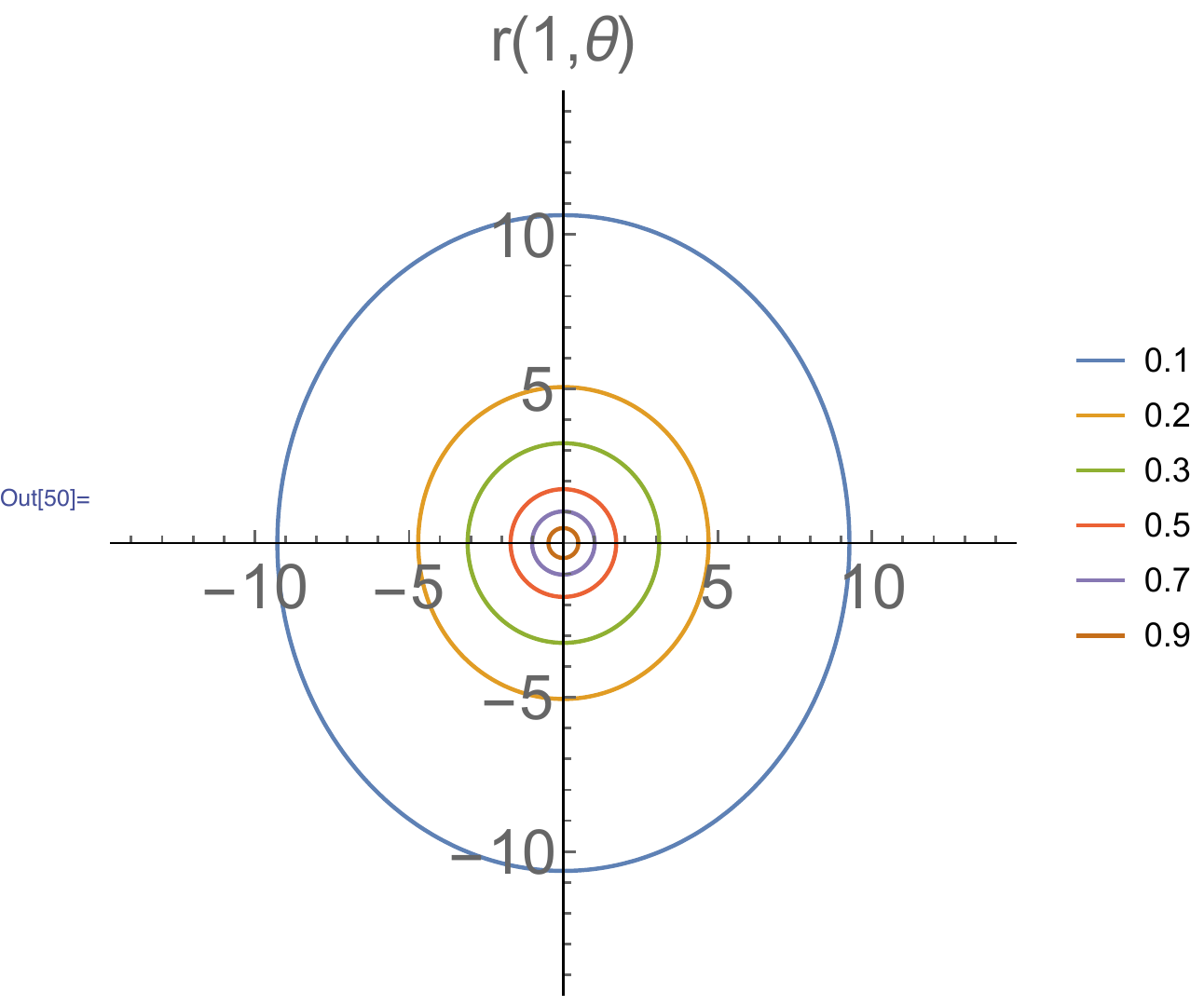}& 
\includegraphics[width=42mm]{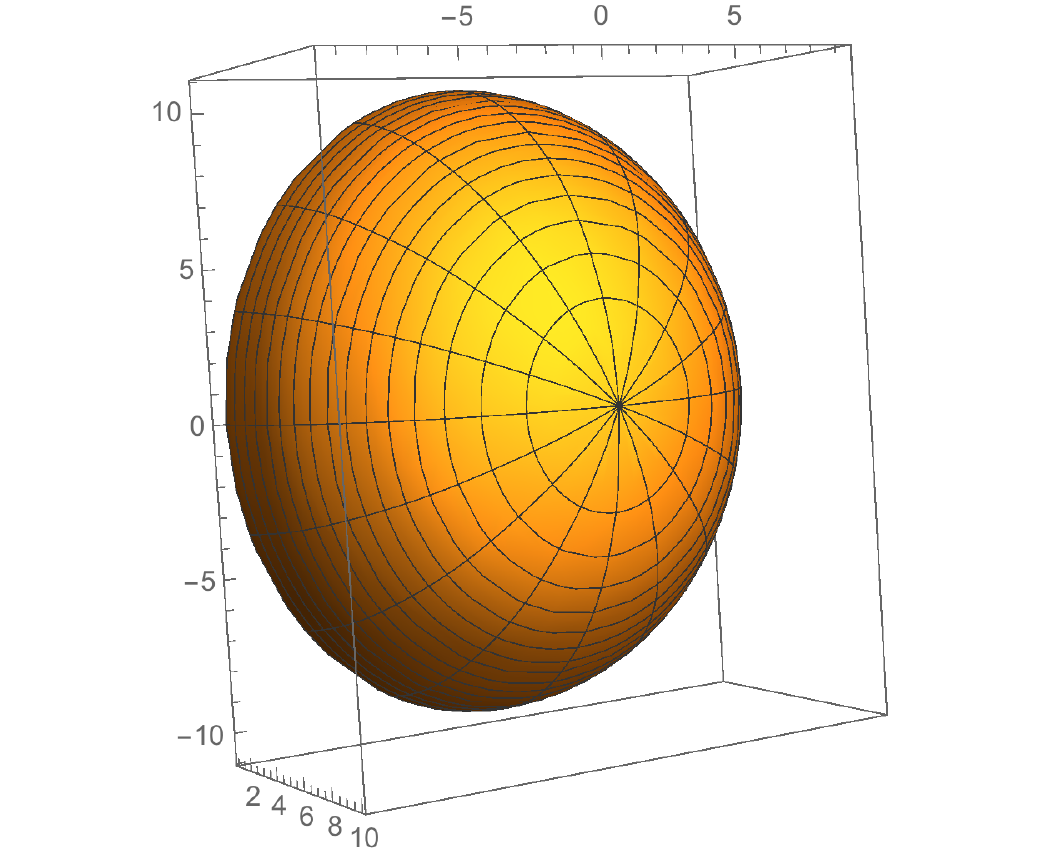}
\end{tabular}}
\caption{{\footnotesize 
Action and solutions for the spatially varying pulse background at various $E$ (shown in the legend) at $k=0.1$. 
The red curve in the action plots shows the action for the catenary solution whilst the blue curve that for the analytic pulse solution. The revolution plot is for $E=0.1$.
}}
\label{holo5}
\end{figure}

\begin{figure}[h!]
\centerline{
\begin{tabular}{ccc}
\epsfxsize=5cm\epsfbox{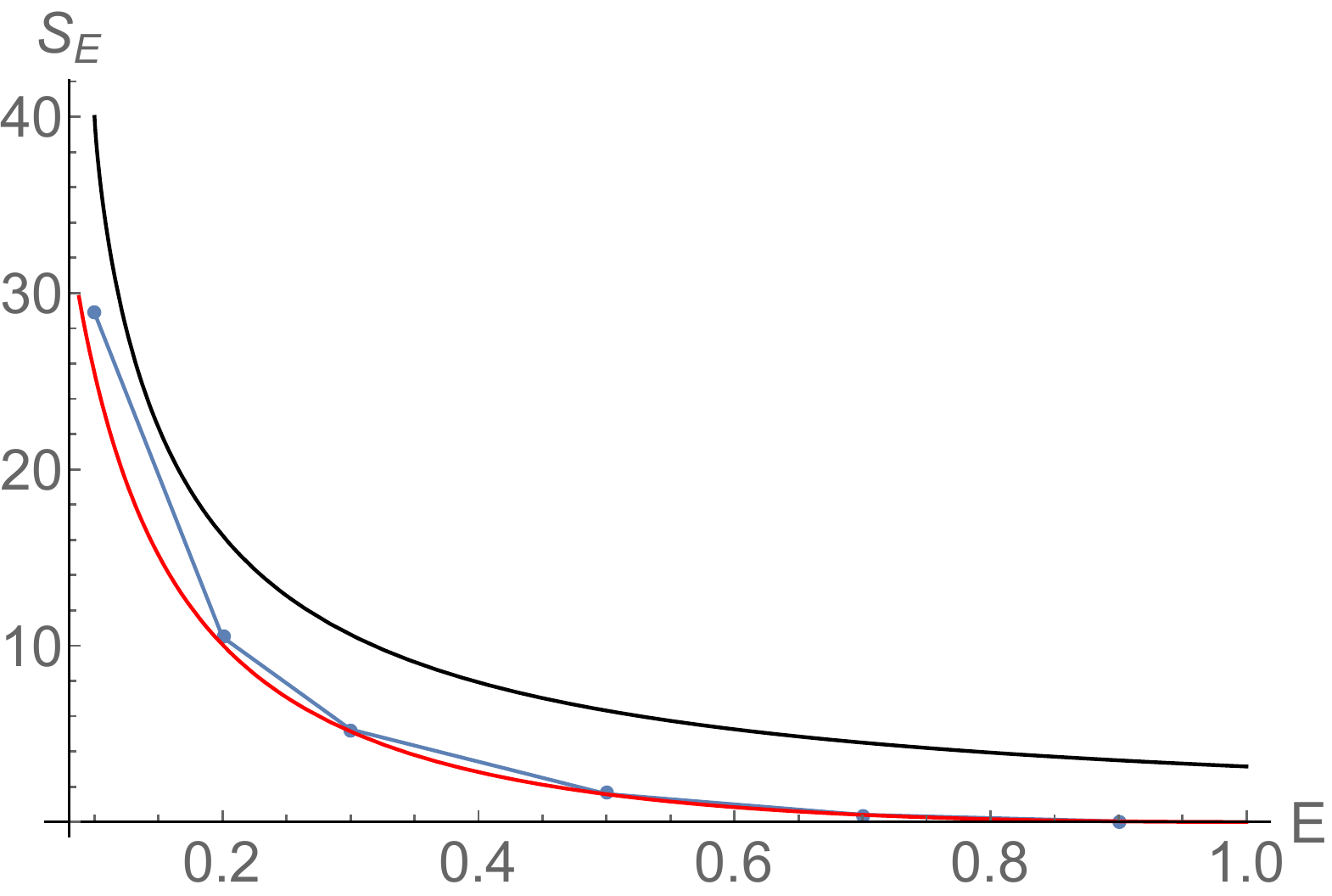}&
\epsfxsize=5cm\epsfbox{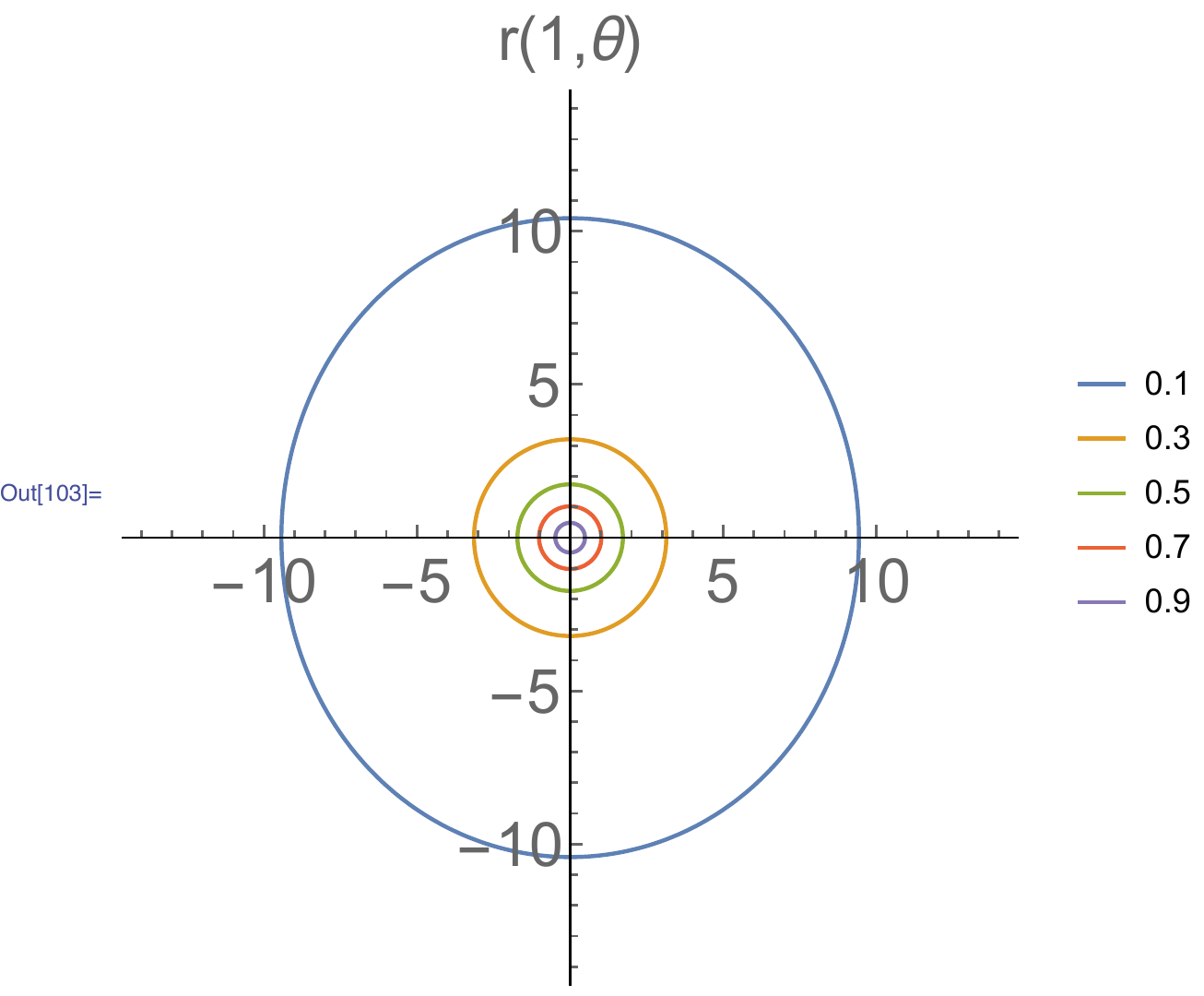}&
\includegraphics[width=34mm]{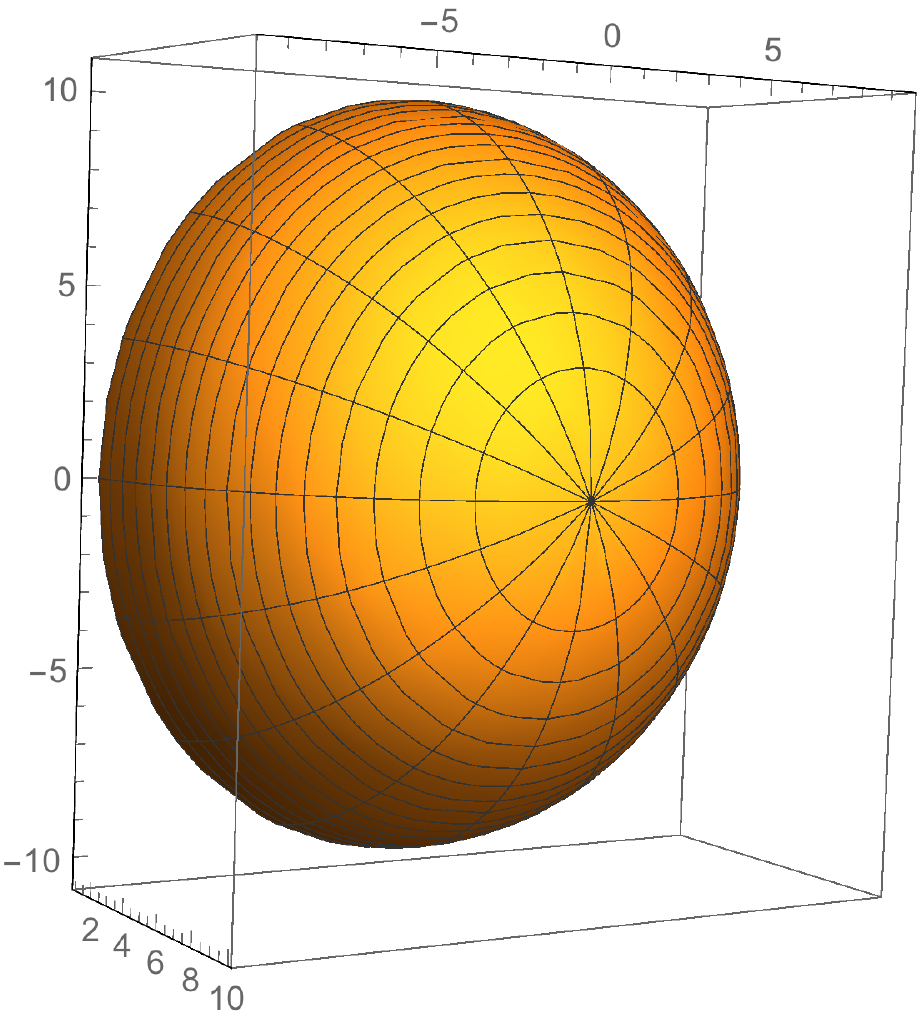}
\end{tabular}}
\caption{{\footnotesize 
Same as Figure \ref{holo5} but this time for the oscillatory background.
}}
\label{holo6}
\end{figure}

\begin{figure}[h!t]
\centerline{
\begin{tabular}{ccc}
\epsfxsize=5cm\epsfbox{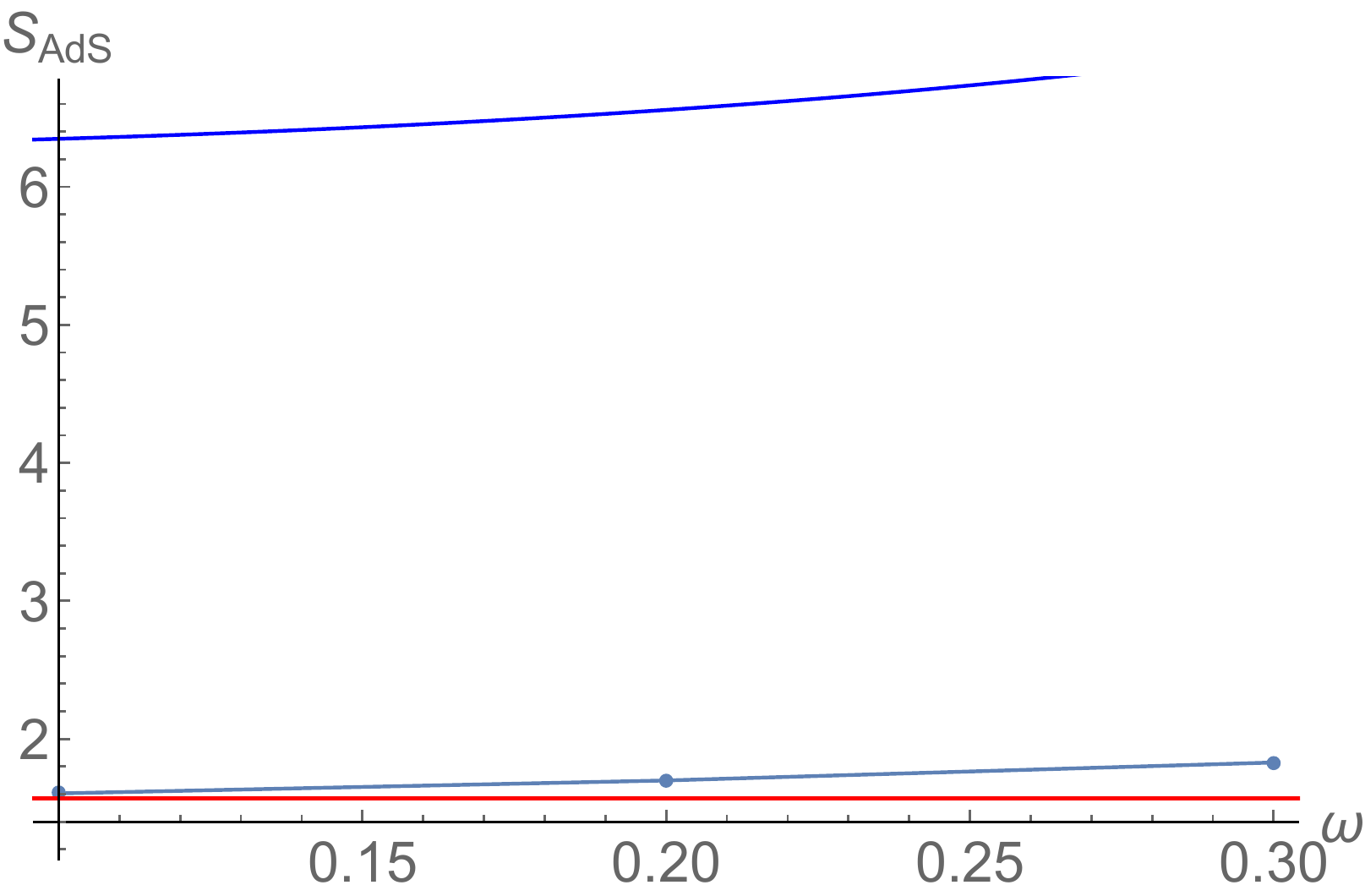}&
\epsfxsize=5cm\epsfbox{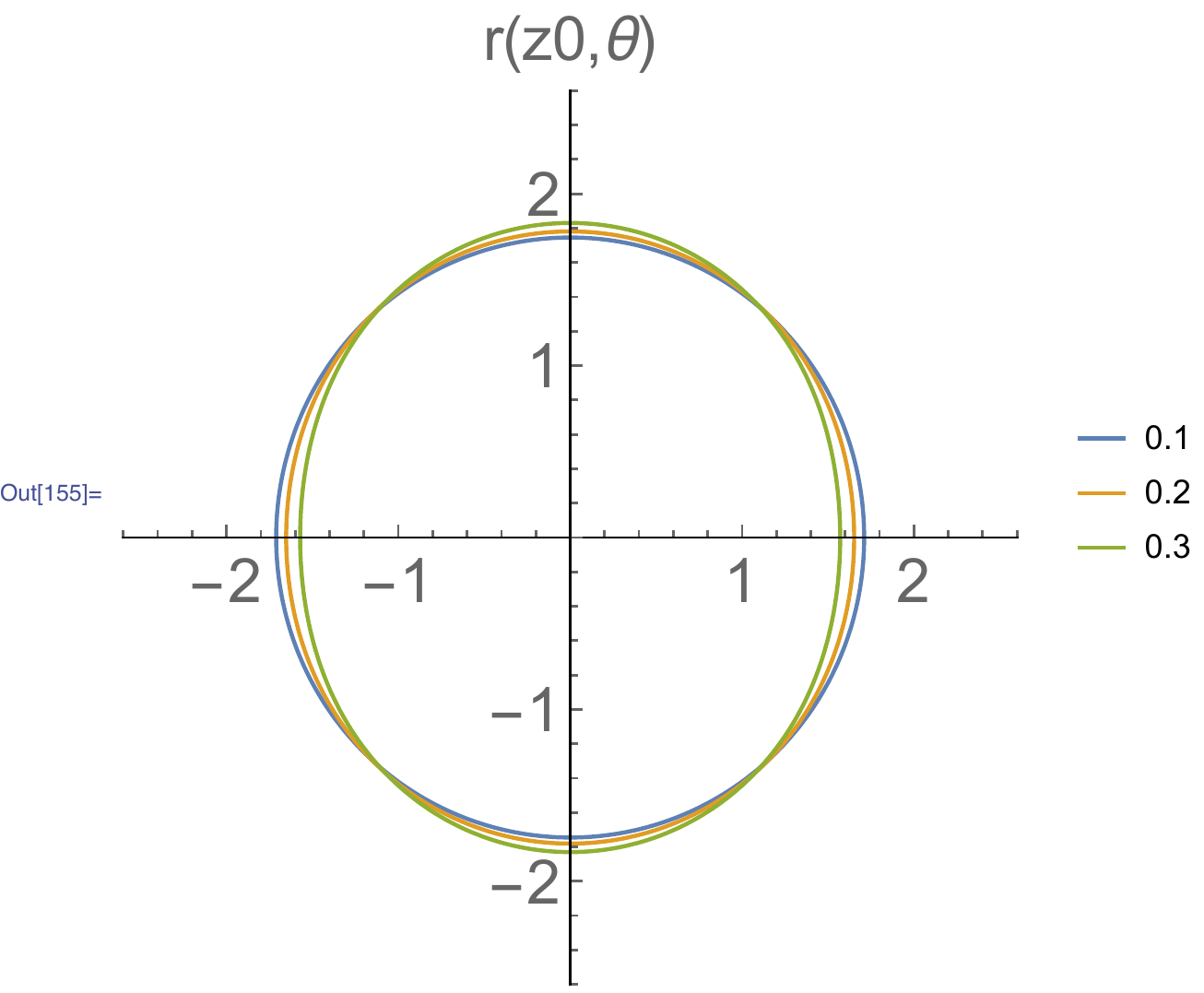}&
\includegraphics[width=35mm]{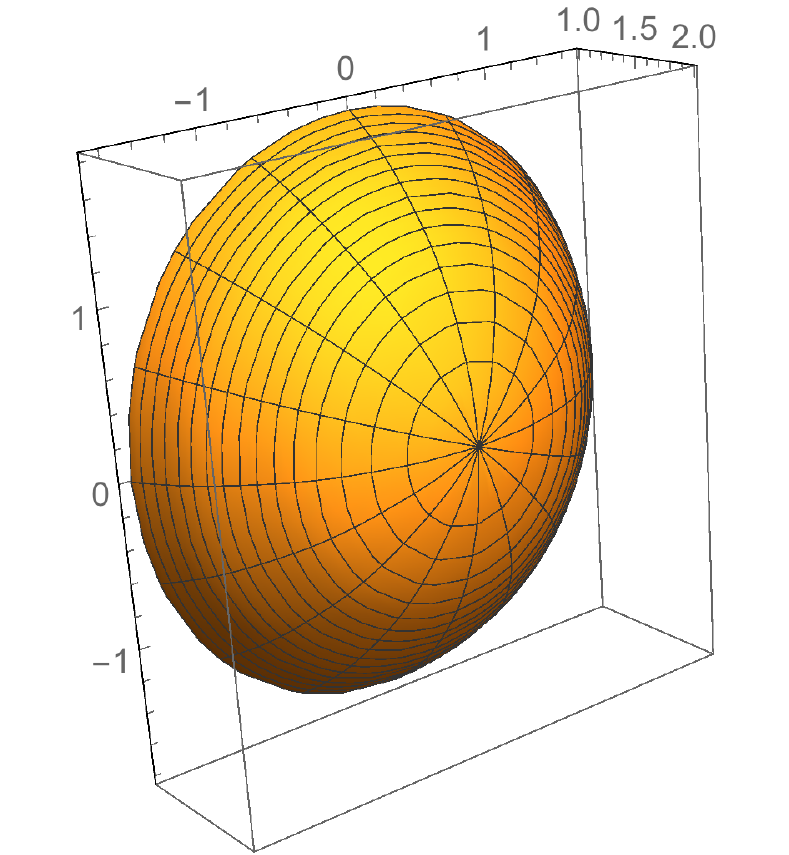}\end{tabular}}
\caption{{\footnotesize Action and solutions at fixed $E=0.5$ for the spatial pulse background as a function of $k$. The revolution plot is for $k=0.2$.}}
\label{holo7}
\end{figure}

\begin{figure}[h!t]
\centerline{
\begin{tabular}{ccc}
\epsfxsize=5cm\epsfbox{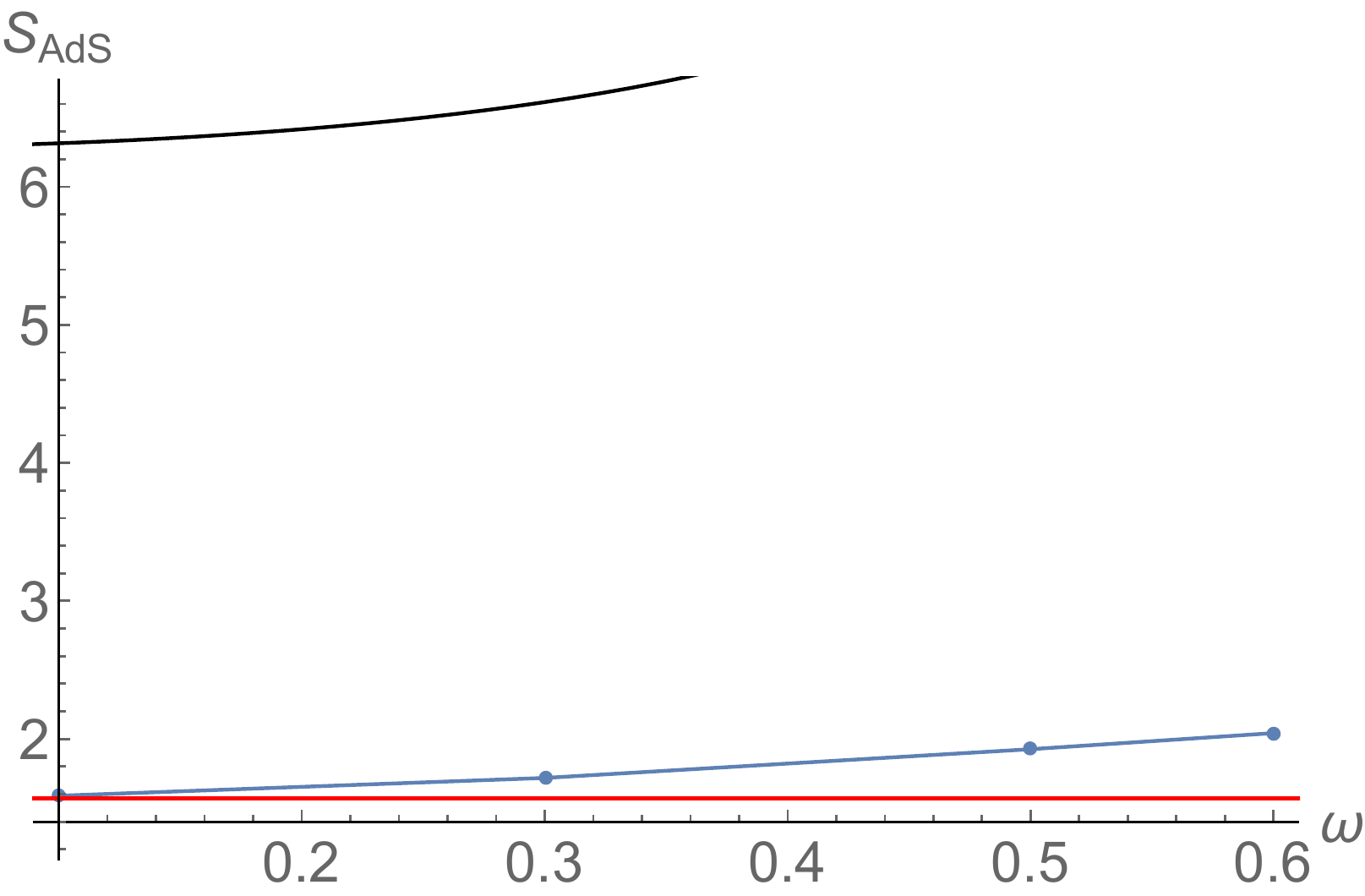}&
\epsfxsize=5cm\epsfbox{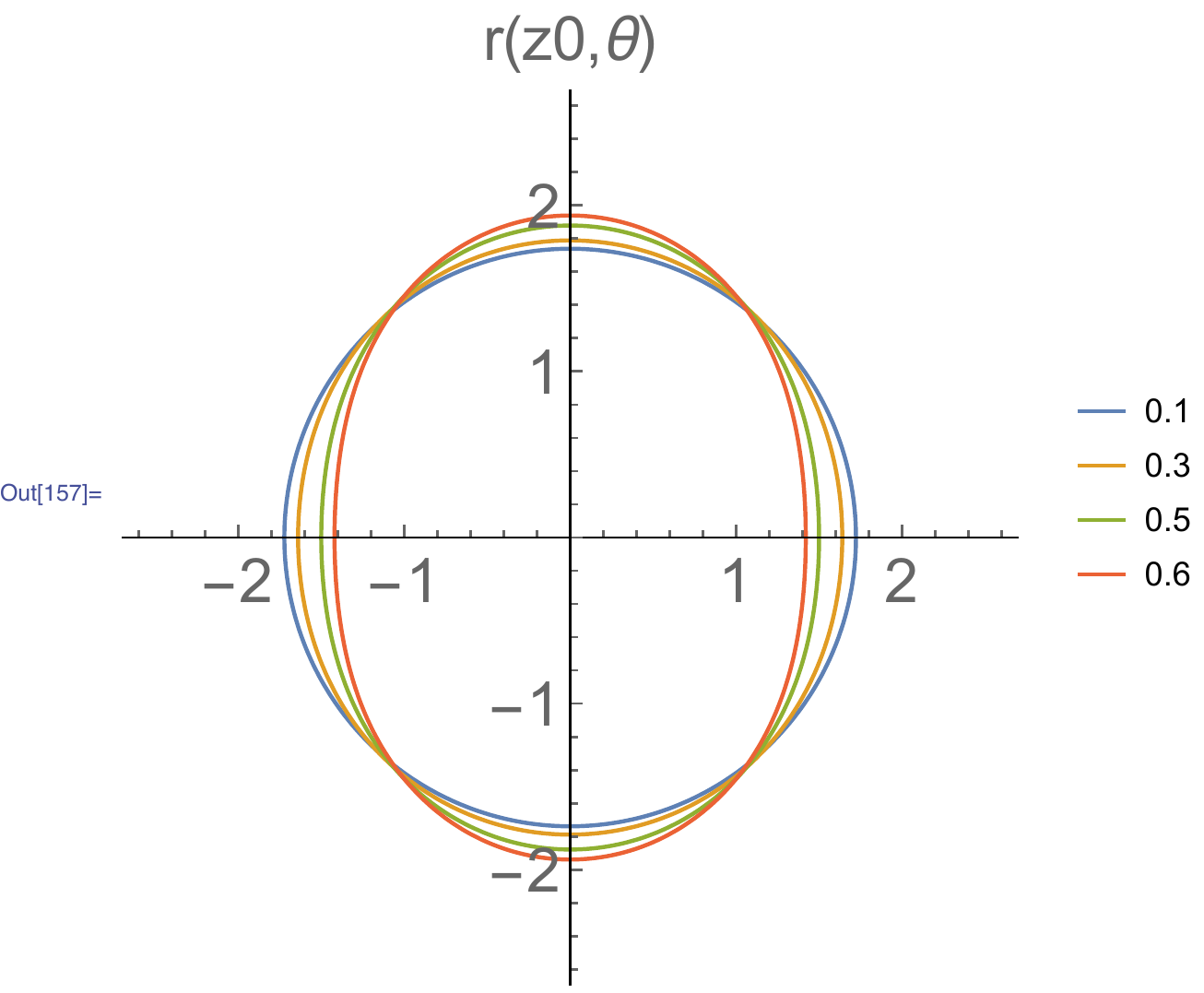}&
\includegraphics[width=35mm]{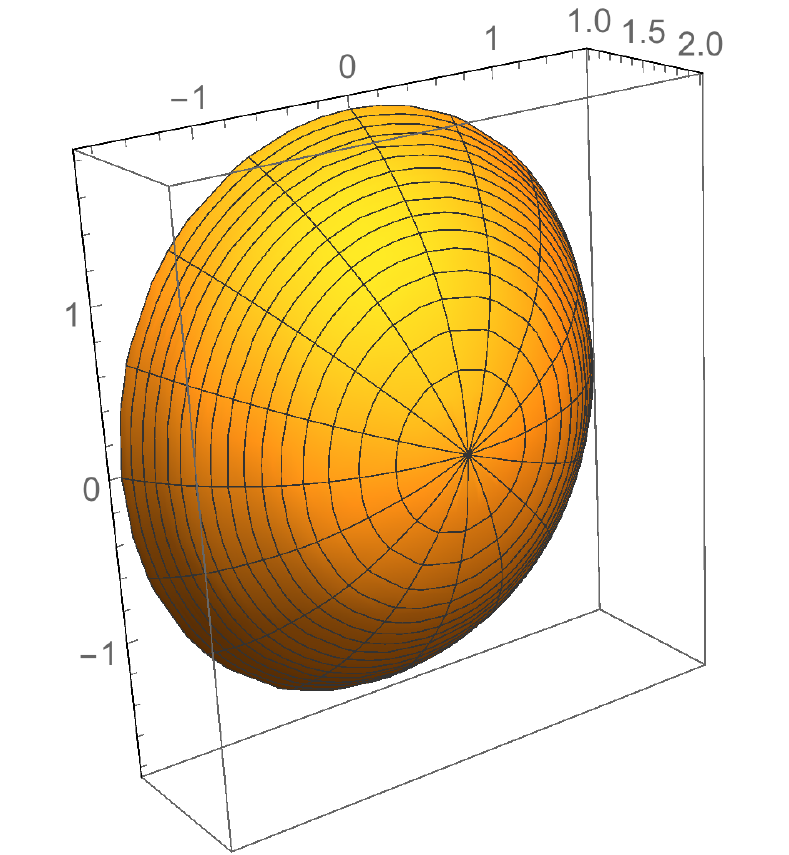}\end{tabular}}
\caption{{\footnotesize 
Same as Figure \ref{holo7} but this time for the oscillatory background as a function of $k$. The revolution plot is for $k=0.3$.}}
\label{holo8}
\end{figure}

Solutions, along with their action values, are shown in Figure \ref{holo1} for a pulse background  and Figure \ref{holo2} for an oscillatory background and $\omega = 0.15$. For a large electric field the solution converges very well to the spherical symmetric solution in constant background. This is also quite visible from the revolution plots at $E=0.9$ which show a circular cap almost entirely tangent to the brane at $z_0$. At smaller values of $E=0.5$ the solution starts to deviate from the circular shape and gives a lower value for the action. In Figures \ref{holo3} and  \ref{holo4}  we plot the variation at fixed $E$ for different $\omega$'s. 
Most of the features we see in the flat case studied in Section \ref{due} are present in the holographic background. In particular the frequency $\omega$ enhances the pair production. We do not have much numerical data in the small $E$ regime, although we expect the feature presented in the last Section to remain the same. One important difference is in the critical limit $E \to E_{\rm cr}$. In the holographic case, in contra distinction to the flat case, the radius of the Euclidean world-sheet goes to zero and this remains true also when time dependence is turned on. So we do not expect any difference with the locally constant approximation in this limit, in particular we do not expect a change in the value  of the critical field due to the non-homogeneity. 
Solutions for space dependent backgrounds are shown in Figure \ref{holo5} for a pulse background  and Figure \ref{holo6} for an oscillatory background and $k = 0.1$. In Figures \ref{holo7} and  \ref{holo8}  we plot the variation at fixed $E$ for different $k$'s. 
The features described above continue to be true with the same distinction between time and space dependent background observed for the flat space-time case. 
Is interesting to note that, unlike for the case of particles and also strings in flat space,  we can find real worldsheet instantons solutions also for $E < k$.


It is interesting to discuss  pair production in confining holographic backgrounds.
For a constant background electric field this phenomenon has been discussed in \cite{Sato:2013hyw,Kawai:2013xya,Kawai:2015mha} for a particular case of confining geometries.
We can here consider the simplest of the confining backgrounds, namely  AdS with a hard-wall located at $z = z_{\rm IR}$. 
In this case we can study the pair production in the constant electric field background and also  infer new results regarding the non-homogeneous electric field background.

In the dual theory, the confinement scale is set by the tension of the confining string. 
The confining string in the bulk corresponds to a string located at the bottom of the gravitational potential $z_{\rm IR}$
\beq
T_{\rm c} = T \frac{L^2}{z_{\rm IR}^2} \ .
\eeq
Charged particles with mass (\ref{mass}) are no longer part of the spectrum of the theory. They are now confined into chargeless states by the confining string.

\begin{figure}[h!t]
\centerline{
\epsfxsize=10cm \epsfbox{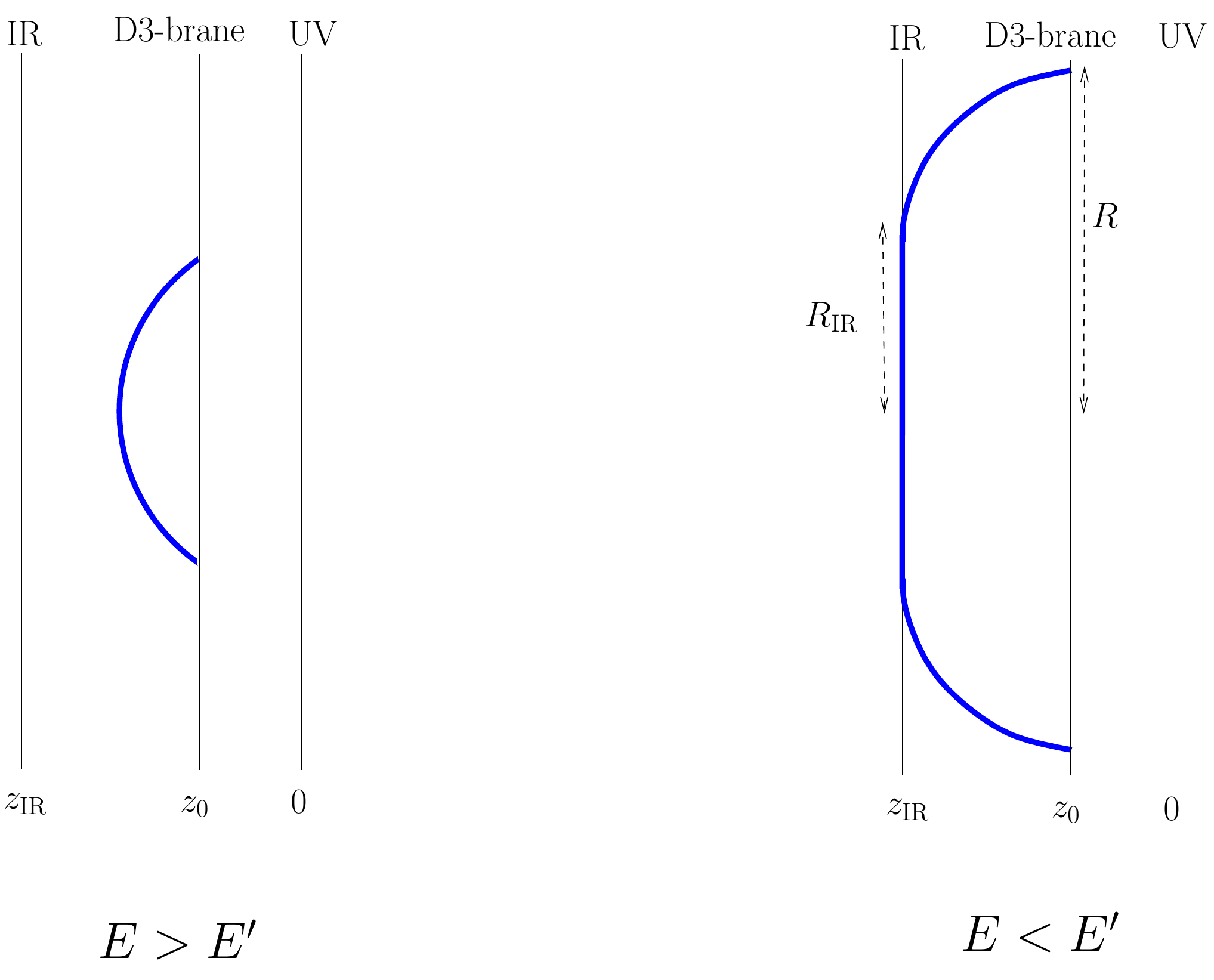} }
\caption{{\footnotesize Euclidean world-sheet for a confining hard-wall background. For $E$ smaller than a $E'$, the spherical cap is not entirely included in the physical space but is substituted by a new type of world-sheet geometry.}}
\label{conf}
\end{figure}

Now we turn on a homogeneous electric field on the D-brane located at $z=z_0$. 
The Euclidean world-sheet solution is given by the same spherical cap found before (\ref{reuclidholo}).
The presence of the infrared hard wall becomes important when the electric field is smaller  than  a certain value that we call $E'$.  For $E$ small enough  the Euclidean spherical cap cannot be entirely  contained in the physical space (see Figure \ref{conf}): the solution must then be  different from the spherical cap.
The new solution consists of two parts smoothly connected. There is a disk located at the infrared brane of radius $R_{\rm IR}$ which is then continuously connected with a minimal surface, tangent at $Z_{\rm IR}$, which end on a circle of radius $R$ on the D-brane at $z_0$.
The value $E'$ can be computed by the condition of the Euclidean world-sheet being tangent to the IR cutoff. 
This happens when $R_{\rm UV} = z_{\rm IR}$ which, using (\ref{rtilde}) and (\ref{reuclidholo}), gives:
\beq
\label{ircritical}
q E'  = \frac{T L^2}{ z_0 z_{\rm IR}}  \ .
\eeq
Note that from the condition $z_{\rm IR} > z_0$ we have the relation between the two critical fields $E' < E_{cr}$.

\begin{figure}[h!t]
\centerline{
\epsfxsize=7.0cm\epsfbox{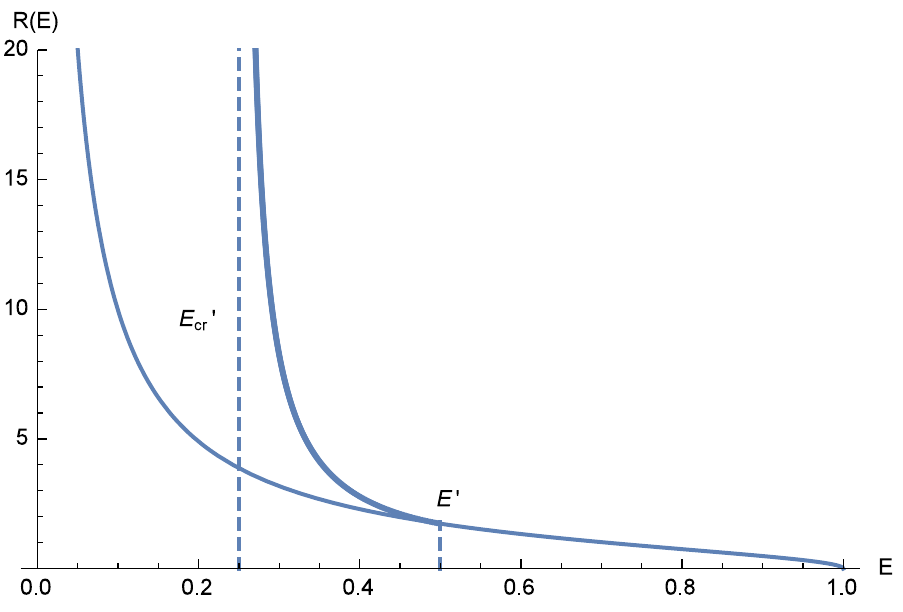} \quad
\epsfxsize=7.0cm\epsfbox{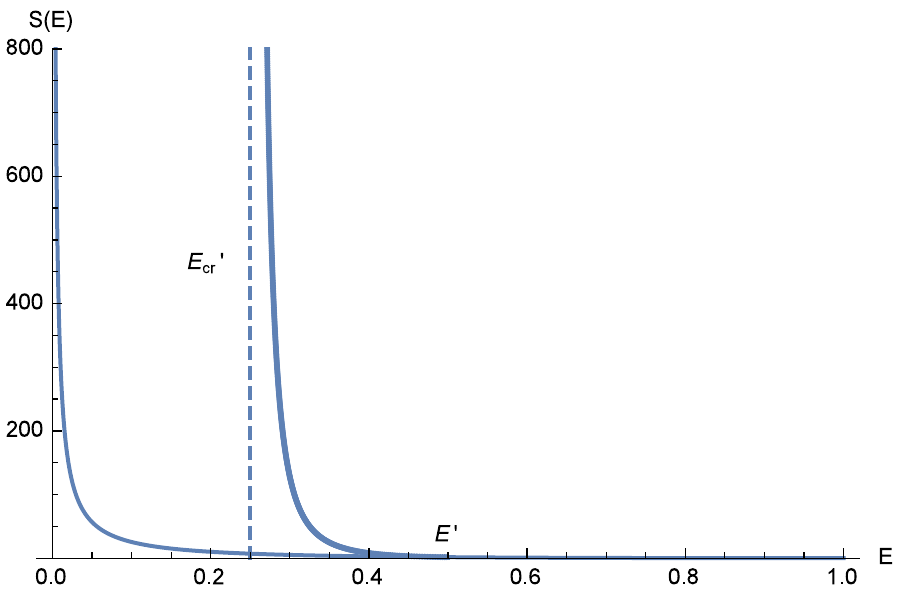}}
\caption{{\footnotesize On the left we present the radius and on the right the action for the solution in the confining hard-wall holographic background for a constant electric field. Note that the radius and the action go to infinity at a low critical field below which there is no pair production. 
We also plot the corresponding curves for the particle case.  The normalization of parameters is $q=T=1$ and $z_0=1$, $z_{\rm IR} = 2$, so that the critical field are  $E_{\rm cr}'=0.25$, $E'=0.5$ and $E_{\rm cr}=1$.
}}
\label{holoconf}
\end{figure}

For $E < E'$ the solution is the right one of Figure \ref{conf}.
The Euclidean action for this type of surface is 
\beq
\label{actiond1string2}
S_{E} = T \pi R_{\rm IR}^2 + T \int_{R_{\rm IR}}^{R} d\rho 2\pi \rho  \left(\frac{L}{z(\rho)}\right)^2  \sqrt{1+ z'(\rho)^2 } -q E \pi R^2
\eeq
We can solve it with the following strategy. We start with a generic value of $R_{\rm IR}$ then solve numerically the minimal surface equation
\beq
2 r \left(z'(r)^2+1\right)+z(r) \left(r z''(r)+z'(r)^3+z'(r)\right)=0
\eeq
with boundary condition $z(R_{\rm IR}) = z_{\rm IR}$ and $z(R_{\rm IR})'=0$. The last condition assures the smooth connection with the disk at $ z_{\rm IR}$.  From the equations
\beq
z(R) = z_0 \ , \qquad  \quad q E = T \sqrt{\frac{1}{1+z'(R)^2}}
\eeq
we determine $R$ and $E$. 
The solution is presented in Figure \ref{holoconf} for $R$ and $S_{E}$ both as function of $E$. 
The spherical cap solution is valid for $E'< E < E_{\rm cr}$ for $E<E'$ we have instead the new type of solution.
The radius and the action diverge for a low critical field $E_{\rm cr}'$
\beq
\label{criticallow}
q E_{\rm cr}' =  \frac{T L^2}{z_{\rm IR}^2} = T_{\rm c} \ .
\eeq 

The interpretation of the critical low field (\ref{criticallow}) is quite natural in the Minkowski formulation of the theory. 
The potential between two charged particles can be approximated as the sum of the bare masses plus the confining term and the electric field potential
\beq
V(R) = T L^2 \left(\frac{1}{z_0}-\frac{1}{z_{\rm IR}}\right) + T_{\rm c} R - q E R 
\eeq
Tunneling can happen only if the background field is larger than the confining force, that is  $q E > T_{\rm c}$.
There is thus a critical low field  (\ref{criticallow})
below which pair production can not happen. The Euclidean world-sheet action goes to infinity as $E \to E_{\rm cr}'$ from above. 
Note that 
\beq
E_{\rm cr}' < E' < E_{\rm cr}
\eeq
The field $E'$ signals a transition in the type of world-sheet solution and its presence is a peculiarity of the hard wall geometry. For smooth confining geometries there would not be a clear analogue of $E'$.

\begin{figure}[h!t]
\centerline{
\epsfxsize=7cm \epsfbox{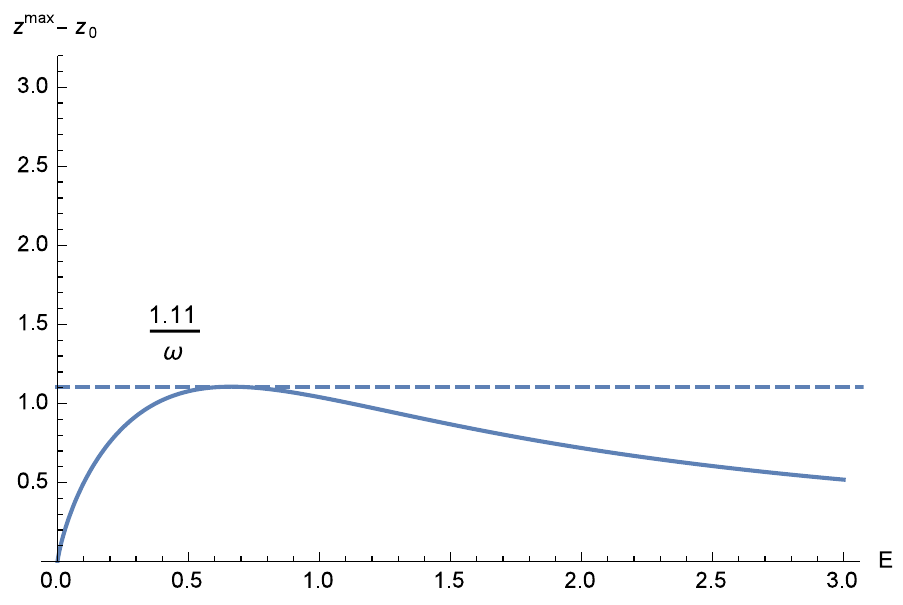} \qquad 
\epsfxsize=7cm \epsfbox{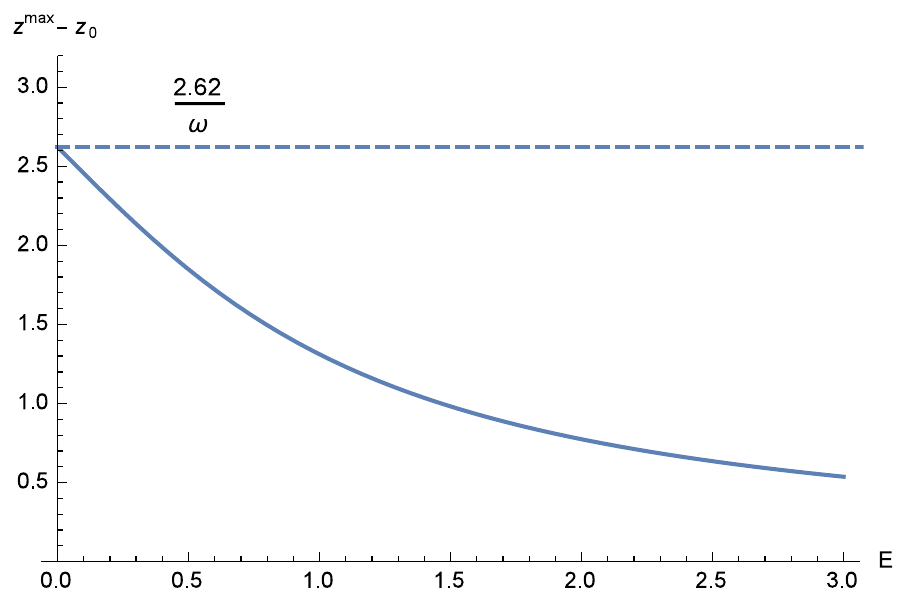}}
\caption{{\footnotesize Maximum holographic $z$ for the Euclidean world-sheet in the limits for the pulse (left panel) and oscillatory (right panel) backgrounds. This is valid in the  regimes  $z_0 \ll 1/\omega$ and $x_3^{\rm max}/ x_4^{\rm max} \ll 1 $.}}
\label{confinement}
\end{figure}

The critical low field $E_{\rm cr}'$ is modified by the frequency $\omega$ of the electric field background. We give an example in which we  can show  $E_{\rm cr}' = 0$, so that pair production can happen at any arbitrarily low electric field, despite the presence of the confining potential.  
We consider the case  $\omega \ll 1/z_0$. The low $E$ limit is described by the particle approximation. The world-sheet geometry consists of  a minimal surface which terminates on the D-brane on a curve like the one obtained in the particle limit, namely (\ref{looppulse}) for the pulse background and (\ref{looposcillatory}) for the oscillatory background. Moreover for small $E$ the curves have the property of being much more extended in $x_4$ than $x_3$, namely $x_3^{\rm max}/ x_4^{\rm max} \to 0 $ as $\gamma \to \infty$. In this limit we can compute how much the minimal surface can dip into the holographic region. The case of a minimal surface which ends on the D-brane on two straight parallel  lines infinitely extended in $x_4$ separated by a distance at $2 \Delta x_3$  can be solved analytically and the maximum depth reached in the holographic direction is\footnote{$K$ and $E$ are complete elliptic integrals of first and second kind.}
\beq
z_{\rm max} - z_0 = c \Delta x_3
\eeq
where $c$ is 
\beq
c =  \frac{1}{E(-1)-K(-1)} \simeq 1.67 \ .
\eeq
The minimal surface dips a bit further than the spherical cap for which $c$ must be replaced by $1$. 
When $x_3^{\rm max}/ x_4^{\rm max} \ll 1 $ we can use this approximation and thus we get 
\beq
z_{\rm max}(E) - z_0= c \, x_3^{\rm max} = \left\{ 
\begin{array}{cc}
  \frac{c}{ \omega \sqrt{1+ \gamma^2}}\,  {\rm arcsinh}{\left( \gamma  \right) }  & {\rm pulse} \\ & \\
 \frac{c}{ \omega }\, \, \sin ^{-1}\left(\frac{\gamma  \text{cd}\left(4 K\left(\frac{\gamma^2 }{{\gamma ^2+1}}\right)|\frac{\gamma^2 }{{\gamma ^2+1}}\right)}{\sqrt{\gamma ^2+1}}\right) & {\rm oscillatory}
\end{array}
\right.
\eeq
These functions are plotted in Figure \ref{confinement}. We note one important feature: $z$ is never bigger than
\bea
&&z_{\rm max} \simeq  z_0 + \frac{c \  0.66}{  \omega } \simeq \frac{1.11}{\omega} \  \qquad {\rm pulse}  \ ,\nonumber \\
&&z_{\rm max} = z_0 + \frac{c \pi}{ 2  \omega } \simeq \frac{2.62}{\omega} \  \qquad {\rm oscillatory} \ ,
\eea
were we also used $z_0 \ll 1/\omega$ as discussed above.
This means that in a confining background with $z_{\rm IR} > z_{\rm max}$ the world-sheet is never deep enough to reach the hard-wall, thus pair production always happens irrespectively of how small is the electric field $E$. This is quite different from what happens for the constant electric field background.  This phenomenon has a physical interpretation. The time-dependent electric field background is composed by photons that have quantum energy $\omega$ and can be absorbed for the pair production. Irrespectively of how large is the charged particles mass, they can be pair-produced and then decay into the lowest energy states, the glueballs of mass $1/z_{\rm IR}$. This means that pair production can always happen if the photon energy is sufficient to produce  glueballs, with the charged particles as intermediary states. 

\section{Conclusions and open issues}
\label{conclusion}

In this paper we studied string pair production in non-homogeneous electric backgrounds using the instanton 
technique. This led us to select two problems which could be 
addressed by this method, strings suspended between two separated D-branes in 
flat space-time, and the holographic Schwinger effect. We have 
found that in those cases the production rate of string pairs has 
been enhanced in the presence of time dependent electric fields 
even more than that of the particle pair production. For spatial 
dependent backgrounds the string pair production rate was 
found to be less suppressed than that of particle pairs. This 
behaviour could be ascribed, this in Euclidean version,  to the 
difference between the geometries of world lines and world 
strings. The Minkowskian aspects were more intuitive.   In the presence of a confining background we found that the IR critical electric field, which is there for a constant electric  field background, disappears for large enough frequencies. 
 It would be interesting to use the instanton technique for the case of strings on coincident D-branes or with Neumann boundary conditions.
 In those cases  we cannnot use the instanton technique as we did in the present paper. A  generalization of this method for constant electric field has been proposed in \cite{Schubert:2010tx} and it may be suitable for the discussion of non-homogeneous backgrounds.
We have 
not gone in this work beyond the semiclassical approximation and have 
not studied this in the direct manner in which it was possible to 
study it in string theory for the case of constant electric \cite{Bachas:1992bh} and 
magnetic fields \cite{Ferrara:1993sq}.  Such methods were discussed in \cite{Durin:2003gj} for time dependent 
backgrounds. The main drawback of such a direct attempt is that 
while the conditions for the absence of a bulk conformal anomaly 
are fulfilled, the fact that the Maxwell equations are not obeyed 
by the background configurations is likely to cause a boundary 
renormalization group flow to an actual solution of Maxwell's 
equations.  
We consider these backgrounds in the spirit of the semiclassical 
approximation hoping to return in the future to find the string 
production rates exactly in an exact string theory background.
For the string sector we discussed in this paper, that is the string 
stretched between the two
branes which has a mass gap, this particular background would 
not create any pathology such as bulk tachyonic instabilities, at 
least as long as  the maximal value of the electric field  is smaller than the critical field.

 In this paper we considered only the part of pair production probability given by the classical action $\Gamma \propto e^{S_E}$ and we have not computed the pre-exponent given by the fluctuation around the classical solution. This computation has been done for the case of the particle pair production in \cite{Dunne:2006st} and for a case of holographic string pair production in constant background \cite{Ambjorn:2011wz}.  The computations in the string and non-homogeneous case remain, as does the disentangling the perturbative part for small fields,
 an open problem.

The electric backgrounds discussed have only one-dimensional 
dependence, either in space or in time. This makes the 
calculation in the instanton approximation tractable (it reduces to a two-dimensional PDE) but the price 
one pays is that these backgrounds
are not proper solutions of the Maxwell equations in the vacuum, 
some charge density distribution (for space dependence) or 
current density (for time dependence) are needed to support 
these solutions.
In fact, this is no different than a calculation in field theory  in a 
given electromagnetic background which does not obey the 
Maxwell equations. 
The time dependent pulse and the oscillatory backgrounds can be considered as approximations for
the more realistic case of two oppositely traveling electro-magnetic waves, which meet at a focal point where the magnetic field cancels and the spatial inhomogeneity  is confined to be transverse to
the time dependent electric field. 
In general, to study the pair production in the background which is a full solution of the Maxwell equations, one has to consider the effects of multidimensional inhomogeneities as well as those of the magnetic field (see for example \cite{Kim:2003qp}). This problem is rather complicated
already for particles and we did not find a rich literature on this subject. In the presence of strings the system may well be even more complicated.
We would like to return to this problem in the future.

One issue which would be of great interest is the fate of the critical electric field for string pair production in presence of inhomogeneities.
For the holographic case of Section \ref{quattro} we saw no deviation from the locally constant approximation at large electric field, close to the critical one. This can be explained by the fact that the instanton world-sheet in this limit becomes very small and its radius approaches zero as $E \to E_{cr}$. The critical field is thus expected to remain the same and not affected by the inhomogeneities.
For the string suspended between the two branes (section \ref{tre}) the critical radius is instead going to infinity as $E$ approaches the critical field.  So we may expect some modification of the locally constant approximation close to the critical field. We believe that the protuberances we see in the numerical analysis are the first signs of what should happen in this regime. 

It is always tempting to challenge bounds. In the case of the critical electric field there could well be an indication that if the bound is violated for a short enough period the system will not destabilize. One interesting problem to explore would be for example to consider background electric fields that are bigger than the critical field but only for a short period of time, such as in a pulse or oscillating fashion.  This would presumably make the theory well defined, at least for some values of parameters (see for a similar problem \cite{roberto}).

\section*{Acknowledgments}

We thank  C. Bachas, Z. Komargodski and C. Schubert for discussions.
S.B. is founded by the program ``Rientro dei Cervelli Rita Levi Montalcini''. G.T. is funded by Fondecyt grant No. 3140122.
E. Rabinovici wants to thank KITP for hospitality during the
KITP Workshop ``Frontiers of Intense Laser Physics'' July 21, 2014 - Sep 19, 2014 as well as the support of the
Simons Distinguished professor Chair at KITP, and the
Chaires Internationales de Recherche Blaise Pascal, financ\'ee par l'Etat et la Region d'Ille-de-France, g\'er\'ee par la Fondation de l'Ecole Normale Superieure. His work is also partially supported by the American-Israeli Bi-National Science Foundation, the Israel Science Foundation Center of Excellence and the I Core Program of the Planning and Budgeting Committee and The Israel Science Foundation
“The Quantum Universe”.

\end{document}